\documentclass[amssymb,nofootinbib,nobibnotes,aps,prd,preprintnumbers]{revtex4}

 \usepackage{here}

\usepackage[dvipdfmx]{graphicx}
\usepackage{amssymb,amsfonts,amsmath,bm,color,multirow,cases,empheq,hyperref}
\usepackage{mathrsfs}

\definecolor{darkgreen}{rgb}{0,0.5,0.3}

\usepackage{enumerate}
\usepackage{ulem}
\usepackage{afterpage}

\hypersetup{%
 setpagesize=false,
 bookmarksnumbered=true,%
 bookmarksopen=true,%
 colorlinks=true,%
 linkcolor=blue,%
 citecolor=red}

\newcommand{\fr}[1]{\frac{1}{#1}}

\newcommand{\ord}[1]{{\mathcal O}(#1)}

\newcommand{\cC}{{\mathcal C}}

\newcommand{\nonum}{\nonumber\\ }

\newcommand{\sR}{{\sf R}}
\newcommand{\sq}{{\sf q}}
\newcommand{\sQ}{{\sf Q}}

\newcommand{\barH}{{\bar{H}}}
\newcommand{\barF}{{\bar{F}}}
\newcommand{\barJ}{{\bar{J}}}
\newcommand{\barD}{{\bar{D}}}

\newcommand{\cout}[1]{}

\newcommand{\arrayL}[1]{\left(\begin{array}{#1}}
\newcommand{\arrayR}{\end{array}\right)}
\newcommand{\arrayLb}[1]{\left[\begin{array}{#1}}
\newcommand{\arrayRb}{\end{array}\right]}

\newcommand{\tbeta}{{\tilde{\beta}}}
\newcommand{\tDelta}{\widetilde{\Delta}}
\newcommand{\tilf}{\tilde{f}}

\begin{document}

\title{New black ring with all independent conserved charges\\
in five-dimensional minimal supergravity
 }

\author{Ryotaku Suzuki}
\email{sryotaku@toyota-ti.ac.jp}
\author{Shinya Tomizawa}
\email{tomizawa@toyota-ti.ac.jp}
\affiliation{\vspace{3mm}Mathematical Physics Laboratory, Toyota Technological Institute\vspace{2mm}\\Hisakata 2-12-1, Tempaku-ku, Nagoya, Japan 468-8511\vspace{3mm}}

\begin{abstract}

We present  a new exact solution for a general non-BPS black ring in the bosonic sector of five-dimensional minimal supergravity. 
This obtained solution carries four independent conserved charges: the mass, two angular momenta, an electric charge, and an additional dipole charge related to other charges.
By employing the Ehlers-Harrison transformation, we derive this solution by transforming a five-dimensional vacuum solution into a charged solution in the theory. 
Previously, our work produced a vacuum doubly rotating black ring solution possessing a Dirac-Misner string singularity by using the Ehlers transformation.
 In this study, we use the singular black ring as the seed for the Harrison transformation.
The resultant solution is regular, free from curvature singularities, conical singularities, orbifold singularities, Dirac-Misner string singularities, and closed timelike curves both on and outside the horizon.
We show that within a specific parameter range, the black ring presents two branches for the same mass, two angular momenta and electric charge but these are distinguished by a dipole charge, which exhibits discontinuous non-uniqueness. 
Furthermore, this newly obtained black ring seamlessly connects to various physically significant solutions, such as the Pomeransky-Sen'kov black ring, the extremal black ring, the supersymmetric black ring, and the charged singly-spinning black ring.

\end{abstract}

\date{\today}
\preprint{TTI-MATHPHYS-31}

\maketitle

\section{Introduction}

When a rotating black ring interacts with the Maxwell field, it induces a type of dipole charge. 
Consequently, the dipole charge, which is not a conserved charge, serves as an additional parameter to characterize the black ring.
The first example of the dipole black ring solution was found by Emparan~\cite{Emparan:2004wy}, which is electrically coupled to a two-form or a dual magnetic one-form field. 
Subsequently, Elvang \textit{et al.}~\cite{Elvang:2004xi} developed further examples of dipole rings within five-dimensional minimal supergravity, based on a seven-parameter family of non-supersymmetric black ring solutions. 
However, this dipole black ring solution does not admit a supersymmetric configuration, additionally the dipole charge in this solution is not an independent parameter because it is related to other conserved charges, among four conserved charges and one dipole charge, yet only three are independent. 
As speculated by the authors in Ref.~\cite{Elvang:2004xi}, a more general non-Bogomol'nyi-Prasad-Sommerfield (BPS) black ring solution is likely to exist, described by its mass, two independent angular momenta, electric charge, and a dipole charge, which remains distinct from other asymptotic conserved charges. 
According to the uniqueness theorem for black rings in minimal supergravity detailed in Ref.~\cite{Tomizawa:2009tb}, assuming a trivial topology of ${\Bbb R} \times \{{\Bbb R}^4 \setminus D^2 \times S^2\}$ for the domain of outer communication, an asymptotically flat, stationary, and bi-axisymmetric black ring with a non-degenerate connected event horizon of cross-section topology $S^1 \times S^2$—if such exists—is uniquely characterized by its mass, electric charge, two independent angular momenta, dipole charge, and additional data on the rod structure like the ratio of the $S^2$ radius to the $S^1$ radius.

\medskip

The sigma model approach has proven to be a highly effective tool for deriving exact solutions to Einstein's equations~\cite{exact}. 
The existence of a single Killing vector simplifies the vacuum Einstein theory into a non-linear sigma model that is coupled with three-dimensional gravity through the process of dimensional reduction~\cite{Ernst:1967wx}. 
In this framework, the target space is represented by the coset $SL(2, \mathbb{R})/SO(2)$, enabling the generation of new solutions by executing group transformations on the coset representation of a seed solution. 
In a similar context, in four-dimensional Einstein-Maxwell theory possessing a single Killing vector, the corresponding coset space is $SU(2,1)/S[U(2) \times U(1)]$\cite{Ernst:1967by}. 
This $SU(2,1)$ symmetry introduces two important transformations: the Ehlers transformation~\cite{Ehlers1957} which introduces a nut charge, and the Harrison transformation~\cite{Harrison1968} which imparts an electric or magnetic charge,  to a given vacuum solution. 
The $SU(2,1)$ symmetry does not simply add rotation to a non-rotating solution while maintaining asymptotic flatness but in cases of stationary and axisymmetric solutions, this can be accomplished by the transformation using an appropriate linear combination of these two Killing vectors. 
For instance, the Kerr solution can be derived from the Schwarzschild solution by using such transformation~\cite{Clement:1997tx,Clement:1999bv}. 
Moreover, in $d$-dimensional ($d > 4$) vacuum Einstein theory with $(d-3)$ commuting Killing vectors, the coset structure is given by $SL(d-2,\mathbb{R})/SO(2,d-4)$, as shown by Maison~\cite{Maison:1979kx}. 
Particularly, for $d=5$, Giusto and Saxena~\cite{Giusto:2007fx} demonstrated that the subgroup of the $SL(3,\mathbb{R})$ can convert a static solution into a stationary solution possessing angular momentum while preserving asymptotic flatness.  
This Ehlers type transformation can actually transform the five-dimensional Schwarzschild solution into the five-dimensional Myers-Perry solution.

 \medskip
 Furthermore, for five-dimensional minimal supergravity, which corresponds to the Einstein-Maxwell-Chern-Simons theory with a particular coupling constant, the coset space can be represented as $G_{2(2)}/[SL(2,{\mathbb R}) \times SL(2,{\mathbb R})]$ or $G_{2(2)}/SO(4)$, depending on whether the chosen Killing vectors are one timelike and one spacelike, or two spacelike, respectively~\cite{Mizoguchi:1998wv,Mizoguchi:1999fu}. 
As developed by Bouchareb et al.~\cite{Bouchareb:2007ax}, this $G_{2(2)}$ symmetry can yield an electric Harrison transformation which converts a five-dimensional vacuum solution into an electrically charged one within five-dimensional minimal supergravity.
This transformation, when applied to the five-dimensional vacuum rotating black hole~\cite{Myers:1986un}, can derive the five-dimensional charged rotating black hole~\cite{Cvetic:1996xz}. 
However, as pointed out by the authors, a direct application of this Harrison transformation to the vacuum doubly rotating black ring~\cite{Pomeransky:2006bd} fails to produce a regular charged doubly spinning black ring solution, since an unavoidable Dirac-Misner string singularity appears.
In our prior work~\cite{Suzuki:2024coe}, we addressed the issue of Dirac-Misner string singularities and successfully constructed an exact solution for a charged rotating black ring. By applying the electric Harrison transformation to a vacuum rotating black ring solution with a Dirac-Misner string singularity, we obtained a charged rotating black ring. We then adjusted parameters to eliminate the Dirac-Misner string singularity, resulting in a regular charged rotating dipole black ring free from curvature singularities, conical singularities, Dirac-Misner string singularities, orbifold singularities, and closed timelike curves (CTCs).
Moreover, this transformation facilitated the generation of the first non-BPS exact solution that represents an asymptotically flat, stationary spherical black hole, referred to as a capped black hole, which has a domain of outer communication (DOC) with a nontrivial topology, specifically, on a timeslice, the topology $[{\mathbb R}^4 \# {\mathbb C}{\mathbb P}^2] \setminus {\mathbb B}^4$~\cite{Suzuki:2023nqf, Suzuki:2024phv}.

 \medskip
The purpose of this paper is to present a new exact solution describing a charged rotating black ring with all independent conserved charges in five-dimensional minimal supergravity, using the electric Harrison transformation~\cite{Bouchareb:2007ax}. In our previous work~\cite{Suzuki:2024eoz}, we constructed a vacuum solution of a doubly rotating black ring with a Dirac-Misner string singularity in five-dimensional Einstein theory using the Ehlers transformation~\cite{Giusto:2007fx}. We then eliminated the singularity by appropriately adjusting the solution's parameters, ultimately deriving the Pomeransky-Sen'kov black ring solution.
In this paper, we use the vacuum singular black ring solution with a Dirac-Misner string singularity as a seed for the Harrison transformation, deriving a charged solution that initially possesses the Dirac-Misner string singularity. 
After the Harrison transformation, we impose appropriate boundary conditions to ensure regularity, eliminating the Dirac-Misner string singularity inside the black ring. The resulting solution is regular, without curvature singularities, conical singularities, Dirac-Misner string singularities, orbifold singularities on and outside the horizon, or CTCs.
The charged rotating black ring solution previously found by Elvang, Emparan, and Figueras~\cite{Elvang:2004xi}, and another black ring solution constructed in our previous work~\cite{Suzuki:2024coe}, have the mass, two angular momenta, an electric charge, and a dipole charge, with only three of these quantities being independent. In contrast, our obtained black ring possesses four independent conserved charges: its mass, two angular momenta, an electric charge, and a dipole charge which is related to the conserved charges.

\medskip
The remainder of this paper is dedicated to constructing the aforementioned black ring solution. 
In Sec.~\ref{sec:setup}, we will briefly review two solution-generation methods, the Ehlers transformation and the Harrison transformation. 
The former and the latter add an angular momentum and an electric charge to a five-dimensional vacuum solution, respectively, preserving asymptotic flatness. 
In Sec.~\ref{sec:seed_Ehlers}, we will briefly explain the construction of the vacuum solution through the Ehlers transformation, which was proven to become a vacuum doubly rotating black ring (Pomeransky-Sen'kov solution) after removing the Dirac-Misner string singularity in our previous work~\cite{Suzuki:2024eoz}.
In Sec.~\ref{sec:seed_Harrison}, we will introduce the constructed vacuum solution before removing the Dirac-Misner string singularity. In Sec.~\ref{sec:sol}, we will apply the Harrison transformation to the singular vacuum solution, leading to a corresponding charged solution in five-dimensional minimal supergravity.
By setting appropriate boundary conditions on the parameters, we will ensure the absence of any singularities—including curvature, conical, and Dirac-Misner string singularities—on the rotational axes and the horizon. This will allow us to derive a charged rotating black ring solution with four independent conserved charges, the mass, two angular momenta, electric charge, and additionally a dipole charge related to other charges. 
In Sec.~\ref{sec:property}, we discuss the phase of the obtained solution.

\section{setup}\label{sec:setup}

Let us begin with a basic setup for  asymptotically flat, stationary and bi-axisymmetric solutions in the bosonic sector of the five-dimensional minimal ungauged supergravity (Einstein-Maxwell-Chern-Simons theory), whose action takes the form
\begin{eqnarray}
S=\frac{1}{16 \pi G_5}  \left[ 
        \int d^5x \sqrt{-g}\left(R-\frac{1}{4}F^2\right) 
       -\frac{1}{3\sqrt{3}} \int F\wedge F\wedge A 
  \right] \,, 
\label{action} 
\end{eqnarray} 
where $F=dA$. The field equation consists of the Einstein equation 
\begin{eqnarray}
 R_{\mu \nu } -\frac{1}{2} R g_{\mu \nu } 
 = \frac{1}{2} \left( F_{\mu \lambda } F_\nu^{ ~ \lambda } 
  - \frac{1}{4} g_{\mu \nu } F_{\rho \sigma } F^{\rho \sigma } \right) \,, 
 \label{Eineq}
\end{eqnarray}
and the Maxwell equation with a Chern-Simons term
\begin{eqnarray}
 d\star F+\frac{1}{\sqrt{3}}F\wedge F=0 \,. 
\label{Maxeq}
\end{eqnarray}
Assuming the existence of one timelike Killing vector $\xi_0 = \partial/\partial t$ and one spacelike axial Killing vector $\xi_1=\partial/\partial\psi$, this theory reduces to the $G_{2(2)}/SL(2,{\Bbb R})\times SL(2,{\Bbb R})$ non-linear sigma models coupled to three-dimensional gravity~\cite{Mizoguchi:1998wv,Mizoguchi:1999fu}.
Under a further assumption of the presence of the third spacelike axial Killing vector $\xi_2=\partial/\partial\phi$, i.e., the existence  of three mutually commuting Killing vectors, the metric can be written in the Weyl-Papapetrou form\footnote{If one chooses two axial Killing vectors for the reduction, then $\lambda_{ab}$ has the Riemanian signature and one has to flip the sign of the line elements $d\rho^2+dz^2$ as in Refs.~\cite{Tomizawa:2009ua,Tomizawa:2009tb}. }
\begin{eqnarray}
ds^2
&=& \lambda_{ab}(dx^a+a^a{}_\phi{}d\phi)(dx^b+a{}^b{}_\phi d\phi)+\tau^{-1}\rho^2 d\phi^2
     +\tau^{-1}e^{2\sigma}(d\rho^2+dz^2) \,, \label{eq:WPform}
\end{eqnarray}
and the gauge potential is written, 
\begin{eqnarray} 
 A = \sqrt{3}\psi_a dx^a + A_\phi d\phi \,, 
\label{pote:gauge}
\end{eqnarray}
where the coordinates $x^a=(t,\psi)$ ($a=0,1$) denote the Killing coordinates, and thus all functions $\lambda_{ab}$, 
$\tau:=-{\rm det}(\lambda_{ab})$, $a^a$, $\sigma$, and $(\psi_a,A_\phi)$ are independent of $\phi$ and $x^a$. 
Note that the coordinates $(\rho,z)$ that span a two-dimensional base space, $\Sigma=\{(\rho,z)|\rho\ge 0,\ -\infty<z<\infty \}$, 
are globally well-defined, harmonic, and mutually conjugate on $\Sigma$. 

\medskip
As discussed in Ref.~\cite{Tomizawa:2009ua}, by using Eqs.~(\ref{Eineq}) and (\ref{Maxeq}), we can introduce the magnetic potential $\mu$ and twist potentials $\omega_a$ by   
\begin{eqnarray} 
d\mu&=&\frac{1}{\sqrt{3}}\star (\xi_0\wedge \xi_1\wedge F) 
    - \epsilon^{ab}\psi_ad\psi_b \,, 
\label{eq:mu} 
\\
 d\omega_a&=&\star (\xi_0\wedge \xi_1 \wedge d\xi_a)+\psi_a(3d\mu+\epsilon^{bc}\psi_bd\psi_c)\,, 
\label{eq:twistpotential} 
\end{eqnarray} 
where $\epsilon^{01}=-\epsilon^{10}=1$.
The metric functions $a^a{}_{\phi}\ (a=0,1)$ and the component $A_\phi$ of the gauge potential are determined by the eight scalar functions $\{\lambda_{ab},\omega_a,\psi_a,\mu\}$ from Eqs.~(\ref{eq:mu}) and (\ref{eq:twistpotential}), and the function $\sigma$ is also determined by these scalar functions up to a constant factor. 
Then, the action~(\ref{action}) reduces to the nonlinear sigma model  for the eight scalar functions $\{\lambda_{ab},\omega_a,\psi_a,\mu\}$ invariant under the $G_{2(2)}$-transformation.

\subsection{Ehlers transformation}\label{sec:GStr}

When one constructs a new asymptotically flat vacuum solution from a certain vacuum seed by using the $SL(3,\mathbb{R})$ transformation, it is convenient to use a special $SL(3,\mathbb{R})$ transformation  that preserves the asymptotic metric at infinity.  
More precisely, choosing the two Killing vectors $(\xi_0, \xi_1) = (\partial / \partial t, \partial / \partial \psi)$ or $(\xi_0, \xi_1) = (\partial / \partial t, \partial / \partial \phi)$ results in only trivial transformations but selecting other Killing vectors $(\xi_0, \xi_1) = (\partial / \partial t, \partial / \partial \phi_+)$ or $(\xi_0, \xi_1) = (\partial / \partial t, \partial / \partial \phi_-)$, where $\phi_\pm := \psi \pm \phi$, yields a physical transformation that adds angular momentum to a static seed solution. 
Under this transformation, the set of the scalar fields $\Phi^A = \{\lambda_{ab},\omega_a,\psi_a=0,\mu=0\}$ is changed  to the set of different  ones $\Phi'^A = \{ \lambda_{ab}',\omega_a',\psi'_a=0,\mu'=0\}$ as follows
\begin{align}\label{eq:scalar-trans}
\begin{split}
&\lambda'_{00} = \tilde{D}^{-1}\left[ \lambda _{00}-2 \tbeta  \lambda _{01}+\tbeta ^2 \left(\lambda _{11}+\epsilon^{ij}\omega_i \lambda_{j0}\right)-\tbeta ^3 \epsilon^{ij} \omega_i \lambda_{j1} 
+\frac{\tbeta ^4 \left(\tau 
   \left(-\omega _0^2+\tau  \lambda _{00}\right)+\left(\epsilon^{ij}\omega_i \lambda_{j0}\right){}^2\right)}{4 \lambda _{00}}\right],\\
&\lambda'_{01}=\tilde{D}^{-1}\left[ \lambda_{01}-\tbeta  \left(\lambda_{11}+\epsilon^{ij}\omega_i \lambda_{j0}\right)+\frac{3}{2} \tbeta ^2 \epsilon^{ij}\omega_i \lambda_{j1}
-\frac{\tbeta ^3 \left(\tau    \left(-\omega _0^2+\tau  \lambda_{00}\right)+\left( \epsilon^{ij} \omega_i \lambda_{j0}\right){}^2\right)}{2 \lambda_{00}} \right],\\
&\lambda'_{11}=\tilde{D}^{-1} \left[\lambda_{11}-2 \tbeta \epsilon^{ij} \omega_i \lambda_{j1} +\frac{\tbeta ^2 \left(\tau  \left(-\omega _0^2+\tau  \lambda
   _{00}\right)+\left(\epsilon^{ij}\omega_i \lambda_{j0}\right){}^2\right)}{\lambda_{00}}\right],\\
&\omega'_0=\tilde{D}^{-1}\left[\omega _0+\tbeta  \left(-\omega _1-\omega _0^2+\tau  \lambda
   _{00}\right)+\frac{3}{2} \tbeta ^2 \left(\omega _0 \omega _1-\tau  \lambda
   _{01}\right)+\frac{1}{2} \tbeta ^3 \left(-\omega _1^2+\tau\lambda _{11} 
   \right)\right],\\
&\omega'_1=\tilde{D}^{-1}\left[\omega _1+\tbeta  \left(-\omega _0 \omega _1+\tau  \lambda
   _{01}\right)+\frac{1}{2} \tbeta ^2 \left(\omega _1^2-\tau\lambda _{11} 
   \right)\right],\\
& \tau' = \tau/\tilde{D},   
\end{split}
\end{align}
with
\begin{align}
\tilde{D} = 1-2 \tbeta  \omega _0+\tbeta ^2 \left(\omega _1+\omega _0^2-\tau  \lambda
   _{00}\right)-\tbeta ^3 \left(\omega _0 \omega _1-\tau  \lambda
   _{01}\right)+\frac{1}{4} \tbeta ^4 \left(\omega _1^2- \tau\lambda _{11}
   \right),
\end{align}
where the new parameter $\tilde{\beta}$ is physically related to the angular momentum with respect to $\partial/\partial{\phi_+}$ or $\partial/\partial{\phi_-}$, depending on the choice of the Killing vector $\xi_1 = \partial/\partial{\phi_+}$ or $\partial/\partial{\phi_-}$. 
The one-forms $a^{\prime a}{}_i dx^i$ $(a = 0,1)$ for the transformed solution are determined by the five scalar functions ${\lambda'_{ab}, \omega'_a}$ from Eq.~(\ref{eq:twistpotential}), after replacing ${\lambda_{ab}, \omega_a}$ with ${\lambda'_{ab}, \omega'_a}$ and substituting Eq.~(\ref{eq:scalar-trans}).

\medskip
Thus, one can obtain the new metric describing the rotating solution to the vacuum Einstein equations with the same Killing isometries.
As shown previously in Ref.~\cite{Bouchareb:2007ax}, applying this transformation to the five-dimensional static black hole solution (the five-dimensional Schwarzschild-Tangherlini solution~\cite{T-schwarzschild}) generates the five-dimensional rotating black hole solution (the five-dimensional Myers-Perry solution~\cite{Myers:1986un}). 
As will be seen in the following section, however, when applied to vacuum black rings, such as the static black ring and the singly rotating black ring, a Dirac-Misner string singularity inevitably appears on the disk inside the ring, similar to the Harrison transformation studied in Ref.~\cite{Bouchareb:2007ax}.
In the following section, to solve this undesirable problem, we will choose the vacuum rotating black ring having a Dirac-Misner string singularity on the disk inside the ring as the seed of the transformation, as performed in our previous work~\cite{Suzuki:2024coe} for the Harrison transformation.

\subsection{Harrison transformation}
In particular, utilizing the $G_{2(2)}$ symmetry, Ref.~\cite{Bouchareb:2007ax} constructed the electric Harrison transformation preserving asymptotic flatness that transforms a five-dimensional vacuum solution $\{\lambda_{ab},\omega_a,\psi_a=0,\mu=0\}$ into a charged solution $\{\lambda'_{ab},\omega'_a,\psi'_a,\mu'\}$ in the five-dimensional minimal supergravity, which is given by
\begin{align}
\begin{split}\label{eq:ctrans-metric}
&\tau' = D^{-1} \tau,\quad
\lambda'_{00} = D^{-2} \lambda_{00},\quad
 \lambda'_{01} =D^{-2} (c^3 \lambda_{01}+s^3 \lambda_{00} \omega_0),\\
& \lambda'_{11} = -\frac{\tau D}{\lambda_{00}} + \frac{(c^3 \lambda_{01}+s^3 \omega_0 \lambda_{00})^2}{D^2 \lambda_{00}},\\
&\omega'_0 = D^{-2}\left[c^3(c^2+s^2+2s^2 \lambda_{00})\omega_0-s^3 (2c^2+(c^2+s^2)\lambda_{00}) \lambda_{01}\right],\\
& \omega'_1 = \omega_1 + D^{-2} s^3 \left[-c^3 \lambda_{01}^2+s(2c^2-\lambda_{00})\lambda_{01} \omega_{0}-c^3 \omega_0^2\right],\\
& \psi'_0 = D^{-1} sc\, (1+\lambda_{00}),\quad
\psi'_1 = D^{-1} sc\, (c\lambda_{01}-s \omega_0),\\
&\mu' = D^{-1}sc\,(c\,\omega_0-s\lambda_{01}),
\end{split}
\end{align}
with
\begin{align}
 D  = c^2+ s^2 \lambda_{00} = 1 + s^2 (1+\lambda_{00}),
\end{align}
where the new parameter $\alpha$ in $(c,s):=(\cosh\alpha,\sinh\alpha)$ is related to the electric charge.
The functions $a^{\prime a}{}_{\phi}\ (a=0,1)$ and the component $A'_\phi$ for the charged solution are determined by the eight scalar functions $\{\lambda'_{ab},\omega'_a,\psi'_a,\mu'\}$ from  Eqs.~(\ref{eq:mu}) and (\ref{eq:twistpotential}) after the replacement of $\{\lambda_{ab},\omega_a,\psi_a,\mu\}$ with $\{\lambda'_{ab},\omega'_a,\psi'_a,\mu'\}$, 
and thus one can obtain the new metric and gauge potential that describe the charged solution for Eqs.~(\ref{Eineq}) and (\ref{Maxeq}).
This transformation adds the electric and dipole charges to a vacuum black ring solution while keeping the asymptotic flatness and Killing isometries.
However, as mentioned in Ref.~\cite{Bouchareb:2007ax}, when one performs the Harrison transformation for the regular vacuum black ring such as the Pomeransky-Sen'kov solution, a Dirac-Misner string singularity inevitably appears on the disk inside the ring, though the transformation can generate the regular Cveti\v{c}-Youm charged black hole for the vacuum black hole such as the Myers-Perry solution.  
In the following, to solve this undesirable problem, we will use the vacuum rotating black ring having a Dirac-Misner string singularity on the disk inside the ring as the seed of the Harrison transformation.

\section{ Vacuum seed solution for Ehlers transformation}\label{sec:seed_Ehlers}

In our earlier work~\cite{Suzuki:2024eoz}, utilizing the Ehlers transformation, reviewed briefly in Sec.~\ref{sec:GStr}, the exact solution for a doubly rotating black ring in five-dimensional vacuum Einstein theory was derived. 
As previously demonstrated by Giusto and Saxena~\cite{Giusto:2007fx},  this Ehlers transformation converted the five-dimensional Schwarzschild black hole into the Myers-Perry black hole.
Nevertheless, in contrast to the black hole case, a simple, direct application of this method to either the static black ring or the Emparan-Reall black ring failed to produce a regular rotating black ring, due to the emergence of a Dirac-Misner string singularity. 
To rectify this issue in Ref.~\cite{Suzuki:2024eoz}, a singular vacuum solution of a rotating black ring/lens, already containing a Dirac-Misner string singularity, was employed as the seed solution for the Ehlers transformation. 
After applying the transformation to the seed, we chose the solution's parameters  appropriately to ensure that the resulting solution is free from Dirac-Misner string singularity. 
Consequently, the final solution represents a doubly rotating black ring with its regularity—it does not have curvature singularities, conical singularities, Dirac-Misner string singularities, and orbifold singularities on the rods. 
It was demonstrated that this solution, derived through the Ehlers transformation, aligned precisely with the Pomeransky-Sen'kov solution.
Thus, in this previous paper, we succeeded in the construction of a vacuum, regular doubly rotating black ring.  In this paper, however, we do not eliminate the Dirac-Misner string singularity.
Instead, in the following section, we will use the vacuum doubly rotating black ring with the Dirac-Misner string singularity as a new seed for the Harrison transformation.

\section{ Vacuum seed solution for Harrison transformation}\label{sec:seed_Harrison}

In our previous work~\cite{Suzuki:2024coe}, as a vacuum seed for the Harrison transformation, we chose the vacuum rotating black ring which initially has a Dirac-Misner string singularity. 
We did not remove this singularity before the Harrison transformation but we eliminated it by appropriately adjusting the parameters of the solution after the transformation.
Now, to construct a new charged black ring solution, we follow the same procedure as in Refs.~\cite{Suzuki:2024coe,Suzuki:2024phv,Suzuki:2023nqf}. 
We use the vacuum doubly rotating black ring with a Dirac-Misner string singularity~\cite{Suzuki:2024eoz} as the seed for the Harrison transformation.

\medskip
The vacuum solution in Ref.~\cite{Suzuki:2024eoz} is given by
\begin{align}
&ds^2 = - \frac{\barH(y,x)}{\barH(x,y)} (dt+\bar{\Omega}_\psi(x,y) d\psi +\bar{\Omega}_\phi(x,y) d\phi)^2+\frac{1}{\barH(y,x)}\left[\barF(y,x)d\psi^2-2\barJ(x,y)d\psi d\phi-\barF(x,y)d\phi^2\right]\nonum
&\quad + \frac{\ell^2  \barH(x,y)}{4(1-\gamma)^3(1-\nu^2)(1-a^2)\Delta(x-y)^2}\left(\frac{dx^2}{G(x)}-\frac{dy^2}{G(y)}\right),
\label{eq:metricsol-vac}
\end{align}

where the metric functions become
\begin{align}
\label{eq:defGx}
G(u) &= (1-u^2)(1+\nu u),\\
\barH(x,y) &=(1+y)^2 \biggr[ \nu  d_1 f_7^2 (1-\nu )  \left(1-x^2\right) \left(2 (\gamma -1) \nu +c_3\right)+ 2 c_3 c_1^2 f_6^2 (\gamma -\nu) (\nu  x+1)^2 \nonum
&\quad  -(\gamma -\nu ) (1-\nu )  \left(c_1^2 c_3 f_6^2-2 (1-\gamma ) \nu  \left(c_1-b (1-\nu   )\right){}^2 f_8^2\right) (1-x) (1+x \nu )\biggr]\nonum
&  -(1+y)  \biggr[ d_5 g_6 (1-x)^2  +d_6 g_7 \left(1-x^2\right)    +2(1-\nu)^{-1} d_6  g_8(1+x)  (1+x \nu )  \biggr] \nonum
& + 4 \left(1-a^2\right) (1-x) (1-\gamma )^3 (1-\nu )^4 \Delta-4 (1+x) (1-\gamma ) (1-\nu )^2 (1+\nu ) d_1 f_2^2+2 \left(1-x^2\right) (1-\gamma ) (1-\nu
   )^2 c_2^2 f_5^2,  \label{eq:Hxy}\\
\barF(x,y)&=\frac{\ell^2 \tilde{v}_0^2(1-\gamma)}{(\gamma^2-\nu^2) (x-y)^2}\biggr[
4 (1+y \nu ) G(x) \left\{ \left(1-a^2\right)^2 (-1+y) (1-\gamma )^3 (1-\nu )^3 \Delta^2-(1+y) d_1^2 f_1^2 f_2^2\right\}\nonum
&+4 (1+x) (1+x \nu )  G(y) \left((1-a b) (\gamma -\nu ) (1+\nu ) c_1 -(1-\nu ) c_2 \right){}^2 g_{5}^2\nonum
&+ \nu^{-1} (1-\nu )^3 (\gamma -\nu ) \biggr\{\left(y^2-1\right) G(x) c_3^2 f_3^2 f_4^2+\left(1-x^2\right) G(y) d_3^2 g_2^2\biggr\}\nonum
& + G(x) G(y) \biggr\{\frac{x (\gamma -\nu ) (c_1 c_3f_3 f_6-b d_1f_1f_7)^2 }{1-\gamma }-(a-b)^2 y (1-\gamma ) (\gamma -\nu ) c_2^2
   f_5^2 f_8^2+\frac{(1-a^2) d_4 \Delta g_4}{\nu }\biggr\}   
\biggr],\\
\barJ(x,y)& =  \frac{ \ell^2 \tilde{v}_0^2 (1-\gamma)(1+x)(1+y)}{(\gamma^2-\nu^2)(x-y)}
\biggr[4 d_1 f_1 f_2 g_{5} \biggr\{(1-a b) (\gamma -\nu ) (1+\nu ) c_1 -(1-\nu ) c_2 \biggr\} (1+x \nu ) (1+y \nu ) \nonum
& - (1-x) (1-y) (\gamma -\nu ) \biggr\{(a-b) (1+x \nu ) (1+y \nu ) c_2 f_5 f_8 \left(c_1 c_3 f_3 f_6 -b d_1 f_1 f_7 \right) 
+(1-\nu )^3 c_3   d_3 f_3 f_4 g_2\biggr\}
\biggr],\\
\bar{\Omega}_\psi (x,y)&= \frac{\ell \tilde{v}_0 (1+y)}{ \barH(y,x)}\biggr[
\nu^{-1}c_2 f_5 \left(c_1 c_3 f_4 f_{16}-b d_1 f_2 f_{13}\right)  (1-x) (1-\nu ) (1+x \nu ) (1+y \nu ) \nonum
&+ (1+y \nu ) \left\{(1+x) (1+x \nu ) c_1 c_3 f_2 f_9 f_{10}-\frac{1}{2} b \left(1-x^2\right) (\gamma -\nu ) (1-\nu ) c_3 f_4 f_{10} f_{12}\right\}\nonum
&-2 (1+x) (1-\nu ) (1+x \nu ) d_1 f_2 f_{11}  f_{15} \left(c_3-(1-a b) (1-\gamma ) (1+\nu ) \right)\nonum
&+\nu^{-1} c_3 \left(d_1-d_2\right) f_4 g_3(1-x) (1-\nu )^2 (1+x \nu ) +c_3 d_2 f_4 g_{10} (1-\nu )^2\left(1-x^2\right) 
 \biggr],\\
\bar{\Omega}_\phi(x,y) &= \frac{\ell \tilde{v}_0 (1+x)}{\barH(y,x)}\biggr[
b (1+x) (1+y) (1+y \nu ) d_1 d_2 f_1 f_{13} f_{14}\nonum
& +\frac{(1+x) \left( 1-y^2\right) (1-\nu ) \nu  c_3 \left(b (1+\nu ) d_1-2 (a-b) (1-\gamma )^2 (1-\nu ) \nu \right) f_3 g_{11}}{1+\nu }\nonum
&+\frac{(y-1) (1-\gamma ) (1-\nu ) \nu  (1+y \nu + x (1+y (4-3 \nu ))) c_3 f_1 f_3 f_{10}}{1+\nu }\nonum
&+\frac{2 (a-b) (1-\gamma )^2 \left((y-1) (x+y) (1-\nu )^3 \nu  c_3 f_3 g_{9}+2 (1+x \nu ) (1+y \nu )^2 d_1 f_1 g_{1}\right)}{1+\nu }
\biggr],
\end{align}

where 
\begin{align}\label{eq:def-cfs}
\begin{split}
&\tilde{v}_0 := \frac{v_0}{\sqrt{\Delta}} := \sqrt{\frac{2(\gamma^2-\nu^2)}{(1-a^2)(1-\gamma)\Delta}},\\
&c _1 :=(1-\gamma) a+(\gamma-\nu) b ,\\
&c _2 := 2 a (1-\gamma ) \nu +b (\gamma -\nu ) (1+\nu ),\\
&c _3 := 2   (1-\gamma ) \nu +b^2 (\gamma -\nu ) (1+\nu ),\\
&d_1 := (\nu +1) c _1^2-(1-\gamma ) (1-\nu )^2,\\
&d_2:= b (\nu +1) c _1 (\gamma -\nu )+2\nu (1-\gamma )   (1-\nu ) ,\\
&  d_3 := \left(a^2-1\right) b (\gamma -1) (\nu +1)-a c _3,\\
&d_4 := b^2 (\gamma -\nu )   \left[(\nu +1)^2 c _1^2 \left(-3 (1-\gamma ) \nu -\nu ^2+1\right)-(1-\gamma ) (1-\nu )^4 (2 \nu +1)\right]\\
&\quad +(1-\gamma   ) \left[\left((1-\nu ) c _2-2 \nu ^2 c _1\right)^2-4 \nu ^2 c _1^2 \left(-\gamma  (\nu +2)+3 \nu
   ^2+1\right)\right],\\
& d_5 := (1-\gamma ) (1-\nu )^3 [(\gamma -3 \nu ) \left(b^2 (\nu -1) (\gamma -\nu )-c_1^2\right)
-2 b c_1 (3 \nu -1) (\gamma -\nu )],\\
& d_6 := c_1(1-\gamma ) (1-\nu) \left(1-\nu ^2\right) \left(c_2-(1-\gamma ) (a-b) (\gamma -\nu )\right),\\
&\Delta := f_8^2 - \frac{d_1(f_1^2-f_8^2)}{(1-a^2)(1-\gamma)(1-\nu^2)}+\frac{c_3 (f_3^2-f_8^2)}{(1-a^2)(1-\gamma)(1+\nu)},
   \end{split}
\end{align}
and the metric depends on the parameter $\beta:= \ell v_0 \tilde{\beta}/4$  through coefficients $\{ f_i\}_{i=1,\dots 16}$ and $\{ g_i\}_{i=1,\dots 11}$, which are quadratic and quartic polynomials of $\beta$ presented in Appendix~\ref{sec:fi-gi}, respectively.

\medskip
We assume that the coordinates $(t,\psi,\phi,x,y)$ run in the ranges,
\begin{eqnarray}
-\infty < t <\infty, \quad 0\leq \psi\le 2\pi, \quad 0\le \phi \leq 2\pi, \label{eq:tphipsirange}
\end{eqnarray}
and
\begin{align}
 -1 \leq x \leq 1,\quad -1/\nu \leq y \leq -1. \label{eq:xyrange}
\end{align}
The solution is described by six parameters $(\ell,\nu,\gamma,a,b,\beta)$, among which the parameters, $\ell,\nu,\gamma$, are assumed to lie within the following range 
\begin{align}\label{eq:rangenugam}
\ell>0,\quad  0 < \nu < \gamma < 1.
\end{align}
Furthermore, the metric functions take real values only if $\tilde{v}_0^2$ is positive.
This imposes another restriction on the parameters, which is less obvious than the above two
\begin{align}\label{eq:necessary-1}
(1-a^2)\Delta >0.
\end{align}

\subsection{Asymptotic infinity and Rod structure}

The boundaries in the C-metric coordinates $(x,y)$ correspond to the rods and asymptotic infinity, whose structures are given as follows:

\begin{enumerate}[(i)]
\item

$\phi$-rotational axis : 
$\partial \Sigma_\phi=\{(x,y)|x=-1,-1/\nu <y<-1 \}$ with the rod vector 
$v_\phi:=(0,0,1)=\partial/\partial \phi$, where  we impose the periodicity of the coordinate $\phi$ as $\phi\sim \phi +2\pi$ to ensure the absence of the conical singularities on $\partial \Sigma_\phi$, 
\item Horizon: 
$\partial \Sigma_{\cal H}=\{(x,y)|-1<x<1,y=-1/\nu \}$ with the rod vector $
 v_{\cal H} := (1, \omega^{\rm vac}_\psi, \omega^{\rm vac}_\phi),
$ 
where the constants $\omega^{\rm vac}_\psi$, $\omega^{\rm vac}_\phi$ are expressed by
\begin{align}
\omega^{\rm vac}_\psi= \frac{(\gamma-\nu)}{ \ell \tilde{v}_0 f_4 g_3 (1-a^2+a(a-b)\gamma-(1-ab)\nu)},
\quad \omega^{\rm vac}_\phi = -\omega^{\rm vac}_\phi \frac{g_2 d_3}{f_3 c_3},
\end{align}

\item

Inner axis: 
$\partial \Sigma_{\rm in}=\{(x,y)|x=1,-1/\nu <y<-1 \}$ 
with the rod vector 
\begin{align}
 v_{\rm in}=(v_{\rm in}^{\rm vac,t},v_{\rm in}^{\rm vac,\psi},1),
\end{align}
where
\begin{align}
v_{\rm in}^{\rm vac,t} &=\frac{\ell \tilde{v}_0 (a-b) g_1'}{f_1},\\
 v_{\rm in}^{\rm vac,\psi} &=\frac{\Delta}{f_1f_2}\left(\frac{ad_1+(1-\gamma)(1+\nu)(1-a^2)c_1}{d_1}-1\right)+1,
\end{align}
and $g_1'$ are defined by
\begin{align}\label{eq:def-g1dash}
g_1' :=  g_{1}+\frac{\nu c_3 f_3 f_{10} }{(1-\gamma ) d_1 (\nu +1) (a-b)}= f_{13} f_{14}+ \frac{c_3 f_3 f_{10}}{2 (a-b) (1-\gamma ) d_1}-\frac{c_1 c_3\left(f_{13} f_{14}-f_{16}   f_{19}\right)}{2 (a-b) (1-\gamma )^2 \nu }.
\end{align}
Here $v^{\rm vac,t}_{\rm in} \neq 0$ implies the existence of the Dirac-Misner string singularity on the axis. 
We do not remove this singularity before the Harrison transformation, but we do so afterward.

\item

$\psi$-rotational axis: 
$\partial \Sigma_\psi=\{(x,y)|-1<x<1,y=-1 \}$
with the rod vector $v_\psi :=(0,1,0)=\partial/\partial \psi$, where 
we impose the periodicity of the coordinate $\psi$ as $\psi\sim \psi +2\pi$ to ensures the absence of the conical singularities on $\partial \Sigma_\psi$.

\item Infinity:  
$\partial \Sigma_\infty =\{(x,y)|x\to y \to -1 \}$:  
By
introducing the coordinates $(r,\theta)$ as
\begin{align}
 x = -1 + \frac{4 \ell^2 (1-\nu) \cos^2\theta }{r^2},\quad y = -1 - \frac{4\ell^2 (1-\nu) \sin^2\theta}{r^2},\label{eq:asym-xy}
\end{align}
one can see that the metric asymptotes to the five-dimensional Minkowski metric at $x\to y\to -1$ as
\begin{align}\label{eq:asym-charges}
ds^2 &= -\left(1-\frac{8G_5 M^{\rm vac}}{3\pi r^2}\right)dt^2-\frac{8G_5 J^{\rm vac}_\psi \sin^2\theta}{\pi r^2}dt d\psi-\frac{8G_5 J^{\rm vac}_\phi \cos^2\theta}{\pi r^2}dt d\phi\nonum
& +dr^2 + r^2\sin^2\theta d\psi^2+r^2 \cos^2\theta d\phi^2+r^2d\theta^2,
\end{align}
where the ADM mass $M^{\rm vac}$ and two ADM angular momenta $J^{\rm vac}_\psi$, $J^{\rm vac}_\phi$ can be written as
\begin{align}
 M^{\rm vac} &=  \frac{3  \pi \ell^2 \tilde{v}_0^2  \left(d_1-(1-\nu ) c_3\right)}{8  G_5 (1-\gamma ) (1+\nu )(\gamma-\nu)} \biggr[
f_8^2+\frac{(1+\nu ) c_2^2 \left(f_8^2-f_5^2\right)}{(1-\nu ) (\gamma +\nu ) \left(d_1-(1-\nu )   c_3\right)}\nonum
   &\quad +\frac{(1+\nu ) d_1 \left((1-\gamma ) \left(f_8^2-f_1^2\right)-(1+\nu )   \left(f_8^2-f_2^2\right)\right)}
    {(1-\nu ) (\gamma +\nu ) \left(d_1-(1-\nu ) c_3\right)} \biggr], \label{eq:mass-vac}\\
  J^{\rm vac}_\psi &= -\frac{ \pi \ell^3\tilde{v}^3_0 \left(c_3 f_4 g_3\left(d_1-d_2\right) +c_2 f_5 \left(c_1 c_3 f_4 f_{16}-b d_1 f_2 f_{13}\right)\right)}
  {8 G_5 \nu  (1-\gamma   )^2 (\gamma^2-\nu^2)  }, \label{eq:Jpsi-vac}\\
  J^{\rm vac}_\phi &= -\frac{\pi  \ell^3\tilde{v}^3_0 \left(  2 \nu  c_3 f_3 f_1 f_{10}-(2 \nu  c_3 f_3 g_{9}+ d_1 f_1   g_{1})(a-b) (1-\gamma ) (1-\nu )\right)}{4 G_5 (1-\gamma ) (1-\nu^2 ) (\gamma^2-\nu^2)}.\label{eq:Jphi-vac}
\end{align}

\end{enumerate}

\subsection{Conditions for a regular vacuum black ring}
To obtain the horizon with the ring topology, one must impose  
\begin{align}
 v_{\rm in}^{\rm vac,\psi} &=\frac{\Delta}{f_1f_2}\left(\frac{ad_1+(1-\gamma)(1+\nu)(1-a^2)c_1}{d_1}-1\right)+1 =0.\label{eq:topology-con}
\end{align}
The absence of the Dirac-Misner string singularity requires
\begin{align}
 (a-b) g_1' = 0.
\end{align}
With these two conditions, $x=1$ becomes a regular $\phi$-rotational axis without the conical singularity if 
\begin{align}
\left(\frac{\Delta \phi}{2\pi}\right)^2 = \frac{d_1^2 f_1^2 f_2^2}{(1-a^2)^2(1-\gamma)^3(1-\nu)^2(1+\nu)\Delta^2} =1. \label{eq:conifree}
\end{align}
In Ref.~\cite{Suzuki:2024eoz}, we showed that the metric~(\ref{eq:metricsol-vac}) reduces to the Pomeransky-Sen'kov metric under these three conditions.
Below, we will see how these conditions are changed by the Harrison transformation.

\section{Charged rotating black ring with all independent conserved charges}\label{sec:sol}

Applying the Harrison transformation, as outlined in Eq.~(\ref{eq:ctrans-metric}), to the vacuum seed solution given in Eq.~(\ref{eq:metricsol-vac}), we can derive the corresponding charged solution presented below in Eq.~(\ref{eq:metricsol}). 
As explained in the below subsections, we demand the absence of Dirac-Misner string singularities, conical singularities, orbifold singularities, and curvature singularities on the rotational axes and the ring topology of the horizon for the charged solution.
Furthermore, it can be shown under these boundary conditions that the spacetime admits neither CTCs nor and curvature singularities on and outside the horizon. 
The metric and gauge potential for the resulting asymptotically flat, charged black ring solution, which is regular in the context mentioned above, are written respectively as

\begin{align}
&ds^2 = - \frac{H (y,x)}{D^2 H (x,y)}(dt + \Omega'_\psi(x,y)d\psi+ \Omega'_\phi(x,y)d\phi)^2 + \frac{D}{H (y,x)}\left[F(y,x) d\psi^2 - 2 J(x,y) d\psi d\phi-F(x,y) d\phi^2\right]\nonum
&\quad + \frac{\ell^2 D H (x,y)}{8h_2(1-\lambda ) \lambda  \nu ^2 \left(1-\nu ^2\right) \left(1-\sigma ^6\right)   (x-y)^2 }\left(\frac{dx^2}{G(x)} - \frac{dy^2}{G(y)}\right),\label{eq:metricsol-new}
\end{align}
\begin{align}\label{eq:gaugesol-new}
&A = \frac{\sqrt{3}cs}{DH (x,y)} \left[ (H (x,y)-H (y,x))dt \right.\nonum
&\hspace{2cm} \left.- (c H (y,x) \Omega_\psi(x,y)-s H (x,y) \Omega_\phi(y,x))d\psi
- (c H (y,x) \Omega_\phi(x,y)-s H (x,y) \Omega_\psi(y,x))d\phi\right],
\end{align}
where
\begin{align}
D &:= \frac{c^2 H (x,y)- s^2 H (y,x)}{ H (x,y)},\\
\Omega_\psi'(x,y) &:= c^3 \Omega_\psi(x,y)-s^3 \Omega_\phi(y,x), \label{eq:def-Omdashpsi}\\
\Omega_\phi'(x,y) &:= c^3\Omega_\phi(x,y)-s^3\Omega_\psi(y,x),\label{eq:def-Omdashphi}
\end{align}
and the metric functions are given by
\begin{align}
H (x,y)&= (1+y)^2 \biggr[ \left(1-x^2\right) (1-\nu ) \nu ^2 h_{-4}^2 \left(2 \sigma ^6 h_1+(1-\nu ) \left(1-\sigma ^6\right) h_{-4}^2\right)\nonum
&\quad +2 (1+\nu ) h_1   \biggr\{(1-\nu ) h_{-4}-(1+x \nu ) \left(h_{-4}+\nu  h_{-6}\right)\biggr\}^2\biggr]\nonum
   &+(-1-y) \biggr[ 2 \left(1-x^2\right) (1-\nu ) \nu ^2 \left(2 (1-\lambda ) \lambda  (1-\nu ) \left(1-\nu ^2\right) \left(1-\sigma ^6\right) h_{-4}^2+h_1
   h_{-5}^2\right)\nonum
&\quad   +(1+x)^2 \nu ^2 (1+\nu ) \left(2 (1-\nu ) \left(1-\sigma ^6\right) h_2 h_{-4}^2+h_1 h_{-5}^2\right)\nonum
&\quad+(1-x)^2 (1-\nu ) \nu ^2 \left(1-\nu
   ^2\right) \left(2 (1-\lambda ) \lambda  (1-\nu ) \left(1-\sigma ^6\right) h_{-4}^2+h_1 h_{-6}^2\right)\biggr]\nonum
& +   8 (1-x) (1-\lambda ) \lambda  (1-\nu )^2 \nu ^2 \left(1-\nu ^2\right) \left(1-\sigma ^6\right) h_2\nonum
&+8 (1+x) \nu ^2 \left(1-\nu ^2\right) \left(1-\sigma
   ^6\right) h_2^2+2 \left(1-x^2\right) (1-\nu ) \nu ^2 h_1 h_{-5}^2,\label{eq:solHtilde}\\
J(x,y) &=\frac{\ell^2 \left(1-x^2\right) \left(1-y^2\right) \nu  h_1 \left((1+x \nu ) (1+y \nu ) h_{-5} h_{-6} h_{-7}+\left(1-\nu ^2\right) h_{+5} h_{+6}
   h_{+7}\right)}{2 (x-y) (1-\lambda ) \lambda  \left(1-\sigma ^6\right) h_2}   ,\label{eq:solJtilde}\\
F(x,y) &=\frac{\ell^2}{2 (x-y)^2} \biggr[-64h_2 (1-\lambda ) \lambda  \nu ^2 (1+y \nu )^2 \left(1-\nu ^2\right) \left(1-\sigma ^6\right) G(x) -\frac{\left(1-y^2\right) \left(1-\nu
   ^2\right) G(x) h_1^2 h_{+5}^2 h_{+6}^2}{(1-\lambda ) \lambda  \left(1-\sigma ^6\right) h_2}\nonum
&\quad   -G(x) G(y) \left(4 \left(1-\nu ^2\right) \left(\left(1-\sigma ^6\right) h_{-4}^2 h_{+4}^2+\nu 
   h_1 h_{-6}^2\right)-4 \nu  h_1 h_{-5}^2+\frac{y \nu  h_1^2   h_{-5}^2 h_{-6}^2}{(1-\lambda ) \lambda  \left(1-\sigma ^6\right) h_2} \right)\nonum
   & +\frac{  G(y)\left(G(x) x \nu  h_{-7}^2     +\left(1-x^2\right) \left(1-\nu ^2\right) h_{+7}^2\right)  }{(1-\sigma^6)(1-\lambda ) \lambda h_2} \biggr],
   \label{eq:solFtilde}
   \end{align} 
\begin{align}   
\Omega_\psi(x,y) &=\sqrt{\frac{\left(1-\nu ^2\right) \left(1-\sigma ^6\right) }{\lambda(1-\lambda)h_2}} \frac{\ell  (1+y)}{H (y,x)} \biggr[
\left(1-x^2\right) \nu  h_1 h_{+4} h_{+5} \left((1+y \nu ) h_{-4}^2-(1+\nu ) h_1\right)\nonum
&\quad -(1-x) \nu  h_1h_{+5} \biggr\{ h_1 \biggr(\frac{\nu  \left(1+\sigma   ^3\right)}{1-\sigma ^3}+\frac{\left(1-\sigma ^3\right) h_2}{1+\sigma ^3}\biggr)+(1+\nu ) h_3 h_{+4}\biggr\} \nonum
   &+ \frac{\nu   h_1 h_{-5}(1+x \nu ) (1+y \nu )(1-x)}{1-\nu ^2} \biggr\{ h_1 \left(\frac{\nu  \left(1+\sigma ^3\right)}{1-\sigma ^3}-\frac{\left(1-\sigma ^3\right)
   h_2}{1+\sigma ^3}\right)-(1-\nu ) h_3 h_{-4}\biggr\} \nonum
   &-2 \nu  (1+x \nu ) (1+y \nu ) h_1 (1+x) h_2 h_{-4} h_{-6}-2 (1+x) \nu  (1+\nu ) (1+x \nu ) h_1 h_2 h_{+4} h_{+6}
\biggr],\label{eq:omegapsi-newsol}\\
\Omega_\phi(x,y) &= \sqrt{\frac{\left(1-\nu ^2\right) \left(1-\sigma ^6\right) }{\lambda(1-\lambda)h_2}}\frac{\ell  (1+x)}{H (y,x)} \biggr[
8 (1-\lambda ) \lambda  \nu ^2 (1+x \nu ) (1+y \nu )^2 h_1 \left((1-\lambda ) \left(1-\sigma ^3\right) h_2-\lambda  \left(1+\sigma ^3\right)\right)\nonum
& +2 (1+x)
   (1+y) (1-\lambda ) \lambda  \nu ^2 (1+\nu ) (1+y \nu ) h_1 h_{-4} h_{+5}+\frac{2 (1-y) \nu ^2 (x+y+\nu +x y \nu ) h_1^2 h_{-6} h_{+6}}{1-\sigma
   ^6}\nonum
&   +2 (1-y) (1-\lambda ) \lambda  (1-\nu ) \nu ^2 h_1 h_{-4} h_{+6} \left(6-2 \nu ^2+2 x (1+\nu )+y   \left(3+2 \nu -\nu ^2\right)+x y \left(3+\nu ^2\right)\right) \nonum
&+(1+x) \left(1-y^2\right) \nu ^2 h_1 h_{-4} h_{+6}\left(\nu  h_1+h_2-(1-\lambda ) \lambda 
   (1-\nu ) \left(7+\nu -2 \nu ^2\right)\right) 
\biggr], \label{eq:omegaphi-newsol}
\end{align}

and the coefficients are defined by
\begin{align}\label{eq:newcoeffs}
\begin{split}
   h_1&:= \lambda ^2+\left(1-\lambda ^2\right) \nu ^2,\\
      h_2 &:= \lambda   +(1-\lambda ) \nu ^2,\\
      h_3 &:= (2-\lambda ) \lambda +(1-\lambda )^2 \nu ^2,\\
   h_{\pm 4}&:= \lambda \pm(1-\lambda ) \nu ,\\
   h_{\pm 5} &:= \nu  \left(1+\sigma ^3\right) \mp  \left(\lambda +(1-\lambda ) \nu ^2\right) \left(1-\sigma ^3\right),\\
   h_{\pm 6}&:= \nu (1-\lambda )   \left(1-\sigma   ^3\right)\pm \lambda  \left(1+\sigma ^3\right),\\
 h_{\pm 7} &:= \lambda  \nu  \left(1+\sigma ^3\right)^2 h_1-(1-\lambda ) \nu  \left(1-\sigma ^3\right)^2 h_1 h_2\pm \left(1-\sigma
   ^6\right) h_3 h_{-4} h_{+4}.
   \end{split}
\end{align}
This solution is described by the four independent parameters $(\ell,\nu,\lambda,\sigma)$, whose parameter region is given as follows:
\begin{align}
 \ell>0,\quad 0< \nu < 1,\quad 0<\lambda <1,\quad -1<\sigma<1, \label{eq:paramregion-extend-new}
\end{align}
and the coordinates $(t,\psi,\phi,x,y)$ run the ranges~(\ref{eq:tphipsirange}) and (\ref{eq:xyrange}).
The parameter $\sigma$ is connected to $\alpha$, introduced in the Harrison transformation~(\ref{eq:ctrans-metric}), via the relationship
\begin{align}
\sigma = \tanh\alpha,\label{eq:abr-t3}
\end{align}
and the other parameter $\lambda$ is defined in relation of  the parameters $(\nu,\lambda)$ of the vacuum seed by
\begin{align}
  \gamma = \nu (1-\lambda^2) +\frac{2-\nu^2}{\nu} \lambda^2. \label{eq:def-lambda}
\end{align}

\medskip

The boundaries in the C-metric coordinates $(x,y)$ correspond to the following rods and asymptotic infinity:

\begin{enumerate}[(i)]
\item

$\phi$-rotational axis : 
$\partial \Sigma_\phi=\{(x,y)|x=-1,-1/\nu <y<-1 \}$ with the rod vector 
$v_\phi:=(0,0,1)=\partial/\partial \phi$, where  we impose the periodicity of the coordinate $\phi$ as $\phi\sim \phi +2\pi$ to ensures the absence of the conical singularities on $\partial \Sigma_\phi$, 
\item Horizon: 
$\partial \Sigma_{\cal H}=\{(x,y)|-1<x<1,y=-1/\nu \}$ with the rod vector $
 v_{\cal H} := (1, \omega_\psi, \omega_\phi),
$ 
where the constants $\omega_\psi$, $\omega_\phi$ are expressed by
\begin{align}
(\omega_\psi,\omega_\phi) = \frac{\kappa}{4 (1-\lambda ) \lambda  \nu  \sqrt{1-\nu ^2} \left(1-\sigma ^6\right) h_2}\left(h_1 h_{+5} h_{+6},h_{+7}\right),\label{eq:omegai}
\end{align}
with the surface gravity
\begin{align}
\kappa &= \frac{2\nu(1-\nu^2)\sqrt{(1-\sigma^6)(1-\lambda)^3\lambda^3 h_2^3}}{\ell h_1 (c^3 h_2 h_{+4} h_{+5}+s^3(1-\lambda ) \lambda  (1-\nu )^2 (\nu +1)  h_{-4}
   h_{+6})}, \label{eq:kappa}
\end{align}

\item

Inner axis: 
$\partial \Sigma_{\rm in}=\{(x,y)|x=1,-1/\nu <y<-1 \}$ 
with the rod vector 
\begin{align}\label{eq:inner-rodvector-new}
v_{\rm in} = (\tilde{\ell} (s^3 -c^3 \sigma^3),0,1),\quad \tilde{\ell} := \frac{2 h_1 \ell}{\sqrt{\lambda(1-\lambda)(1-\sigma^6)(1-\nu^2)h_2}}.
\end{align}

Thus, from the condition~(\ref{eq:abr-t3}), the Dirac-Misner string singularity does not appear, and $\partial \Sigma_{\rm in}$ becomes the $\phi$-rotational axis that is absent of the conical singularity with the periodicity of $\phi \sim \phi+2\pi$.

\item

$\psi$-rotational axis: 
$\partial \Sigma_\psi=\{(x,y)|-1<x<1,y=-1 \}$
with the rod vector $v_\psi :=(0,1,0)=\partial/\partial \psi$, where 
we impose the periodicity of the coordinate $\psi$ as $\psi\sim \psi +2\pi$ to ensures the absence of the conical singularities on $\partial \Sigma_\psi$.

\item Infinity:  
$\partial \Sigma_\infty =\{(x,y)|x\to y \to -1 \}$:  

In the the standard spherical coordinates $(r,\theta)$ introduced in Eq.~(\ref{eq:asym-xy}),
one can see that the metric asymptotes to the five-dimensional Minkowski metric at $x\to y\to -1$.
From the asymptotic expansion as in Eq.~(\ref{eq:asym-charges}), the ADM mass $M$ and two ADM angular momenta $J_\psi$, $J_\phi$ are obtained as
\begin{align}\label{eq:mass-new}
 M &= \frac{3  \pi \ell^2 (c^2+s^2) h_1 \left((1-\lambda ) \lambda  \left(1-\nu ^2\right) \left(1-\sigma ^6\right)+h_1\right)}{2G_5 (1-\lambda ) \lambda  \left(1-\nu ^2\right) \left(1-\sigma   ^6\right) h_2},\\
J_\psi &= \frac{\ell^3 \pi  h_1 (c^3 j_1 + s^3 j_2)}{2G_5 (1-\lambda )^{3/2} \lambda ^{3/2} \left(1-\nu ^2\right)^{3/2} \left(1-\sigma   ^6\right)^{3/2} h_2^{3/2}},\label{eq:jpsi-new}\\
 J_\psi &= \frac{\ell^3 \pi  h_1 (c^3 j_2 + s^3 j_1)}{2G_5 (1-\lambda )^{3/2} \lambda ^{3/2} \left(1-\nu ^2\right)^{3/2} \left(1-\sigma   ^6\right)^{3/2} h_2^{3/2}},\label{eq:jphi-new}
\end{align}
where
\begin{align}\label{eq:j1j2}
\begin{split}
j_1 &=  \nu ^2 \left(1+\sigma ^3\right)^2 h_1+\left(1-\sigma ^3\right)^2 h_1 h_2^2\\
&\quad +\left(1-\sigma ^6\right)\left[ \left((1-\lambda ) \lambda  \left(1-\nu ^4\right)+2   h_1\right) h_3-(1-\lambda ) \lambda  \left(1-\nu ^2\right)^2 \sigma ^3  h_{-4} h_{+4}\right],\\
j_2&=  (1-\nu^2) \left[ \lambda ^2 \left(1+\sigma ^3\right)^2 h_1-(1-\lambda )^2 \nu ^2 \left(1-\sigma ^3\right)^2 h_1\right.\\
&\quad\left.
 -(1-\lambda ) \lambda  \left(1-\nu ^2\right) \sigma ^3
   \left(1-\sigma ^6\right) \left(4 (1-\lambda ) \lambda -h_3\right)   +(1-\lambda ) \lambda  \left(1-\nu ^2\right) \left(1-\sigma ^6\right) h_{-4} h_{+4}\right].
   \end{split}
\end{align}
From the asymptotic behavior of the gauge field $A$, one can determine the electric charge
\begin{align}\label{eq:Qe}
 Q& := \fr{8\pi G_5} \int_S \left( \star F + \fr{\sqrt{3}} F\wedge A\right)\nonum
 & = \fr{8\pi G_5} \int_{S_\infty} \star F\nonum
 & = - \frac{2M \tanh(2\alpha)}{\sqrt{3}},
\end{align}
where $S$ is an $S^3$-surface that encloses both $\partial \Sigma_{\cal H}$ and $\partial \Sigma_{\rm in}$, and $S_\infty$ represents spatial infinity. In the second line above, we used the fact that the Chern-Simons term decays much faster than the first term and does not contribute to the integral on $S_\infty$.

\end{enumerate}

\medskip

We note that, from Eq.~(\ref{eq:inner-rodvector-new}), the Dirac-Misner string singularity is unavoidable unless Eq.~(\ref{eq:abr-t3}) is imposed.
In particular, if we assume $\sigma$ and $\alpha$ as independent parameters and set $\alpha=0$, the above metric reduces to the vacuum doubly-spinning black ring with the Dirac-Misner string singularity, where $\sigma$ serves as the parameter for the Dirac-Misner string singularity as $v_{\rm in} = (-\tilde{\ell}\sigma^3,0,1)$. 
This singular vacuum black ring solution can be seen as the vacuum seed for the above charged black ring solution
since the Harrison transformation changes the rod vector of the inner axis from $v_{\rm in} = (-\tilde{\ell}\sigma^3,0,1) $ to $v_{\rm in} = (0,0,1)$ under the condition~(\ref{eq:abr-t3}).
 This is similar to the three parameter charged black ring solution in our previous work~\cite{Suzuki:2024coe}, which can be obtained from the Harrison transformation of the vacuum singly-spinning black ring solution with the Dirac-Misner string singularity.

\medskip
Below, we illustrate the derivation of the above solution and demonstrate the absence of the singularities.

\subsection{Derivation of the charged rotating black ring solution}

As in the previous works~\cite{Suzuki:2024coe,Suzuki:2024phv,Suzuki:2023nqf}, 
the charged solution obtained by the Harrison transformation~(\ref{eq:ctrans-metric}) from the vacuum seed solution in the C-metric coordinates~(\ref{eq:metricsol-vac}) can be expressed  in terms of the metric functions in the seed solution, respectively, as
\begin{align}
&ds^2 = - \frac{\barH(y,x)}{\barD^2\barH(x,y)} (dt+\bar{\Omega}'_\psi(x,y) d\psi + \bar{\Omega}'_\phi(x,y)d\phi)^2+\frac{\barD}{\barH(y,x)}\left[\barF(y,x)d\psi^2-2\barJ(x,y)d\psi d\phi-\barF(x,y)d\phi^2\right]\nonum
&\quad + \frac{\ell^2 \barD \barH(x,y)}{4(1-\gamma)^3(1-\nu^2)(1-a^2)\Delta (x-y)^2}\left(\frac{dx^2}{G(x)}-\frac{dy^2}{G(y)}\right),
\label{eq:metricsol}
\end{align}
and
\begin{align}\label{eq:gaugesol}
&A = \frac{\sqrt{3}cs}{\barD\barH(x,y)} \left[ (\barH(x,y)-\barH(y,x))dt \right.\nonum
&\hspace{2cm} \left.- (c \barH(y,x) \bar{\Omega}_\psi(x,y)-s \barH(x,y) \bar{\Omega}_\phi(y,x))d\psi
- (c \barH(y,x) \bar{\Omega}_\phi(x,y)-s \barH(x,y) \bar{\Omega}_\psi(y,x))d\phi\right],
\end{align}
with
\begin{align}
\bar{D} &:= \frac{c^2 \barH(x,y)-s^2 \barH(y,x)}{\barH(x,y)},\\
\bar{\Omega}_\psi'(x,y) &:= c^3 \bar{\Omega}_\psi(x,y)-s^3 \bar{\Omega}_\phi(y,x), \label{eq:def-Omdashpsi}\\
\bar{\Omega}_\phi'(x,y) &:= c^3\bar{\Omega}_\phi(x,y)-s^3\bar{\Omega}_\psi(y,x).\label{eq:def-Omdashphi}
\end{align}

\subsubsection{Asymptotic infinity and rod structure}
\label{sec:chargedrods}

Let us examine the changes in the rod structure resulting from the Harrison transformation. 
The boundaries of the C-metric coordinates $(x,y)$ for the charged solution mentioned above can be described as follows:

\begin{itemize}

\item[(i)] $\phi$-rotational axis $\partial \Sigma_\phi=\{(x,y)|x=-1,-1/\nu<y<-1\}$ with the rod vector $v_\phi=(0,0,1)$\,.

\item[(ii)] Horizon:  $\partial \Sigma_{\cal H}=\{(x,y)|-1<x<1,y=-1/\nu\}$, accompanied by the rod  vector $v_{\cal H}=(1, \omega_\psi, \omega_\phi)$. 
Here, $\omega_{\psi}$ and $\omega_{\phi}$ represent the horizon angular velocities along the directions $\partial/\partial \psi$ and $\partial/\partial \phi$, respectively, 
\begin{align}
&\omega_{\psi} := -\frac{(\gamma -\nu )  f_3
   f_4}{\ell \tilde{v}_0 \left(c^3 \left(a   c_1-1+\nu \right) f_3 g_3+s^3(a-b)    (\gamma -\nu ) f_4 g_{13}\right)  },\\
&    \omega_{\phi} :=\frac{(\gamma -\nu ) d_3 
   g_2}{\ell \tilde{v}_0 c_3 \left(c^3 \left(a   c_1-1+\nu \right) f_3 g_3+s^3(a-b)    (\gamma -\nu ) f_4 g_{13}\right) }, 
   \end{align}
with
   \begin{align}
   g_{13} := f_8 f_{18}-\frac{f_1 f_{10}}{2 (1-\gamma ) (1-\nu ) (b-a)}.
   \end{align}

\item[(iii)]  
Inner axis: $\partial \Sigma_{\rm in}=\{(x,y)|x=1,-1/\nu<y<-1\}$ with the rod  vector 
\begin{align}
v_{\rm in} = \left( \ell \tilde{v}_0 \left( c^3 f_1^{-1} g_1'(a-b) - s^3f_2^{-1}g_{12}(1-ab)\right), v_{\rm in}^{\rm vac,\psi}  ,1 \right),
\end{align}
with
\begin{align}
g_{12} := f_{5}f_{13}+\frac{c_3(f_4 f_{17}-f_{5}f_{13})}{2\nu (1-\gamma)(1-ab)}.
\end{align}
As noted in Refs.~\cite{Suzuki:2024phv, Suzuki:2024coe}, the Harrison transformation only affects the $t$-component  in $v_{\rm in}$. 
Consequently, the conditions for the ring topology~(\ref{eq:topology-con}) and the absence of the conical singularity~(\ref{eq:conifree}) remain unchanged from the vacuum case.
The absence of the Dirac-Misner string singularity now specifies the parameter $\alpha$ as 
\begin{align}\label{eq:def-t3}
\tanh^3\alpha = \frac{(a-b)f_2 g_1'}{(1-ab)f_1 g_{12}},
\end{align}
where the absolute value of the right-hand side must be less than or equal to one.

\item[(iv)] $\psi$-rotational axis $\partial \Sigma_\psi=\{(x,y)|-1<x<1,y=-1\}$ with the rod  vector $v_\psi=(0,1,0)$\,.

\item[(v)]
Infinity:  
$\partial \Sigma_\infty 
= \{(x,y)|x\to y\to -1 \}$, where the mass and angular momenta are expressed in terms of their counterparts from the vacuum case:
\begin{align}
 M &= (c^2+s^2) M^{\rm vac} \label{eq:mass},\\
  J_\psi &= c^3 J^{\rm vac}_\psi+s^3 J^{\rm vac}_\phi , \label{eq:Jpsi}\\
  J_\phi &= c^3 J^{\rm vac}_\phi + s^3 J^{\rm vac}_\psi.\label{eq:Jphi}
\end{align}

\end{itemize}

As demonstrated in Ref.~\cite{Suzuki:2024phv}, the vacuum seed metric must possess the Dirac-Misner string singularity to allow for its elimination after the Harrison transformation. 
It is evident that the Harrison transformation, specifically when $(a-b)g_1'=0$, which has been shown to yield the Pomeransky-Sen'kov black ring~\cite{Suzuki:2024eoz}, results in a charged solution that inevitably retains the Dirac-Misner string singularity.

\subsubsection{Parameter region for a regular solution}

We require the resulting charged solution to be regular, meaning it should be free from curvature singularities, conical singularities, orbifold singularities, and Dirac-Misner string singularities both on and outside the horizon.
Since the last three of these conditions are ensured by Eqs.~(\ref{eq:topology-con}), (\ref{eq:conifree}) and (\ref{eq:def-t3})---as these singularities, if present, would appear only on the $\psi$ and $\phi$-rotational axes---we will focus primarily on the curvature singularities in the following discussion.
The Kretchmann scalar $R^{\mu\nu\rho\sigma}R_{\mu\nu\rho\sigma}$ is proportional to $\barH(x,y)^{-6}\barD^{-6}$.
Therefore, to avoid curvature singularity, $\barH(x,y)>0$ and $\barD>0$ must be positive for $-1 \leq x \leq 1$ and $-1/\nu \leq y \leq -1$, as evidenced by the positivity of $\barH(x,y)$ and $\barD$ at infinity $x \to y \to -1$ guaranteed by Eq.~(\ref{eq:necessary-1})
\begin{align}
\barH(x=-1,y=-1) = 8(1-\gamma)^3(1-\nu)^4(1-a^2) \Delta >0 , \quad \barD(x=-1,y=-1) =1. \label{eq:necessary-1b}
\end{align}
In particular, ensuring the positivity of $\barH(x,y)$ and $\barD$ at $(x,y)=(1,-1)$ requires, respectively,  
\begin{align}
\barH(x=1,y=-1) &= -8(1-\gamma)(1-\nu)^2(1+\nu)d_1 f_2^2 >0  \quad \Longleftrightarrow  \quad d_1<0,\label{eq:necessary-2}
\end{align}
and
\begin{align}\label{eq:necessary-3}
 \barD(x=1,y=-1) = c^2 - s^2 \frac{(1-\gamma)f_1^2}{(1+\nu) f_2^2} >0  \quad \Longleftrightarrow \quad \tanh^2\alpha < \frac{(1+\nu)f_2^2}{(1-\gamma)f_1^2}.
\end{align}

As mentioned previously, the two conditions, the topology condition~(\ref{eq:topology-con}) of a black ring and the regularity condition~(\ref{eq:conifree})  for the absence of conical singularities remain unchanged after the Harrison transformation.
Hence, we  can use the same analysis as the vacuum case performed in Ref.~\cite{Suzuki:2024eoz}, where 
Eqs.~(\ref{eq:topology-con}) and (\ref{eq:conifree}) admit four different solutions for $b$:
\begin{align}
 b =\frac{\tilde{\gamma}(\tilde{\gamma}\tilde{\nu}+2-\tilde{\nu}^2)}{\tilde{\nu}(2-\tilde{\gamma}^2-\tilde{\nu}^2)},\quad \frac{\tilde{\gamma}(\tilde{\gamma}\tilde{\nu} - (2-\tilde{\nu}^2))}{\tilde{\nu}(2-\tilde{\gamma}^2-\tilde{\nu}^2)},\quad
 \frac{\tilde{\gamma}(a \tilde{\gamma}\tilde{\nu} + 2-\tilde{\nu}^2)}{\tilde{\nu}(2-\tilde{\gamma}^2-\tilde{\nu}^2)},\quad  \frac{\tilde{\gamma}(a \tilde{\gamma}\tilde{\nu} - (2-\tilde{\nu}^2))}{\tilde{\nu}(2-\tilde{\gamma}^2-\tilde{\nu}^2)}.\label{eq:n0sol-b}
 \end{align}
 Here we introduced the new parameters: 
\begin{align}
 \tilde{\gamma} := \sqrt{1-\gamma} ,\quad \tilde{\nu} := \sqrt{1+\nu}, \label{eq:defnewparams}
\end{align}
or conversely,
\begin{align}
\gamma = 1-\tilde{\gamma}^2,\quad \nu = \tilde{\nu}^2-1.
\end{align}
The bound for $\nu$ and $\gamma$ in Eq.~(\ref{eq:rangenugam}) is rewritten as
\begin{align}
0<\tilde{\gamma}<\sqrt{2-\tilde{\nu}^2}, \quad 1<\tilde{\nu}<\sqrt{2}.
\end{align}
In Eq.~(\ref{eq:n0sol-b}), the latter two solutions are not regular  due to $d_1=0$, whereas the first two solutions are mutually related  through the transformation $(\tilde{\nu},\tilde{\gamma}) \to (-\tilde{\nu},\tilde{\gamma})$ or $(\tilde{\nu},\tilde{\gamma})\to (\tilde{\nu},-\tilde{\gamma})$, 
and therefore can be merged to a single solution
\begin{align}\label{eq:n0sol-b1}
b =\frac{\tilde{\gamma}(\tilde{\gamma}\tilde{\nu}+2-\tilde{\nu}^2)}{\tilde{\nu}(2-\tilde{\gamma}^2-\tilde{\nu}^2)},
\end{align}
by letting $\tilde{\nu}$ and $\tilde{\gamma}$ to take negative values within the range
\begin{align}
 0 < |\tilde{\gamma}| < \sqrt{2-\tilde{\nu}^2},\quad 1< | \tilde{\nu}| < \sqrt{2}.
\end{align}

 Additionally, substituting Eq.~(\ref{eq:n0sol-b1}) into Eq.~(\ref{eq:topology-con}),
one can obtain the product of two quadratic equation with respect to $\beta$. 
Taking one root from each equation, we obtain
\begin{align}\label{eq:n0sol-beta1}
\beta = \beta_1:=\frac{\tilde{\nu } \left(2-\tilde{\gamma }^2-\tilde{\nu }^2\right)}{{\cal B}_0}\biggr[(1+a) \tilde{\gamma }^3 \tilde{\nu }^2+\tilde{\nu } \left(2-\tilde{\nu   }^2\right)^2+\tilde{\gamma } \left(2-\tilde{\nu }^2\right) \left(2-3 \tilde{\gamma }   \tilde{\nu }-(1+a) \tilde{\nu }^2\right)\nonum
 - (1-a) \tilde{\gamma } \tilde{\nu }
   \sqrt{2\left(2-\tilde{\gamma }^2-\tilde{\nu }^2\right) \left(-2+3 \tilde{\nu
   }^2-\tilde{\nu }^4\right)}\biggr],
      \end{align}
  with 
\begin{align}   
&\quad  {\cal B}_0 = (1+a)^2 \tilde{\gamma }^5 \tilde{\nu }^3-\tilde{\nu }^2 \left(2-\tilde{\nu
   }^2\right)^3+\tilde{\gamma }^3 \tilde{\nu } \left(2-\tilde{\nu }^2\right)   \left(10-\left(5+4 a-a^2\right) \tilde{\gamma } \tilde{\nu }   -(2+a)^2 \tilde{\nu   }^2\right)\nonum
&\quad\quad+\tilde{\gamma } \left(2-\tilde{\nu }^2\right)^2 \left(-2 \tilde{\gamma }-4
   \tilde{\nu }+\left(4+4 a-a^2\right) \tilde{\gamma } \tilde{\nu }^2+(1+2 a) \tilde{\nu
   }^3\right),   
\end{align}
and 
\begin{align}\label{eq:n0sol-beta2}
\beta=\beta_2: =\frac{\tilde{\nu } \left(2-\tilde{\gamma }^2-\tilde{\nu }^2\right)
   \left(\left(\tilde{\gamma }+\tilde{\nu }\right) \left(2-\tilde{\nu   }^2\right)-\tilde{\gamma } \tilde{\nu } \left((1+a) \tilde{\gamma }-(1-a)
   \sqrt{2-\tilde{\nu }^2}\right)\right)}{(1+a)^2 \tilde{\gamma }^4 \tilde{\nu }^2-2
   (1+a) \tilde{\gamma }^3 \tilde{\nu } \left(2-\tilde{\nu }^2\right)+2 \tilde{\gamma }
   \tilde{\nu } \left(2-\tilde{\nu }^2\right)^2+\tilde{\nu }^2 \left(2-\tilde{\nu   }^2\right)^2+\tilde{\gamma }^2 \left(2-\tilde{\nu }^2\right)   \left(2-\left(4+a^2\right) \tilde{\nu }^2\right)},
\end{align}
and another root for each equation is simply given by Eqs.~(\ref{eq:n0sol-beta1}) and (\ref{eq:n0sol-beta2}) with the transformation $(\tilde{\gamma},\tilde{\nu}) \to (-\tilde{\gamma},-\tilde{\nu})$, respectively. As we did for $b$ in Eq.~(\ref{eq:n0sol-b1}), one can merge the latter two roots into the first two in Eqs.~(\ref{eq:n0sol-beta1}) and (\ref{eq:n0sol-beta2}) by extending $\tilde{\nu}$ and $\tilde{\gamma}$ into the negative domain. Henceforth, the solution for $b$ and $\beta$ is given by Eqs.~(\ref{eq:n0sol-b1}) and (\ref{eq:n0sol-beta1}), or Eqs.~(\ref{eq:n0sol-b1}) and (\ref{eq:n0sol-beta2}) defined in the following four domains
\begin{description}
\item[(i)] $0< \tilde{\gamma} < \sqrt{2-\tilde{\nu}^2},   \quad 1<\tilde{\nu} <\sqrt{2}$,
\item[(ii)] $- \sqrt{2-\tilde{\nu}^2}< \tilde{\gamma} <0,  \quad 1<\tilde{\nu} <\sqrt{2}$,
\item[(iii)] $0< \tilde{\gamma} < \sqrt{2-\tilde{\nu}^2}, \quad -\sqrt{2}<\tilde{\nu} <-1$,
\item[(iv)] $- \sqrt{2-\tilde{\nu}^2}< \tilde{\gamma} <0 ,  \quad -\sqrt{2}<\tilde{\nu} <-1$.
 \end{description}

 \medskip
Next, we consider the regularity condition~(\ref{eq:def-t3}) for the absence of the Dirac-Misner string singularity, which  leads to
 \begin{align}\label{eq:n0sol-alpha1}
 \tanh^3\alpha = \frac{2 \tilde{\nu } \left(-2+\tilde{\nu }^2\right)+\tilde{\gamma
   }^2 \tilde{\nu } \left(-2+\tilde{\nu }^2\right)+\tilde{\gamma }
   \left(4-2 \tilde{\nu }^2+\tilde{\nu }^4\right)+a \tilde{\gamma }
   \tilde{\nu } \left(\tilde{\nu }^3+\tilde{\gamma }
   \left(-2+\tilde{\nu }^2\right)\right)}{\tilde{\gamma }^2
   \tilde{\nu }^3-2 \tilde{\nu } \left(-2+\tilde{\nu }^2\right)+a
   \tilde{\gamma } \tilde{\nu }^2 \left(-2+\tilde{\gamma }
   \tilde{\nu }+\tilde{\nu }^2\right)+\tilde{\gamma }
   \left(-4+\tilde{\nu }^4\right)},
 \end{align}
 for the solution~(\ref{eq:n0sol-beta1})
 and 
 \begin{align}\label{eq:n0sol-alpha2}
&\tanh^3\alpha=\frac{1}{{\cal B}_1} \biggr[-\left((1+a) \tilde{\gamma } \tilde{\nu }^2 \sqrt{2-\tilde{\nu
   }^2} \left(\tilde{\gamma }^2-\tilde{\nu }^2\right) \left((1+a)
   \tilde{\gamma } \tilde{\nu }+2 \left(-2+\tilde{\nu
   }^2\right)\right)\right)\nonum
&   - 2 \left(\tilde{\nu }^2-1\right)   \left(
\tilde{\nu }^2 \left(2-\tilde{\nu   }^2\right)^2
-\tilde{\gamma } \tilde{\nu } \left(2-\tilde{\nu }^2\right) \left(4-(1-a) \tilde{\nu }^2\right)
-(1+a)   \tilde{\gamma }^3 \tilde{\nu } \left(2+a \tilde{\nu   }^2\right)
+ \tilde{\gamma }^2 \left(2-\tilde{\nu }^2\right)   \left(2+(1+2 a) \tilde{\nu }^2\right)\right)\biggr],
\end{align}
 for the solution~(\ref{eq:n0sol-beta2}), respectively, where 
 \begin{align}
&\quad {\cal B}_1 := 
(1+a)^2
   \tilde{\gamma }^4 \tilde{\nu }^2 \left(-2+\tilde{\nu }^2\right)
   +2   (1+a) \tilde{\gamma }^3 \tilde{\nu } \left(2-\tilde{\nu   }^2\right)^2+2 \tilde{\nu }^2 \left(2-\tilde{\nu }^2\right)^2   \left(-1+\tilde{\nu }^2\right)\nonum
&\quad   +2 \tilde{\gamma } \tilde{\nu }
   \left(-2+\tilde{\nu }^2\right) \left(-4+6 \tilde{\nu }^2+(-1+a)
   \tilde{\nu }^4\right)+\tilde{\gamma }^2 \left(-8-8 (-1+a)
   \tilde{\nu }^2+2 (1+6 a) \tilde{\nu }^4+\left(-1-2 a+a^2\right)
   \tilde{\nu }^6\right).
\end{align}
From Eqs.~(\ref{eq:n0sol-alpha1}) and (\ref{eq:n0sol-alpha2}), one can show the parameters must be restricted in  the following ranges, respectively, 
\begin{align}
&-1 <  \tanh \alpha < 1 \Longleftrightarrow   
\left\{ \begin{array}{cc} 
\displaystyle -1 < a < \frac{4-\tilde{\gamma} \tilde{\nu} -2 \tilde{\nu}^2}{\tilde{\nu }\tilde{\gamma}} & {\rm for}\quad  \tilde{\gamma} \tilde{\nu}>0 \\ 
\displaystyle  \frac{4-\tilde{\gamma} \tilde{\nu} -2 \tilde{\nu}^2}{\tilde{\nu }\tilde{\gamma}} <  a < -1&{\rm for} \quad \tilde{\gamma} \tilde{\nu}<0
 \end{array} \right.,  \label{eq:tanh-con-n0-a}\\
&-1 <  \tanh \alpha < 1 \Longleftrightarrow   \left\{ \begin{array}{cc} \displaystyle a<-1  \quad {\rm or} \quad \frac{4-\tilde{\gamma} \tilde{\nu} -2 \tilde{\nu}^2}{\tilde{\nu }\tilde{\gamma}} <a & {\rm for}\quad  \tilde{\gamma} \tilde{\nu}>0 \\ \displaystyle  a< \frac{4-\tilde{\gamma} \tilde{\nu} -2 \tilde{\nu}^2}{\tilde{\nu }\tilde{\gamma}} \quad {\rm or } \quad -1<a&{\rm for} \quad \tilde{\gamma} \tilde{\nu}<0 \end{array} \right. .\label{eq:tanh-con-n0-b}
\end{align}
Meanwhile, utilizing Eq.~(\ref{eq:n0sol-b1}), we can derive
\begin{align}
d_1 = \tilde{\nu} \tilde{\gamma}^3 (1+a) \left[(1+a) \tilde{\gamma} \tilde{\nu} - 2 \left(2-\tilde{\nu}^2\right)\right].
\end{align}
From this equation, it is straightforward to observe that $d_1<0$ in the range specified by Eq.~(\ref{eq:tanh-con-n0-a}) and $d_1>0$ according to Eq.~(\ref{eq:tanh-con-n0-b}). 
Therefore, the solution associated with Eq.~(\ref{eq:tanh-con-n0-b}) results in a singular spacetime.

\medskip
Finally, we confirm whether the necessary conditions for regularity outlined in Eqs.~(\ref{eq:necessary-1}) and (\ref{eq:necessary-3}) is met for the solution specified by Eqs.~(\ref{eq:n0sol-b1}) and (\ref{eq:n0sol-beta1}).
Using Eqs.~(\ref{eq:n0sol-b1}) and (\ref{eq:n0sol-beta1}), we can straightforwardly show that  $(1+\nu) f_2^2/(1-\gamma)f_1^2$  in Eq.~(\ref{eq:necessary-3}) can be written as
\begin{align}\label{eq:necessary-3a}
  \frac{(1+\nu) f_2^2}{(1-\gamma)f_1^2} = 
  \left[ 
  \frac{2 \left(2-\tilde{\gamma }^2-\tilde{\nu }^2\right)+\sqrt{2} \tilde{\nu } \left(\tilde{\nu}^2-\tilde{\gamma }^2\right) \sqrt{\left(2-\tilde{\gamma }^2-\tilde{\nu }^2\right) \left(3 \tilde{\nu }^2-2-\tilde{\nu}^4\right)}}
   {\left(2-\tilde{\gamma }^2-\tilde{\nu }^2\right)\left(2-\tilde{\gamma }^2+\tilde{\nu }^2+\tilde{\gamma }^2 \tilde{\nu }^2-\tilde{\nu }^4\right)}
     \right]^2,
\end{align}
where the right hand side is demonstrably  larger than $1$ within the domains (i) and (ii) mentioned above, leading to 
\begin{align}\label{eq:necessary-3b}
  \frac{(1+\nu) f_2^2}{(1-\gamma)f_1^2} >1,
\end{align}
which implies the condition~(\ref{eq:necessary-3}) is automatically satisfied and thus does not affect on the parameter region.  
On the other hand, it can be found that within domains (iii) and (iv), the condition~(\ref{eq:necessary-3})  is not necessarily satisfied, and numerical evidence indicates that $D$ exhibits a negative region near $x=1$. 
Therefore, in the subsequent discussion, we will concentrate exclusively on domains (i) and (ii).

\medskip
Furthermore, we can also confirm that within the domains (i) and (ii), the condition~(\ref{eq:necessary-1}) can be satisfied as follows.
Using Eqs.~(\ref{eq:n0sol-b1}) and (\ref{eq:n0sol-beta1}), we can derive
\begin{align}\label{eq:delta-f2}
(1-a^2)\Delta =\frac{\left(2-\tilde{\gamma }^2-\tilde{\nu }^2\right) \left(\tilde{\gamma }^2 \left(\tilde{\nu }^2-1\right)+\left(2-\tilde{\nu
   }^2\right) \left(1+\tilde{\nu }^2\right)\right)}{\left(2-\tilde{\nu }^2\right) \left(2 \left(2-\tilde{\gamma }^2-\tilde{\nu
   }^2\right)+\sqrt{2} \tilde{\nu } \left(\tilde{\nu }^2-\tilde{\gamma }^2\right) \sqrt{\left(2-\tilde{\gamma }^2-\tilde{\nu
   }^2\right) \left(-2+3 \tilde{\nu }^2-\tilde{\nu }^4\right)}\right)}\frac{(-d_1) f_2^2}{\tilde{\gamma}^4}.
\end{align}
Having shown $d_1<0$, one can easily see the positivity of this quantity in the domains (i) and (ii).
Therefore, the condition~(\ref{eq:necessary-1}) is satisfied within these domains.

\medskip
From Eq.~(\ref{eq:n0sol-alpha1}), the parameter $a$ is expressed in terms of $(\tilde{\nu},\tilde{\gamma},\alpha)$
\begin{align}\label{eq:n0sol-a}
a = \frac{\tilde{\gamma }^2 \left(-2 \tilde{\nu }-\left(\left(\sigma^3-1\right) \tilde{\nu
   }^3\right)\right)+\tilde{\gamma } \left(\tilde{\nu }^4-2 \tilde{\nu }^2-\sigma^3 \tilde{\nu }^4+4
   \sigma^3+4\right)+2 \left(\sigma^3+1\right) \tilde{\nu } \left(\tilde{\nu }^2-2\right)}{\tilde{\gamma }
   \tilde{\nu } \left(\tilde{\gamma } \left(\left(\sigma^3-1\right) \tilde{\nu   }^2+2\right)+\left(\sigma^3-1\right) \tilde{\nu }^3-2 \sigma^3 \tilde{\nu }\right)},
\end{align}
where we introduced $\sigma$ by Eq.~(\ref{eq:abr-t3}).
Substituting this into Eq.~(\ref{eq:n0sol-beta1}), one can also express $\beta$ by $(\tilde{\nu},\tilde{\gamma},\sigma)$. 
Moreover, by replacing $1/\sqrt{1-\gamma}$ with $1/\tilde{\gamma}$ in $\tilde{v}_0$ defined in Eq.~(\ref{eq:def-cfs}),
it is straightforward to demonstrate that replacing $\tilde{\gamma} \to -\tilde{\gamma}$ corresponds to the identical metric and gauge potential. 
Therefore, without loss of generality, it is sufficient to consider only the domain (i), and
 the black ring solution, which is regular at least at $(x,y)=(-1,-1)$ and $(x,y)=(1,-1)$, is characterized by the  parameters $(\ell,\tilde{\nu},\tilde{\gamma},\sigma)$  within the range
\begin{align}
  1<\tilde{\nu} <\sqrt{2},\quad 0 <\tilde{\gamma} <\sqrt{2-\tilde{\nu}^2},\quad \ell>0,\quad -1< \sigma < 1. 
  \label{eq:param-region-new}
\end{align}
As shown below, the conditions outlined in Eqs.~(\ref{eq:n0sol-b1}), (\ref{eq:n0sol-beta1}) and (\ref{eq:n0sol-a}) together with Eq.~(\ref{eq:param-region-new}) are also  sufficient to ensure the absence of curvature singularities and CTCs throughout the entire region on and outside the horizon.

\subsubsection{Extension of the parameter beyond $\gamma=1$}

As explained above, the parameter region~(\ref{eq:param-region-new}) ensures that the black rings are physical.
Here, however,  we show that this region~(\ref{eq:param-region-new}) covers only a portion of the phase space for black rings. 
We will see later that the solution for $\tilde{\gamma}=0$ (which corresponds $\gamma=1$), which lies on the boundary of the parameter region~(\ref{eq:param-region-new}), does not correspond to either the extremal or the singular limit.
Indeed, it describes a regular black ring as shown in Figs.~\ref{fig:jjplot-ex1} and \ref{fig:n0-jjplots}.
This suggests a potential expansion of the parameter region beyond $\tilde{\gamma}=0$. 

\medskip

\medskip

To explore this extension, we reinstate the initial parameters $\nu$ and $\gamma$ when defining $\tilde \nu$ and $\tilde \gamma$ in Eq.~(\ref{eq:defnewparams}),  subsequently extending the solution to $\gamma \geq 1$ beyond the initial parameter range~(\ref{eq:rangenugam}).
This procedure is not straightforward, as evident from Eqs.~(\ref{eq:n0sol-beta1}) and (\ref{eq:n0sol-a}), where the parameters $\beta$ and $a$ becomes complex values for $\gamma > 1$, and so do the metric functions in Eq.~(\ref{eq:defGx}) as they include odd powers of $\tilde{\gamma}$. 
However, after eliminating $a,b,\beta$, using Eqs.~(\ref{eq:n0sol-b1}), (\ref{eq:n0sol-beta1}) and (\ref{eq:n0sol-a}),
one observes that the terms with odd powers of $\tilde{\gamma}$ can be factored out in the metric functions as a common factor, which cancels out in the resultant metric~(\ref{eq:metricsol}) and gauge potential~(\ref{eq:gaugesol}).

\medskip
To ensure the metric functions become real functions even for $\gamma>1$, 
one can remove the common factor that contains odd powers of $\tilde{\gamma}$ in the metric functions by the following redefinition
\begin{align}\label{eq:functilde}
\begin{split}
 H (x,y) &:=  \cC\, [-8(1-\gamma)(1-\nu)^2(1+\nu) d_1 f_2^2 ]^{-1} \barH(x,y),\\
  F(x,y) &:=  \cC\, [-8(1-\gamma)(1-\nu)^2(1+\nu) d_1 f_2^2]^{-1} \barF(x,y),\\
   J(x,y) &:=  \cC\, [-8(1-\gamma)(1-\nu)^2(1+\nu) d_1 f_2^2]^{-1} \barJ(x,y),
 \end{split}
\end{align}
while $\bar{\Omega}_\psi(x,y)$ and $\bar{\Omega}_\phi(x,y)$ yield only even powers of $\tilde{\gamma}$ without rescaling. $\cC>0$ is a constant that is chosen later to simplify the resulting metric functions.
One can confirm that the redefined functions contain only even powers of $\tilde{\gamma}$ by rewriting them in terms of $\tilf_i$ in Appendix~\ref{sec:prooffi}.

\medskip
Using these rescaled functions, one can confirm the conditions outlined in Eqs.~(\ref{eq:necessary-1b}), (\ref{eq:necessary-2})  and (\ref{eq:necessary-3}) are also satisfied for $\gamma>1$.
Given that $H (x=1,y=-1)= \cC>0 $ by definition, the absence of the curvature singularity requires $H (x,y)> 0$, thereby establishing the necessary condition at infinity, 
\begin{align} \label{eq:necessary-1-tilde}
H (x=-1,y=-1) = \cC\tDelta :=\cC \frac{(1-\gamma)^2(1-\nu)^2(1-a^2)\Delta}{(1+\nu) (-d_1)f_2^2}>0.
\end{align}
Using Eq.~(\ref{eq:delta-f2}), one can show that this implies
\begin{align}
\tDelta=
 \frac{(1-\nu)(\gamma -\nu ) \left(2-\gamma  \nu -\nu ^2\right)}{(1+\nu) \left(2 (\gamma -\nu )+ \sqrt{2\nu(1-\nu^2 )    (\gamma -\nu )} (\gamma +\nu )\right)}>0 \quad \Leftrightarrow
\quad  \nu< \gamma < \frac{2-\nu^2}{\nu}.
\end{align}
Moreover, from Eq.~(\ref{eq:necessary-3a}), it is straightforward to show Eq.~(\ref{eq:necessary-3b}) is also satisfied in this extended range.
Consequently, the parameter region meeting the necessary conditions outlined in Eqs.~(\ref{eq:necessary-1b}), (\ref{eq:necessary-2})  and (\ref{eq:necessary-3})  can now be extended to:
\begin{align}\label{eq:paramregion-extend}
0<\nu<1,\quad \nu < \gamma < \frac{2-\nu^2}{\nu},\quad \ell>0,\quad -1< \sigma < 1.
\end{align}
One can show that this condition actually is the sufficient condition for the black ring to be regular both on and outside the horizon.

 \medskip
 To demonstrate the absence of singularities and CTCs, it is convenient to introduce the new parameter $\lambda$ through Eq.~(\ref{eq:def-lambda}), with which the parameter region in Eq.~(\ref{eq:paramregion-extend}) is simplified to that in Eq.~(\ref{eq:paramregion-extend-new}).
By the use of Eqs.~(\ref{eq:n0sol-b1}), (\ref{eq:n0sol-beta1}) and (\ref{eq:n0sol-a}), one can rewrite the redefined metric functions~(\ref{eq:functilde}) in terms of the new parameters $(\ell,\nu,\lambda,\sigma)$, which are presented in Eqs.~(\ref{eq:solHtilde}), (\ref{eq:solJtilde}) and (\ref{eq:solFtilde}),
 with the choice $\cC = 16\nu^2(1-\nu^2)(1-\sigma^6)h_2^2$. 
 By rewriting $\bar{\Omega}_\psi$ and $\bar{\Omega}_\phi$ using $H(x,y)$ instead of $\bar{H}(x,y)$, one can obtain $\Omega_\psi$ and $\Omega_\phi$ presented in Eqs.~(\ref{eq:omegapsi-newsol}) and (\ref{eq:omegaphi-newsol}), respectively.

\medskip

\subsection{Absence of curvature singularities}\label{sec:absence-sing}

Now, we demonstrate that within the parameter range specified in Eq.~(\ref{eq:paramregion-extend-new}), the charged black ring solution~(\ref{eq:metricsol-new})
exhibits no curvature singularities both on and outside the horizon. 
Curvature singularities, if they exist,  appear at points where the metric (\ref{eq:metricsol-new}) or its inverse diverges. 
This divergence occurs only on the surfaces $H (x,y)=0$, $D=0$, and at the boundaries $x=\pm 1$, $y=-1$, $y=-1/\nu$ of the C-metric coordinates.

\medskip

In particular, as previously noted, the surface $H (x,y)=0$ or the surface $D=0$ leads to a curvature singularity due to the divergence of the Kretchmann scalar at these locations. 
However, it is easy to see that every term in Eq.~(\ref{eq:solHtilde}) is nonnegative and does not vanish at the same point  within the range~(\ref{eq:xyrange}).
This proves $H (x,y)>0$ for the range~(\ref{eq:xyrange}).
Additionally, once $H (x,y)>0$ is established, $D>0$ within the ranges (\ref{eq:xyrange}) becomes evident from the equation $D=1+s^2(H (x,y)-H (y,x))/H (x,y)\geq1$ since
\begin{align}\label{eq:proofDpositive-2}
  H (x,y)-H (y,x) & = 2 (x-y) \nu ^2 h_1 \biggr[ 4 (1+x \nu ) (1+y \nu ) \left(1-\sigma ^6\right) h_3\nonum
  &\quad +\left((1-x) (1-y) (1-\nu )+(1+x) (-1-y) (1+\nu )\right) \left(\left(1+\sigma ^6\right) h_1+\nu  \left(1-\sigma   ^6\right) h_3\right)\biggr]\geq 0.
\end{align}

\medskip

Furthermore, at each boundary of the coordinate system $(x,y)$, the absence of curvature singularities can also be demonstrated by introducing regular coordinates in the following manner:

\begin{enumerate}[(a)]

\item

The limit $x\to y \to -1$ corresponds to asymptotic infinity. 
Using the standard spherical coordinates $(r,\theta)$, defined in Eq.~(\ref{eq:asym-xy}),
we can observe that the metric as $r\to\infty$ ($x\to y \to -1$) asymptotically behaves as the Minkowski metric
\begin{align}
ds^2 \simeq -dt^2+dr^2 + r^2 (d\theta^2+\sin^2\theta d\psi^2+\cos^2\theta d\phi^2).\label{eq:aslimds}
\end{align}
Hence, the charged solution describes an asymptotically flat spacetime.

\item

The point $(x,y)=(1,-1)$ corresponds to a center of the five-dimensional Minkowski spacetime, specifically the intersection of the $\psi$-rotational axis and inner $\phi$-rotational axis.
By using  the coordinates $(r,\theta)$  introduced as follows:
\begin{align}
\begin{split}
&x = 1 - \frac{(1+\nu){\cal C}_1}{\ell^2} r^2 \cos^2\theta ,\quad
y = -1 - \frac{(1-\nu){\cal C}_1}{\ell^2} r^2 \sin^2\theta,\nonum
&{\cal C}_1 := \frac{(1-\lambda ) \lambda  h_2}{c^2 h_1 h_3+s^2(1-\lambda )^2 \lambda ^2 \left(1-\nu ^2\right)^2},
\end{split}
\end{align}
where $\cC_1>0$ is evident follows from Eq.~(\ref{eq:newcoeffs}), and
we can show that the metric near $r= 0$ ($(x,y)= (1,-1)$) behaves as
\begin{align}
ds^2 \simeq -dt'^2+dr^2 + r^2 ( d\theta^2 + \sin^2 \theta d\psi^2 + \cos^2 \theta d\phi^2)
\end{align}
where $t' := (1-\nu^2)\cC_1 t$.
Hence, the point $(x,y)=(1,-1)$ represents a regular origin in the five-dimensional Minkowski spacetime.

\item

The boundaries $x=-1$ and $x=1$  ($-1/\nu<y<-1$) correspond to the $\phi$-rotational axes outside and inside the black ring, respectively.
By introducing the radial coordinate $r$ defined by  $x = \pm 1 \mp (1\pm \nu) r^2/2$  for $x=\pm1$,
we can see that the metric as $r\to0$ ($x\to \pm 1$) behaves as
\begin{align}
 ds^2 \simeq \gamma^{\pm}_{tt}(y) dt^2+2 \gamma^{\pm}_{t\psi}(y) dt d\psi+ \gamma^{\pm}_{\psi\psi}(y) d\psi^2 + \alpha_\pm(y) ( dr^2+r^2 d\phi^2-G^{-1}(y) dy^2),
\end{align}
where 
\begin{align}
&\gamma_{tt}^{\pm} =-\frac{D|_{x=\pm1}H (y,\pm1)}{H (\pm1,y)},\quad
\gamma_{t\psi}^{\pm} = -\frac{H (y,\pm1)[c^3 \Omega_\phi(\pm1,y)-s^3 \Omega_\phi(y,\pm1)]}{D^2|_{x=\pm1}H (\pm1,y)},\nonum
&\gamma_{\psi\psi}^{\pm} =\frac{D|_{x=\pm1} F(y,\pm1)}{H (y,\pm1)}-\frac{H (y,\pm1)[c^3\Omega_\psi(\pm1,y)-s^3 \Omega_\phi(y,\pm1)]^2}{D^2|_{x=\pm 1} H (\pm1,y)},\\
&\alpha_\pm = \frac{\ell^2 D|_{x=\pm1} H (\pm1,y)}{8 h_2 (1-\lambda ) \lambda  \nu ^2 \left(1-\nu ^2\right) \left(1-\sigma ^6\right) (1\mp y)^2}.
\end{align}
For the two-dimensional metric $\gamma^\pm_{AB}(A,B=t,\psi)$, one can also show
\begin{align}
{\rm det} \, (\gamma^{\pm}_{AB})
 =\frac{32 \ell^2 h_2 (1-\lambda ) \lambda   (1-\nu^2 ) \nu ^2 (1\pm \nu )^2 \left(1-\sigma ^6\right) (1\pm y) (1+\nu  y)}{ (1\mp y) D|_{x=\pm1} H (\pm1,y)}.
\end{align}
From $H (x,y)>0$ and $D>0$ within the ranges specified by Eq.~(\ref{eq:xyrange}), 
 it is evident that $\alpha_\pm$ is a positive definite function and $\gamma^\pm_{AB}$ is a nonsingular and non-degenerate matrix for $-1/\nu<y<-1$. 
Thus, the metric has no curvature singularities at $x=\pm1$.

\item 

The boundary $y=-1$ ($-1<x<1$) corresponds to the $\psi$-rotational axis. 
By introducing the radial coordinate $r$ by $y=-1-(1-\nu) r^2/2$, 
we observe that the metric as $r \to 0$ ($y\to -1$) behaves as
\begin{align}
 ds^2 \simeq \gamma^0_{tt}(x)dt^2 + 2 \gamma^0_{t\phi}(x) dt d\phi + \gamma^0_{\phi\phi}(x)d\phi^2 + \alpha_0(x)
 (dr^2+r^2 d\psi^2+G^{-1}(x)dx^2),
\end{align}
where 
\begin{align}
&\gamma_{tt}^{0} =-\frac{D|_{y=-1}H (-1,x)}{H (x,-1)},\quad
\gamma_{t\phi}^{0} = -\frac{H (-1,x)[c^3 \Omega_\phi(x,-1)-s^3 \Omega_\psi(-1,x)]}{D^2|_{y=-1}H (x,-1)},\nonum
&\gamma_{\phi\phi}^{0} =-\frac{D|_{y=-1} F(x,-1)}{H (-1,x)}-\frac{H (-1,x)[c^3\Omega_\phi(x,-1)-s^3 \Omega_\psi(-1,x)]^2}{D^2|_{y=-1} H (x,-1)},\nonum
& \alpha_0 =\frac{\ell^2 D|_{y=-1}  H (x,-1)}{8 (1+x)^2 (1-\lambda ) \lambda  \nu ^2 \left(1-\nu ^2\right) \left(1-\sigma ^6\right) h_2}.
\end{align}
For the two-dimensional metric $\gamma^0_{AB}(A,B=t,\phi)$, one can also show
\begin{align}
{\rm det}\,(\gamma^0_{AB})
=-\frac{32 \ell^2 (1-\lambda ) \lambda  (1-\nu )^3 \nu ^2 (1+\nu )  (1-x)(1+x \nu ) \left(1-\sigma ^6\right) h_2}{(1+x) D|_{y=-1} H (x,-1)}.
\end{align}

It is evident from $H (x,y)>0$ and $D>0$ within the ranges~(\ref{eq:xyrange}), 
that $\alpha_0$ is a positive definite function and $\gamma^0_{AB}$ is a nonsingular and non-degenerate matrix for $-1<x<1$. 
Thus, the metric has no curvature singularities at $y=-1$.

\item

The boundary $y=-1/\nu$ $(-1<x<1)$ corresponds to the event horizon with the surface gravity~(\ref{eq:kappa})
and the null generator of the horizon is expressed as $v_{\cal H} = \partial/\partial t + \omega_\psi \partial/\partial\psi + \omega_\phi \partial/\partial \phi$ with the horizon angular velocity given in Eq.~(\ref{eq:omegai}).
It becomes evident that $y=-1/\nu$ is a regular Killing horizon if we introduce the ingoing/outgoing Eddington-Finkelstein coordinates by
\begin{align}
 dx^i = dx'^i \pm v_{\cal H}^i \frac{(1-\nu^2)}{2 \nu \kappa G(y)} dy, 
\end{align}
where $x^i=(t,\psi,\phi)$ and the metric near $y=-1/\nu$ behaves as
\begin{align}
&ds^2 \simeq \alpha_H(x) \left(\frac{4 \nu^2 \kappa^2 G(y) }{(1-\nu^2)^2}dt'^2 \pm \frac{4\nu \kappa}{1-\nu^2} dt' dy+\frac{dx^2}{G(x)} \right)\nonum
&+ \gamma^H_{\psi\psi}(x)(d\psi'-\omega_\psi dt')^2+ 2\gamma^H_{\psi\phi}(x)(d\psi'-\omega_\psi dt')(d\phi'-\omega_\phi dt')
+ \gamma^H_{\phi\phi}(x)(d\phi'-\omega_\phi dt')^2,
\end{align}
with
\begin{align}
\begin{split}
&\gamma_{\psi\psi}^{H} =\frac{D|_{y=-1/\nu} F(-1/\nu,x)}{H (-1/\nu,x)}
-\frac{H (-1/\nu,x)[c^3\Omega_\psi(x,-1/\nu)-s^3 \Omega_\phi(-1/\nu,x)]^2}{D^2|_{y=-1/\nu} H (x,-1/\nu)},\\
&\gamma_{\psi\phi}^{H} =-\frac{D|_{y=-1/\nu} J(x,-1/\nu)}{H (-1/\nu,x)}-\frac{H (-1/\nu,x)[c^3\Omega_\psi(x,-1/\nu)-s^3 \Omega_\phi(-1/\nu,x)][c^3\Omega_\phi(x,-1/\nu)-s^3 \Omega_\psi(-1/\nu,x)]}{D^2|_{y=-1/\nu} H (x,-1/\nu)},\\
&\gamma_{\phi\phi}^{H} =-\frac{D|_{y=-1/\nu} F(x,-1/\nu)}{H (-1/\nu,x)}-\frac{H (-1/\nu,x)[c^3\Omega_\phi(x,-1/\nu)-s^3 \Omega_\psi(-1/\nu,x)]^2}{D^2|_{y=-1/\nu} H (x,-1/\nu)},\\
& \alpha_H = \frac{\ell^2 D|_{y=-1/\nu}  H (x,-1/\nu)}{8 (1-\lambda ) \lambda  (1+x \nu )^2 \left(1-\nu ^2\right) \left(1-\sigma ^6\right) h_2},
\end{split}
\end{align}
where the two-dimensional metric $\gamma^H_{AB}(A,B=\psi,\phi)$ satisfies
\begin{align}
{\rm det}\, (\gamma^H_{AB})
=\frac{8 \ell^2 \left(1-\sigma^6\right) \left(1-x^2\right) (1-\lambda ) \lambda  \nu ^2 \left(1-\nu ^2\right)^3 h_2}{\kappa ^2 (1+x \nu ) D|_{y=-1/\nu}H (x,-1/\nu)}.
\end{align}
It can be shown from $H (x,y)>0$ and $D>0$ within the ranges~(\ref{eq:xyrange}), 
 that $\alpha_H$ is a positive definite function and $\gamma_{AB}^{H}$ is a nonsingular and non-degenerate matrix for $-1<x<1$.
Thus, the spacetime is smoothly continued to $-\infty<y<-1/\nu$ across the horizon $y=-1/\nu$. 

Moreover, in the Eddington-Finkelstein coordinate, the gauge potential also remains regular at the horizon $y=-1/\nu$ under the gauge transformation
\begin{align}
 A' =  A \pm d\left( \frac{(1-\nu^2)\Phi_e}{2\nu \kappa} \int \frac{dy}{G(y)}\right),
\end{align}
where $\Phi_e$ is the electric potential defined by
\begin{align}
&\Phi_e := - (A_t+A_\psi \omega_\psi+A_\phi \omega_\phi)\biggr|_{y=-1/\nu}  = -\frac{\sqrt{3} \ell c s \kappa  h_1}{4 \nu  \left(1-\nu ^2\right)
   \left((1-\lambda ) \lambda  \left(1-\sigma ^6\right) h_2\right){}^{3/2}} \nonum
&\quad \times    \biggr[ s \left((1-\lambda ) \nu  \left(1-\sigma ^3\right)^2 h_1-\lambda  \left(1+\sigma ^3\right)^2 h_1-\left(1-\sigma ^6\right) h_3 h_{-4}\right) h_{+5}\nonum
&\quad \quad +c
   \left(h_1 \left(\nu  \left(1+\sigma ^3\right)^2+\left(1-\sigma ^3\right)^2 h_2\right)+(1+\nu ) \left(1-\sigma ^6\right) h_3 h_{+4}\right) h_{+6}\biggr].\label{eq:phi-e}
\end{align}

\item

 Near the inner and outer rims of the ring horizon at $(x,y)=(\pm1,-1/\nu)$, the spacetime asymptotes to the Rindler spacetime. 
By introducing the coordinates $(r,\theta)$ 
\begin{align}
  x = \pm 1 \mp \frac{(1\pm\nu)\kappa R_{1,\pm}}{2\nu(1\mp\nu) \ell^2} r^2 \sin^2\theta,\quad y =-\fr{\nu} \left(1- \frac{(1\pm\nu)\kappa R_{1,\pm}}{4\nu \ell^2} r^2 \cos^2\theta\right),
\end{align}
the metric near $r=0$ ($(x,y)=(\pm1,-1/\nu)$) behaves as
\begin{align}
 ds^2 \simeq dr^2 + r^2 d\theta^2 + r^2 \sin^2\theta (d\phi-\omega_\phi dt)^2  - r^2 \kappa^2 \cos^2 \theta dt^2 + R_{1,\pm}^2 (d\psi-\omega_\psi dt)^2,
\end{align}
where $R_{1,-}$ and $R_{1,+}$ are the $S^1$-radii of the outer and inner rims, respectively, given by
\begin{align}
& R_{1,+} :=\frac{4 (1-\lambda ) \lambda  \nu  \left(1-\nu ^2\right) \left(1-\sigma ^6\right) h_2}{\kappa  h_1 \left(c^2 h_{+5}^2+s^2 \left(1-\nu ^2\right) h_{+6}^2\right)},
\\
&R_{1,-} := 
\frac{4 (1-\lambda ) \lambda  \nu  \left(1-\nu ^2\right) \left(1-\sigma ^6\right) h_2}{\kappa  h_1 \left(s^2 h_{+5}^2+c^2 \left(1-\nu ^2\right) h_{+6}^2\right)}.
\end{align}
In the Cartesian coordinates $(T, X, Y, Z,W)=(\kappa t, r\cos\theta, r\sin\theta \cos(\phi-\omega_\phi t), r\sin\theta \sin(\phi-\omega_\phi t),R_{1,\pm} (\psi-\omega_\psi t) )$, the asymptotic metric becomes
\begin{align}
 ds^2 \simeq - X^2 dT^2 + dX^2 + dY^2 + dZ^2 +dW^2,
 \end{align}
 where $X=0$ corresponds the Rindler horizon. Hence, the metric is regular at $(x,y)=(\pm1,-1/\nu)$.

\end{enumerate}

\subsection{Absence of closed timelike curves}

The absence of CTCs are demonstrated  if the two-dimensional metric $g_{IJ} (I,J=\psi,\phi)$ is positive definite, except at the axes where $x=\pm1$ and $y=-1$.  
This is equivalent to having the conditions ${\rm det}(g_{IJ})>0$ and ${\rm tr}(g_{IJ})> 0$ for $-1<x<1$ and $-1/\nu \leq y <-1$. 
Following the same reason as in our previous work~\cite{Suzuki:2024coe} on the charged dipole black ring, it is sufficient to  demonstrate only ${\rm det}(g_{IJ})>0$.

\begin{figure}[t]
\includegraphics[width=6cm]{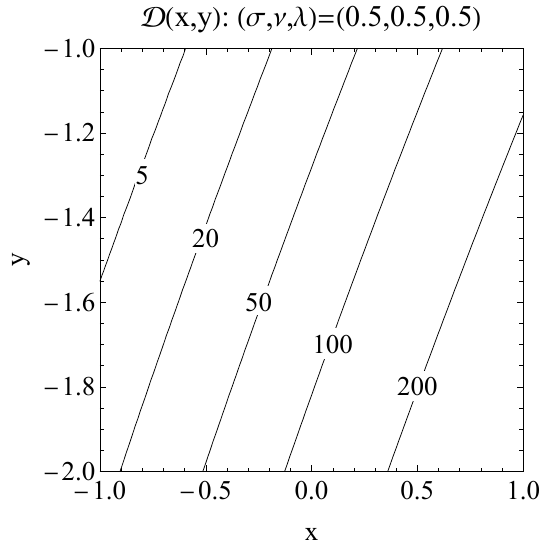}
\caption{
 ${\cal D}(x,y)$ for $\sigma=0.5, \ \nu=0.5,\ \lambda=0.5$. 
Other parameter choices yield similar profiles.
\label{fig:noctc}}
\end{figure}

From Eq.~(\ref{eq:def-t3}), $\mathrm{det}(g_{IJ})$ vanishes on the $\phi$ and $\psi$-rotational axes $x=\pm 1$ and $y=-1$.
To demonstrate the positive definiteness of $\mathrm{det}(g_{IJ})$ away from these, 
it is beneficial to consider the positive definiteness of the alternative function ${\cal D} (x,y)$ for $-1 \leq x \leq 1$ and $-1/\nu \leq y \leq -1$, defined by
\begin{align}
{\cal D} (x,y):= \frac{ (x-y)^4 D H (x,y)}{ \ell^4  (1-x^2)(-1-y)}  {\rm det}(g_{IJ}).
\end{align}
Although analytically proving the positivity of  ${\cal D} (x,y)$ across the entire region is challenging, it is still possible to demonstrate positivity on the horizon:
\begin{align}
{\cal D} (x,-1/\nu) = \frac{8 h_2 (1-\lambda ) \lambda  (1-\nu )^2 (1+\nu )^3 \left(1-\sigma ^6\right) (1+\nu  x)^3}{\nu \ell^2 \kappa ^2  }>0.
\end{align}
For the other region, the positivity both on and outside the horizon has been numerically confirmed for several values of $\nu,\gamma,\sigma$ within the parameter range presented in Eq.~(\ref{eq:paramregion-extend-new})
(see Fig. \ref{fig:noctc}).

\section{Phase of the charged rotating black ring}\label{sec:property}

Now, we study the physical properties of the charged dipole black ring constructed in the previous section. 

\subsection{Thermodynamics}

As discussed in Ref.~\cite{Tomizawa:2009tb},  the charged dipole black ring in five-dimensional minimal supergravity, in general, is characterized by four independent conserved charges : mass, two angular momenta and electric charge as presented in Eqs.~(\ref{eq:mass-new}), (\ref{eq:jpsi-new}), (\ref{eq:jphi-new}) and (\ref{eq:Qe}), and
in addition, a non-conserved quantity called the dipole charge, defined by
\begin{align}
q &:= \fr{4\pi} \int_{S^2} F = - \frac{\sqrt{3} \ell s^2 h_1}{c\sqrt{(1-\nu^2)(1-\sigma^6)(1-\lambda)\lambda h_2}},
\end{align}
As argued  initially for the dipole black ring~\cite{Emparan:2004wy}, although the dipole charge is not a conserved charge such that obtained from the surface integral at infinity, it behaves as the fifth independent parameter in the first law. 
This can be interpreted that the black ring in the five-dimensional minimal supergravity exhibits infinite non-uniqueness or has a `thicker hair', in contrast to the black holes in the same theory that is specified uniquely only by asymptotic conserved charges~\cite{Tomizawa:2009tb,Tomizawa:2009ua}.

\medskip
Together with the surface gravity~(\ref{eq:kappa}) and 
the area of the horizon cross-section given by
\begin{align}
 A_H = 8 \pi^2 \ell^2 \nu \kappa^{-1},
\end{align}
the charged dipole black ring follows
 the Smarr formula and the first law with respect to the five  variables as below~\cite{Copsey:2005se}
\begin{align}
  M& = \frac{3}{16\pi} \kappa  A_H  +\frac{3}{2} \omega_\psi J_\psi +\frac{3}{2}\omega_\phi J_\phi
  + \frac{1}{2} \Phi_e Q+\fr{2}\Phi_m q, \label{eq:smarr}\\
\delta M &= \frac{1}{8\pi} \kappa \delta A_H  + \omega_\psi \delta J_\psi +\omega_\phi \delta J_\phi+ \frac{1}{2} \Phi_e \delta Q+\Phi_m \delta q, \label{eq:1stlaw}
\end{align}
where $\Phi_e$ is the electric potential defined in Eq.~(\ref{eq:phi-e})
 and $\Phi_m$ is the `magnetic dipole potential' defined in Refs.~\cite{Copsey:2005se,Kunduri:2013vka}. 
Due to the cumbersome definition of $\Phi_m$, we instead use the Smarr formula~(\ref{eq:smarr}) to evaluate $\Phi_m$, which yields
 \begin{align}
 \Phi_m =\sqrt{\frac{\left(1-\sigma^6\right) (1-\lambda ) \lambda  h_2}{1-\nu ^2}} \frac{\sqrt{3} \ell \pi c s  (1-\nu )^2 (1+\nu )  h_{-4} h_{+4}}{c^3 (1+\nu ) h_2 h_{+4}^2+s^3 (1-\nu ) h_{-4}
   \left((1-\lambda ) \lambda  \left(1-\nu ^2\right) h_{+6}-h_2 h_{+4}\right)}.\label{eq:dipolepot}
 \end{align}
 With this magnetic dipole potential, one can confirm the first law~(\ref{eq:1stlaw}) by differentiating with the four independent parameters $(\ell, \nu,\lambda,\sigma)$.
Despite the expectation that the most general charged dipole black ring is described by the five independent parameters~\cite{Emparan:2004wy}, our solution has only four independent parameters, implying that the dipole charge is not independent of the asymptotic charges.

\subsection{Phase diagram}

To study the phase diagram of the black ring, it is convenient to introduce the dimensionless variables defined in terms of the mass scale introduced by $r_M :=\sqrt{ 8 G_5 M/3\pi}$ 
\begin{align}
j_\psi := \frac{4 G_5  }{\pi r_M^3} J_\psi ,\quad j_\phi := \frac{4G_5 }{\pi r_M^3}J_\phi ,\quad \bar{q}= \frac{q}{r_M},\quad
a_H := \frac{\sqrt{2}}{\pi^2 r_M^3} A_H, 
\end{align}
and the charge to mass ratio is given by Eq.~(\ref{eq:Qe}), where the solution saturates the BPS bound at $\sigma \to \pm 1$. 
We find that the dimensionless variables are written in the following forms
\begin{align}
&j_\psi =\frac{ j_1 + \sigma^3 j_2}{4(1+\sigma^2)^{3/2} \sqrt{h_1}(h_1+(1-\lambda)\lambda(1-\nu^2)(1-\sigma^6))^{3/2}},\\
&j_\phi =\frac{ j_2 + \sigma^3 j_1}{4(1+\sigma^2)^{3/2} \sqrt{h_1}(h_1+(1-\lambda)\lambda(1-\nu^2)(1-\sigma^6))^{3/2}},
\end{align}
with $j_1$ and $j_2$ given in Eq.~(\ref{eq:j1j2}), and
\begin{align}
\bar{q}&= -\frac{\sqrt{3} \sqrt{h_1} \sigma ^2}{2 \sqrt{\sigma ^2+1} \sqrt{h_1+(1-\lambda ) \lambda  \left(1-\nu ^2\right) \left(1-\sigma ^6\right)}},\\
a_H&=\frac{\left(1-\sigma^6\right) \sqrt{1-\nu ^2} \left(h_2 h_{+4} h_{+5}+\sigma^3 (1-\lambda ) \lambda  (1-\nu )^2 (1+\nu ) h_{-4}
   h_{+6}\right)}{\sqrt{2} \left(1+\sigma^2\right)^{3/2} \sqrt{h_1} \left(h_1+(1-\lambda ) \lambda  \left(1-\nu ^2\right) \left(1-\sigma
   ^6\right)\right){}^{3/2}}.
\end{align}
Additionally, the surface gravity and horizon angular velocities are expressed as
\begin{align}
r_M \kappa &=\frac{4 h_2 (1-\lambda ) \lambda  \nu  \sqrt{1-\nu ^2} \left(1-\sigma ^2\right) \sqrt{\sigma ^2+1} \sqrt{h_1+(1-\lambda ) \lambda  \left(1-\nu
   ^2\right) \left(1-\sigma ^6\right)}}{\sqrt{h_1} \left((1-\lambda ) \lambda  (1-\nu )^2 (\nu +1) \sigma ^3 h_{-4} h_{+6}+h_2 h_{+4}
   h_{+5}\right)},\\
r_M \omega_\psi &=
\frac{h_1 r_M \kappa h_{+5} h_{+6}}{4 h_2 (1-\lambda ) \lambda  \nu  \sqrt{1-\nu ^2} \left(1-\sigma^6\right)},\\
r_M \omega_\phi&=\frac{r_M \kappa h_{+7}}{4 h_2 (1-\lambda ) \lambda  \nu  \sqrt{1-\nu ^2} \left(1-\sigma^6\right)}.
\end{align}

One can observe that  the replacement  $(\sigma,j_\phi) \to (- \sigma, -j_\phi)$ corresponds to a simple coordinate transformation $(t,\psi) \to (-t,-\psi)$.
Hence,  without loss of generality, we restrict our discussion to the case of $\sigma \geq 0$.

\medskip

\begin{figure}
\begin{center}
\includegraphics[width=8.5cm]{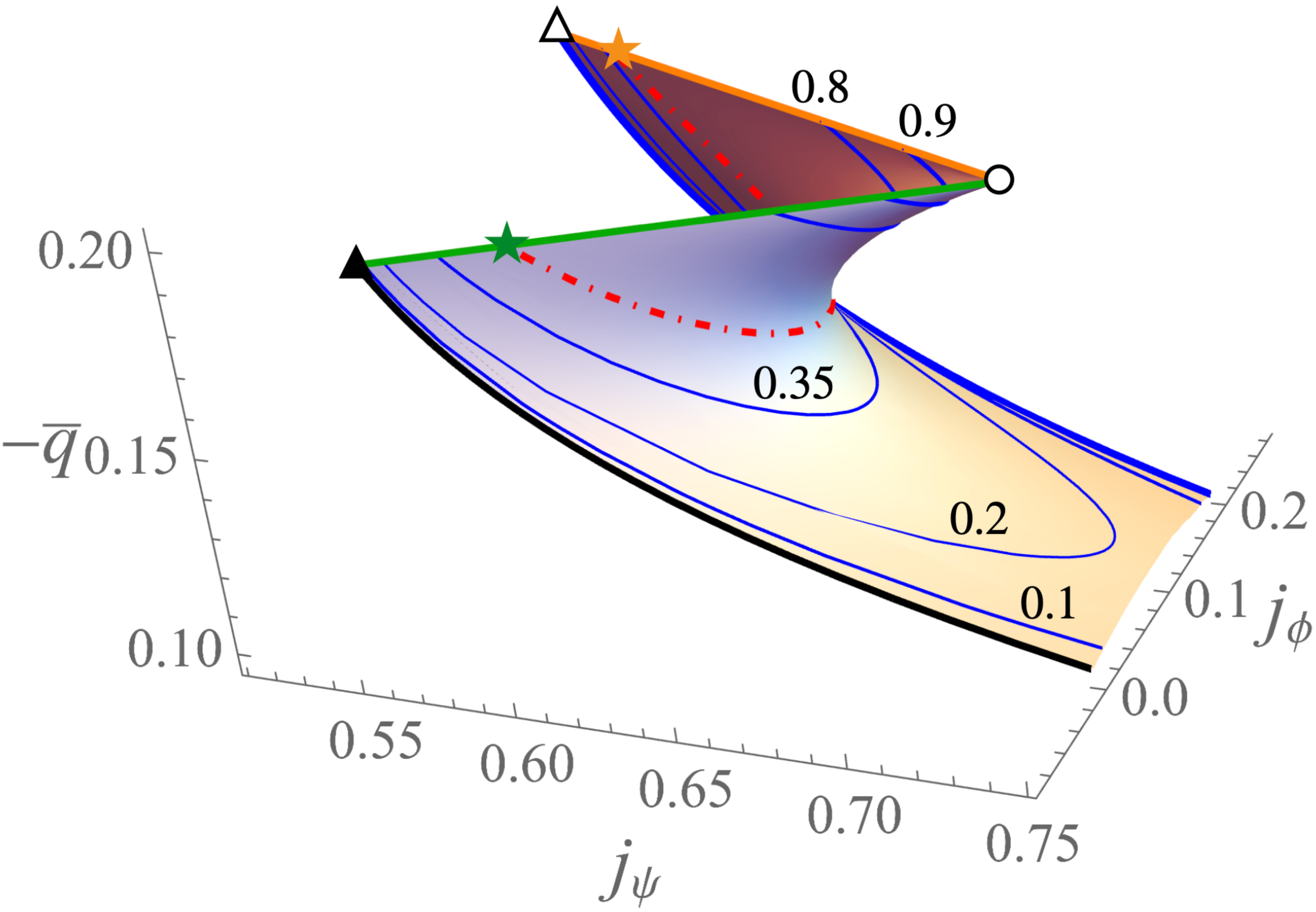}\\
\includegraphics[width=6.5cm]{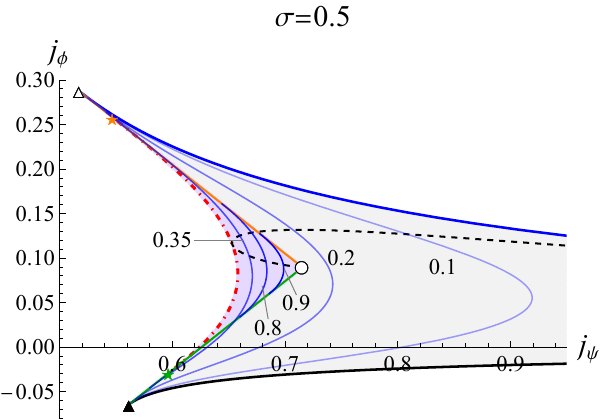}\hspace{0.1cm}
\includegraphics[width=6.7cm]{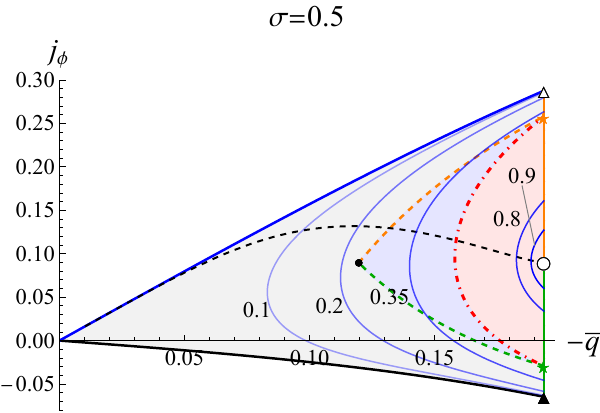}
\caption{
The phase diagram for the charged dipole black ring with $\sigma=0.5$ illustrated in the $(j_\psi,j_\phi)$ plane (bottom left panel), the $(-\bar{q},j_\phi)$ plane (bottom right panel), and as a two-dimensional surface in the $(j_\psi,j_\phi,-\bar{q})$ space (top panel). 
Here, the curves of $\nu={\rm constant}$ for $\tilde{\nu}=0.1, 0.2, 0.35, 0.8, 0.9$ are shown by blue thin curves.
The black dashed curves in bottom panels correspond to $\gamma=1 (\tilde{\gamma}=0)$, above which the parameter region is extended. 
The boundaries of the allowed region are defined by two extremal black ring curves (black thick curve and blue thick curve), two extremal black hole curves (green line for $\lambda=0$ and orange line for
$\lambda=1$), and a critical curve where the fat ring and thin ring branches bifurcate (red dot-dashed curve).
The white circles correspond to the point $(j_{\psi,*},j_{\phi,*})$ at which the two extremal black hole curves intersect, and the white and black triangles corresponds to, respectively, the points $(j_{\psi,\#}^+,j_{\phi,\#}^+)$ and
$(j_{\psi,\#}^-,j_{\phi,\#}^-)$ at which the extremal black hole curves intersect with the extremal black ring curves.
The orange and green stars represent the points $(j_{\psi,c}^+,j_{\phi,c}^+)$ and $(j_{\psi,c}^-,j_{\phi,c}^-)$, respectively at which the critical curves end at the extremal black hole curves.
In the bottom left panel, the two branches exist for the same asymptotic charges within the purple colored region bounded by the critical, $\lambda=0$, and $\lambda=1$ curves, whereas in the bottom right panel, the fat ring branch and its counterpart in the thin ring branch are shown in red and blue regions, respectively. 
The latter region is bounded by the critical curve and the counterparts of the $\lambda=0$ and $\lambda=1$ curves, which are represented by the green dashed and orange dashed curves, respectively. 
The black dot in the bottom right panel corresponds to the counterpart of $(j_{\psi,*},j_{\phi,*})$.
\label{fig:jjplot-ex1}}
\end{center}
\end{figure}

\begin{figure}
\begin{minipage}[b]{0.47\columnwidth}
\includegraphics[width=6.5cm]{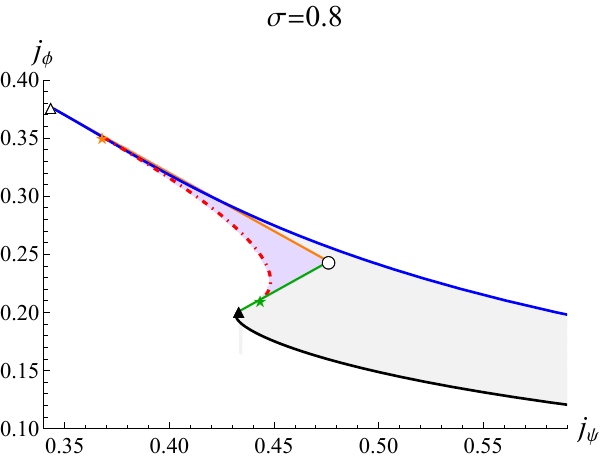}
\includegraphics[width=6.5cm]{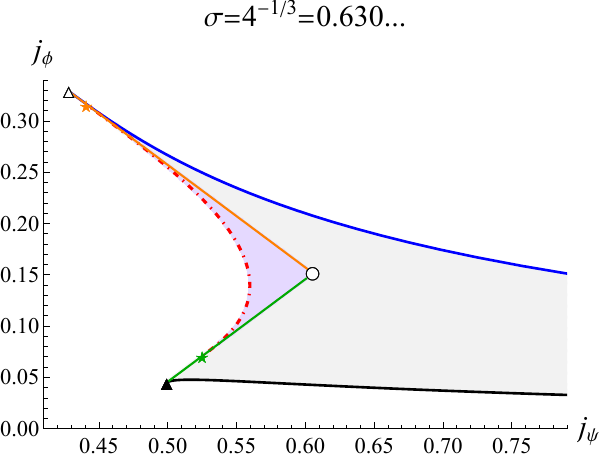}\\
\includegraphics[width=6.5cm]{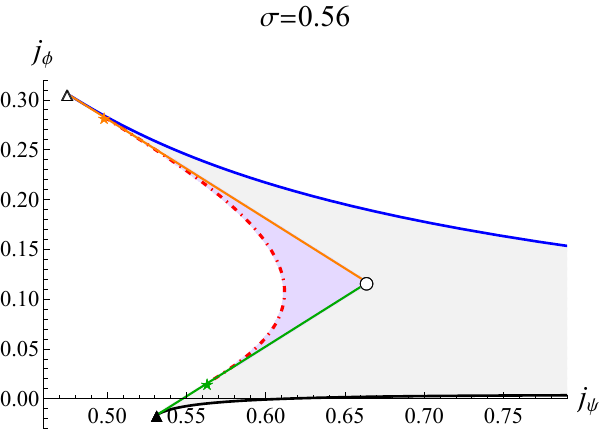}\\
\end{minipage}
\begin{minipage}[b]{0.47\columnwidth}
\includegraphics[width=6.7cm]{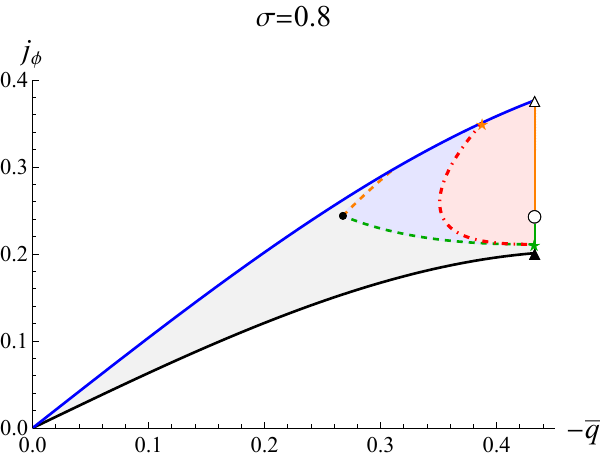}
\includegraphics[width=6.7cm]{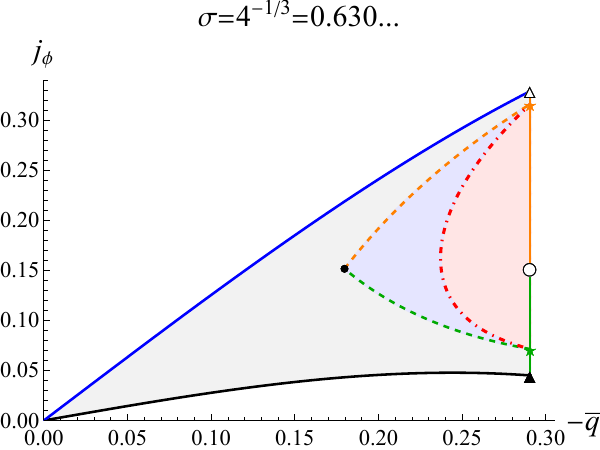}\\
\includegraphics[width=6.7cm]{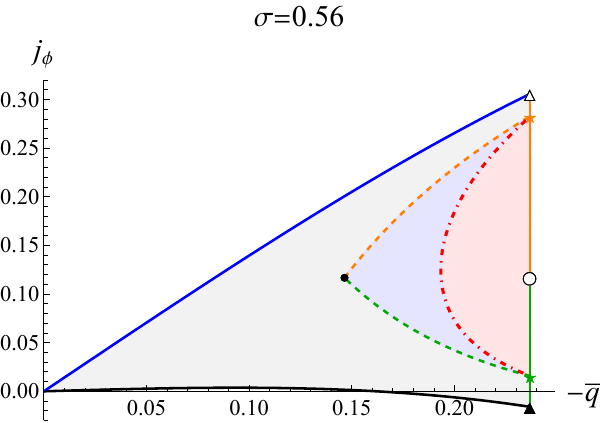}\\
\end{minipage}
\caption{
Phases diagrams in the $(j_\psi,j_\phi)$-plane (left panels) and in the $(-\bar{q},j_\phi)$-plane (right panels) for $\sigma=0.8,4^{-1/3}=0.630\ldots,0.56$.
The boundaries of the allowed region are given by the two extremal black ring curves (thick black curve and blue black curve), the two extremal black hole curves ($\lambda=0$ for green line and $\lambda=1$ for orange line) and the critical curve on which fat ring and thin ring branches bifurcate (red dot-dashed curve). 
The white circles correspond to $(j_{\psi,*},j_{\phi,*})$ at which the two extremal black hole curves intersect.
The white and black triangles corresponds to, respectively, the points  $(j_{\psi,\#}^+,j_{\phi,\#}^+)$ and 
$(j_{\psi,\#}^-,j_{\phi,\#}^-)$ at which the extremal black hole curves intersect with the extremal black ring curves.
The orange and green stars correspond to, respectively, the points $(j_{\psi,c}^+,j_{\phi,c}^+)$ and $(j_{\psi,c}^-,j_{\phi,c}^-)$ at which the critical curves end at the extremal black hole curves or extremal black ring curves. 
The two branches exist for the same asymptotic charges in the purple colored region bounded by critical, $\lambda=0$ and $\lambda=1$ curves. 
In right panels, the fat ring branch and the counterpart corresponding to the thin ring  are shown in the red and blue colored regions, respectively. 
The latter region is bounded by the critical curve and the counterparts of $\lambda=0$ and $\lambda=1$ curves where the latter two are represented by green dashed and orange dashed curves, respectively. 
The black dot in right panels corresponds to the counterpart of $(j_{\psi,*},j_{\phi,*})$.
  \label{fig:n0-jjplots}}
\end{figure}

The phases for $\sigma={\rm constant}$ are  determined simply by two independent parameters, $(\nu,\lambda) \in (0,1) \times (0,1)$.
As a representative example, in the top panel of Fig.~\ref{fig:jjplot-ex1}, we show the phase diagram for $\sigma=0.5$ as a two dimensional surface in the $(j_\psi,j_\phi,-\bar{q})$ space.
We also display the phase diagram in the $(j_\psi,j_\phi)$ and $(-\bar{q},j_\phi)$ planes for $\sigma=0.5$ in Fig.~\ref{fig:jjplot-ex1} and for other values of $\sigma$ in Fig.~\ref{fig:n0-jjplots}.
Additionally,  the curves of $\nu={\rm constant}$ are depicted as blue thin curves in Fig.~\ref{fig:jjplot-ex1}.
In bottom panels of Fig.~\ref{fig:jjplot-ex1}
, we indicate the $\gamma=1\ (\tilde\gamma=0)$ curve by black dashed curves, beyond which the parameter region is extended. 
The initial parameter region~(\ref{eq:rangenugam}) before the extension corresponds to  the shaded region below this curve in  each bottom panel.

\medskip

The boundaries of the phase space for $\sigma={\rm constant}$ are divided into the following limits:

\begin{enumerate}

\item 
The $j_\psi \to \infty$ limit given by $\nu \to 0$  while maintaining $\lambda = \ord{\nu}$:
 This limit also results in $j_\phi \to 0$, $\bar{q} \to 0$ and $a_H \to 0$, with both ring radii $R_{1,\pm}/r_M$ diverging as $\nu^{-1/2}$.

\item 

The non-BPS (non-supersymmetric), extremal black hole limit: 
This limit can be obtained by two limits $\lambda\to 0$ with  $\nu  \ (0<\nu<1)$,  $\sigma(\not= 1)$ fixed  (depicted as green curves in Fig.~\ref{fig:jjplot-ex1} and \ref{fig:n0-jjplots}) and $\lambda\to 1$ with $\nu\ (0<\nu<1)$, $\sigma (\not= 1)$ fixed (depicted as orange curves  in Fig.~\ref{fig:jjplot-ex1} and \ref{fig:n0-jjplots}).
The dimensionless angular momenta for $\lambda\to 0$ is expressed as
\begin{align}
j_\psi = \frac{\hat{l}_1+\sigma^3 \hat{l}_2 }{(1+\sigma^2)^{3/2}},\quad
j_\phi = \frac{\hat{l}_2+\sigma^3 \hat{l}_1 }{(1+\sigma^2)^{3/2}}
\end{align}
with
\begin{align}
\hat{l}_2 =\hat{l}_1-1 = -\fr{4}(1-\nu^2)(1+\sigma^3)^2.
\end{align}
Similarly, for $\lambda\to 1$ they can be in the same form but with different values for $\hat{l}_1$ and $\hat{l}_2$ as
\begin{align}
\hat{l}_2 =1-\hat{l}_1 = \fr{4}(1-\nu^2)(1+\sigma^3)^2.
\end{align}
Both limits correspond to the non-BPS, extremal limits of the Cveti\v{c}-Youm black hole~\cite{Cvetic:1996xz,Cvetic:1996kv}, where $\hat{l}_1$ and $\hat{l}_2$ above correspond to $l_1/\sqrt{2m}$ and $l_2/\sqrt{2m}$ therein, respectively. The similar limit to extremal black holes exists for the Pomeransky-Sen'kov solution, in which case such a limit leads to the extremal Myers-Perry black holes~\cite{Elvang:2007hs}.
One can find that the metric in Eq.~(\ref{eq:metricsol-new}) appears to be divergent both as $\lambda \to 0$ and $\lambda \to 1$. 
However, 
as discussed for the vacuum black hole limit from the vacuum black ring~\cite{Elvang:2007hs}, this divergence is merely apparent in the charged case as well and can be eliminated through the appropriate coordinate transformation and redefinition of the parameter $\ell$.
In fact, one can reproduce the corresponding solution of the extremal Cveti\v{c}-Youm black hole from Eqs.~(\ref{eq:metricsol-new}) and (\ref{eq:gaugesol-new}) by taking either the limit $\lambda \to 0$ with $m:=2\nu^2\ell^2/(\lambda(1-\nu^2)(1-\sigma^6))$ held fixed, or
 the limit $\lambda \to 1$ with $m:=2\ell^2/((1-\lambda)(1-\nu^2)(1-\sigma^6))$ held fixed, using the coordinates $(r,\theta)$ defined by $(x+1,y+1)=4(1-\nu)\ell^2 (\sin^2 \theta,-\cos^2\theta)/(r^2-2m|\hat{l}_1||\hat{l}_2|)$.

\medskip

In the $(j_\psi,j_\phi)$ plane, the limits $\lambda\to1$ and $\lambda\to0$ with $\nu$ fixed are presented by  the straight lines, respectively, 
\begin{align}
j_\psi + j_\phi = \frac{1+\sigma^3}{(1+\sigma^2)^{3/2}},
\end{align}
\begin{align}
j_\psi - j_\phi = \frac{1-\sigma^3}{(1+\sigma^2)^{3/2}}.
\end{align}
These two lines join at the limit $\nu \to 1$ (white circles in Fig.~\ref{fig:jjplot-ex1} and \ref{fig:n0-jjplots}), where
\begin{align}
(j_{\psi,*},j_{\phi,*}) :=  \frac{1}{(1+\sigma^2)^{3/2}}(1,\sigma^3).
\end{align}
At the limit $\nu\to0$, the curves represented by $\lambda\to 1$ and $\lambda\to 0$ have other ends at $(j_{\psi,\#}^+,j_{\phi,\#}^+)$ and  at 
$(j_{\psi,\#}^-,j_{\phi,\#}^-)$, respectively (depicted by white and black triangles in Figs.~\ref{fig:jjplot-ex1} and \ref{fig:n0-jjplots}, respectively), where
\begin{align}
(j_{\psi,\#}^\pm,j_{\phi,\#}^\pm):= \left(\frac{3\mp \sigma^3+\sigma^6\pm \sigma^9}{4\left(1+\sigma^2\right)^{3/2}}, 
\pm \frac{1 \pm 5 \sigma^3 - \sigma^6 \mp \sigma^9}{4\left(1+\sigma^2\right)^{3/2}}\right).
\end{align}

\medskip

Here, it is worth noting  that for both of these limits, the dipole charge has the same nonzero value for $\sigma\neq 0$:
\begin{align}
 \bar{q} = -\bar{q}_{\rm max} := - \frac{\sqrt{3} \sigma^2}{2\sqrt{1+\sigma^2}},
\end{align}
which gives the upper bound for the absolute value of the dipole charge, 
as illustrated in Figs.~\ref{fig:jjplot-ex1} and \ref{fig:n0-jjplots}.

\item

The non-BPS (non-supersymmetric), extremal black ring limit: 
This limit can be obtained by the limits $\nu \to 0$ with $\lambda/\nu^2$ and $\sigma(\not= 1)$ fixed (black thick curves in Figs.~\ref{fig:jjplot-ex1} and \ref{fig:n0-jjplots}) and $\nu \to 0$ with $\lambda$ and $\sigma(\not= 1)$ fixed
 (blue thick curves in Figs.~\ref{fig:jjplot-ex1} and \ref{fig:n0-jjplots}).
 These two limits correspond to the  non-BPS, extremal black ring phase with $r_M \kappa=0$.

\medskip
To obtain the former curve, we take the limit $\nu\to 0$ under the constraint
\begin{align}
 \lambda = \frac{1-\eta^2}{\eta^2(1-\sigma^6)} \nu^2, \quad 0<\eta<1.
\end{align}
where $\eta$ serves as the control parameter for this limit. 
At this limit, the dimensionless physical quantities are expressed as
\begin{align}\label{eq:phase-extremalring}
\begin{split}
j_\psi &= \frac{2+\eta ^2 \left(1+\sigma ^6\right)+\eta ^4 \sigma^3 \left(1-\sigma^6 \right)}{4 \eta  \left(1+\sigma ^2\right)^{3/2}},\\
j_\phi &=-\frac{\eta 
   \left(1-6 \sigma^3 - \sigma^6 + \eta^2 \sigma^3 (1+\sigma^6)\right)}{4   \left(1+\sigma ^2\right)^{3/2}},\\
\bar{q} &= -     \frac{ \sqrt{3} \sigma ^2  }{  2 \sqrt{1+\sigma ^2}}  \eta.
\end{split}
\end{align}
For the latter curve, the limit $\nu\to 0$ with $\lambda$ fixed yields in the following expressions:
\begin{align}
\begin{split}
j_\psi &= \frac{2+\lambda -\lambda ^2 \sigma ^3+\left(-4+4 \lambda +\lambda ^2\right)
   \sigma ^6+\lambda ^2 \sigma ^9+\left(2-5 \lambda +3 \lambda ^2\right) \sigma
   ^{12}}{4 \sqrt{\lambda } \left(1+\sigma ^2\right)^{3/2} \left(1-(1-\lambda )   \sigma ^6\right)^{3/2}},\\
j_\phi &=   \frac{\sqrt{\lambda } 
\left(1+(6-\lambda ) \sigma   ^3-(2-\lambda ) \sigma ^6-(6-5 \lambda ) \sigma ^9+(1-\lambda ) \sigma  ^{12}\right)}{4 \left(1+\sigma ^2\right)^{3/2} \left(1-(1-\lambda ) \sigma   ^6\right)^{3/2}},\\
\bar{q} &=  - \frac{\sqrt{3} \sigma ^2 \sqrt{\lambda}
   }{2\sqrt{\left(1+\sigma ^2\right) \left(1-(1-\lambda ) \sigma ^6\right)}}.
   \end{split}
\end{align}
By the transformation $\sigma \to -\sigma$ and $j_\phi \to - j_\phi$, one can verify that this coincides with
 Eq.~(\ref{eq:phase-extremalring})
 through the correspondence
\begin{align}
\lambda = \frac{(1-\sigma^6)\eta^2}{1-\sigma^6 \eta^2},\quad 0<\eta <1,
\end{align}

meaning that a solution on each curve with $(\sigma,\eta) = (\sigma_1>0,\eta_1)$ is identified with another solution on the other curve with $(\sigma,\eta)=(-\sigma_1<0,\eta_1)$ in the phase space $(j_\psi,j_\phi,\bar{q})$ through the transformation $\sigma \to -\sigma$ and $j_\phi \to - j_\phi$, which stems from the reflection symmetry of the system $(t,\psi) \to (-t,-\psi)$. Moreover, one can also confirm that the metric~(\ref{eq:metricsol-new}),  up to the same symmetry transformation, is identical for the two curves with the same $\eta$.
Henceforth, these two curves describe the same non-supersymmetric extremal black ring phase.

Below, we use $\eta$ as the common parameter for both curves.

\medskip

These two curves join with the non-BPS, extremal black hole limit 
at $(j_\psi,j_\phi,\bar{q})=(j_{\psi,\#}^\pm,j_{\phi,\#}^\pm,-\bar{q}_{\rm max})$ for  $\eta=1$,
For $\eta \to 0$, 
the both curves  extend to the $j_\psi \to \infty$ limit (bottom left panel in Fig.~\ref{fig:jjplot-ex1} and left panels in Fig.~\ref{fig:n0-jjplots}), at which the former curve approaches the $j_\phi=0$ line  with the following asymptotic form

\begin{align}
j_\phi \approx -\frac{1-6 \sigma^3-\sigma^6}{8 (1+ \sigma^2){}^3} \frac{1}{j_\psi},
\end{align}
and the latter with the form
\begin{align}
j_\phi \approx \frac{1+6 \sigma^3-\sigma^6}{8 (1+ \sigma^2){}^3} \frac{1}{j_\psi}.
\end{align}
In terms of $\eta$, the entropy and horizon angular velocities on each curve can be written as
\begin{align}
a_{H}^\pm  &=   \frac{\eta  \left(\eta ^2 \sigma^3 \left(\mp 1-2 \sigma ^3
\pm \sigma ^6\right)+\sigma ^6+1\right)}{\sqrt{2} \left(\sigma
   ^2+1\right)^{3/2}},\\
r_M \omega_{\psi}^\pm  &=  \frac{(1-\sigma^2)^2(1+\sigma^2+\sigma^4)\eta^2}{\sqrt{2}(1+t^2)a_{H}^\pm  },\\
r_M \omega_{\phi}^\pm &= \pm \frac{(1-\sigma^2)(2-(1+\sigma^6)\eta^2)}
{\sqrt{2}(1+t^2)a_{H}^\pm },
\end{align}
where the minus signature is for the former curve and the plus signature for the latter.
Note that, at the neutral limit $\sigma \to 0$, this phase reduces to the extremal limit of the Pomeransky-Sen'kov solution studied in Ref.~\cite{Elvang:2007hs}.

\item

The limit to the critical curve: 
This is defined by the envelope curves for  constant $\lambda$ or constant $\nu$ (illustrated as red dot-dashed curves in Figs.~\ref{fig:jjplot-ex1} and \ref{fig:n0-jjplots}), derived from the condition 
\begin{align}\label{eq:crit-con}
\partial_\nu j_\psi \partial_\lambda j_\phi - \partial_\lambda j_\psi \partial_\nu j_\phi =0,
\end{align}
which we solve numerically. 
This curve forms a part of the lower bound for $j_\psi$ in the $(j_\psi,j_\phi)$ plane, at which two distinct branches with the same $(j_\psi,j_\phi)$  bifurcate. 
As illustrated as the purple regions in the bottom left panel of  Fig.~\ref{fig:jjplot-ex1} and left panels of Fig.~\ref{fig:n0-jjplots}, the phase space for these two branches overlaps in the $(j_\psi,j_\phi)$-plane, which implies that  there exist two different black rings that share the same $(j_\psi,j_\phi)$, or equivalently,  the same conserved charges $(M,Q,J_\psi.J_\phi)$.
 However, these branches can be differentiated by the values of their dipole charge  in the $(-\bar{q},j_\phi)$ plane, as indicated  by the red and blue colored regions in the bottom right panel of  Fig.~\ref{fig:jjplot-ex1} and right panels of Fig.~\ref{fig:n0-jjplots}.  
In the $(-\bar{q},j_\phi)$ plane, this critical curve is not the boundary of the phase space but just gives a ridge curve for $j_\psi=j_\psi(\bar{q},j_\phi)$. 

\medskip
The critical curve has endpoints on the curve corresponding to the extremal black holes or the extremal black rings, depending on the value of $\sigma$. 
For $0<\sigma \leq 3^{-1/3}$, it terminates at the point $(j_{\psi,c}^-,j_{\phi,c}^-)$ corresponding to the intersection point of the extremal black hole curve $\lambda= 0$ with  $\nu=\nu_{c,-}$ (green stars in Figs.~\ref{fig:jjplot-ex1} and \ref{fig:n0-jjplots})  and at the point $(j_{\psi,c}^+,j_{\phi,c}^+)$ corresponding to the intersection point of another extremal black hole curve $\lambda=1$ with $\nu=\nu_{c,+}$ (orange stars in Fig.~\ref{fig:jjplot-ex1} and middle and bottom panels of Fig.~\ref{fig:n0-jjplots}),  where
\begin{align}
\nu_{c,\pm} = \sqrt{\frac{1\mp 2 \sigma ^3-3 \sigma ^6}{5\mp 2 \sigma ^3-3 \sigma ^6}},
\end{align}
and
\begin{align}
\begin{split}
(j_{\psi,c}^\pm,j_{\phi,c}^\pm) = \left(\frac{4\pm \sigma ^3-\sigma ^6}{\left(1+\sigma
   ^2\right)^{3/2} \left(5\pm 3 \sigma ^3\right)}, \pm \frac{1\pm 7 \sigma
   ^3+4 \sigma ^6}{\left(1+\sigma ^2\right)^{3/2} \left(5\pm 3 \sigma
   ^3\right)} \right).
   \end{split}
\end{align}
For $\sigma> 3^{-1/3}$, the endpoint $(j_{\psi,c}^+,j_{\phi,c}^+)$ transitions from the extremal black hole curve given by $\lambda=1$ to the adjacent extremal black ring curve, as illustrated by the orange star in the top panels of Fig.~\ref{fig:n0-jjplots}. 
This eventually approaches the $j_\psi \to \infty$ limit as $\sigma \to 1$. 
In this range, we numerically determine $(j_{\psi,c}^+,j_{\phi,c}^+)$ by solving Eq.~(\ref{eq:crit-con}) with $\nu=0$. 
The other endpoint remains on the extremal black hole curve $\lambda=0$.

\end{enumerate}

As in the case of the vacuum black ring~\cite{Pomeransky:2006bd}, the charged black ring admits (discrete) non-uniqueness in terms of the asymptotic charges $(M,Q,J_\psi,J_\phi)$, i.e. in a certain parameter region, two different branches exist for the same asymptotic charges $(M,Q,J_\psi,J_\phi)$.
 In Fig.~\ref{fig:jjplot-ex1} (bottom left panel), we have two branches bifurcating from the critical curve for the same $(j_\psi,j_\phi)$ in the purple colored region, where we call one extending to $j_\psi\to \infty$ as the `thin ring' and the other as the `fat ring'. In contrast to the vacuum case, 
these two branches are distinguished by labeling with the dipole charge as in Fig.~\ref{fig:jjplot-ex1} (bottom right panel), where the red and blue colored regions correspond to the fat ring and  the thin ring branches, respectively. 

\medskip

\begin{figure}
\begin{center}
\begin{minipage}[c]{0.3\columnwidth}
\includegraphics[width=4.9cm]{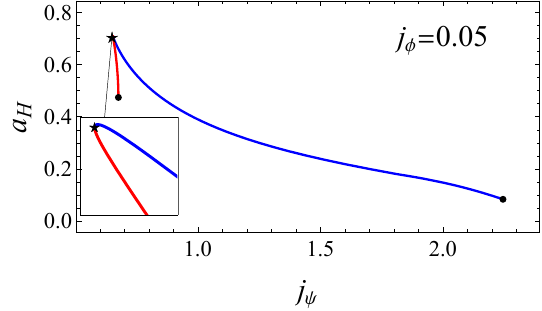}\\
\includegraphics[width=4.9cm]{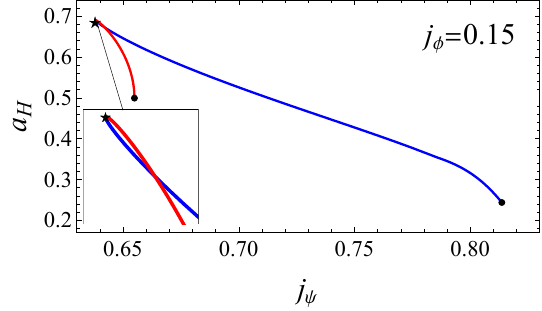}
\end{minipage}
\begin{minipage}[c]{0.3\columnwidth}
\includegraphics[width=4.9cm]{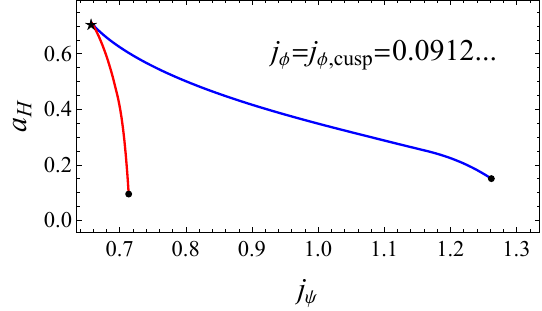}\\
\includegraphics[width=4.9cm]{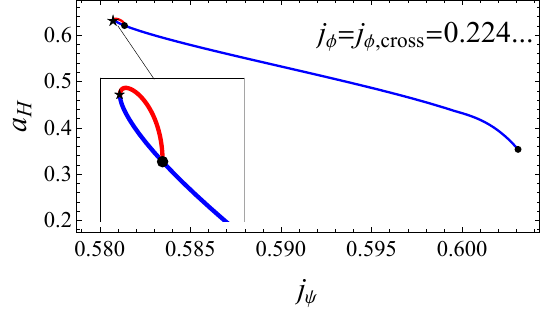}
\end{minipage}
\begin{minipage}[c]{0.38\columnwidth}
\includegraphics[width=6.2cm]{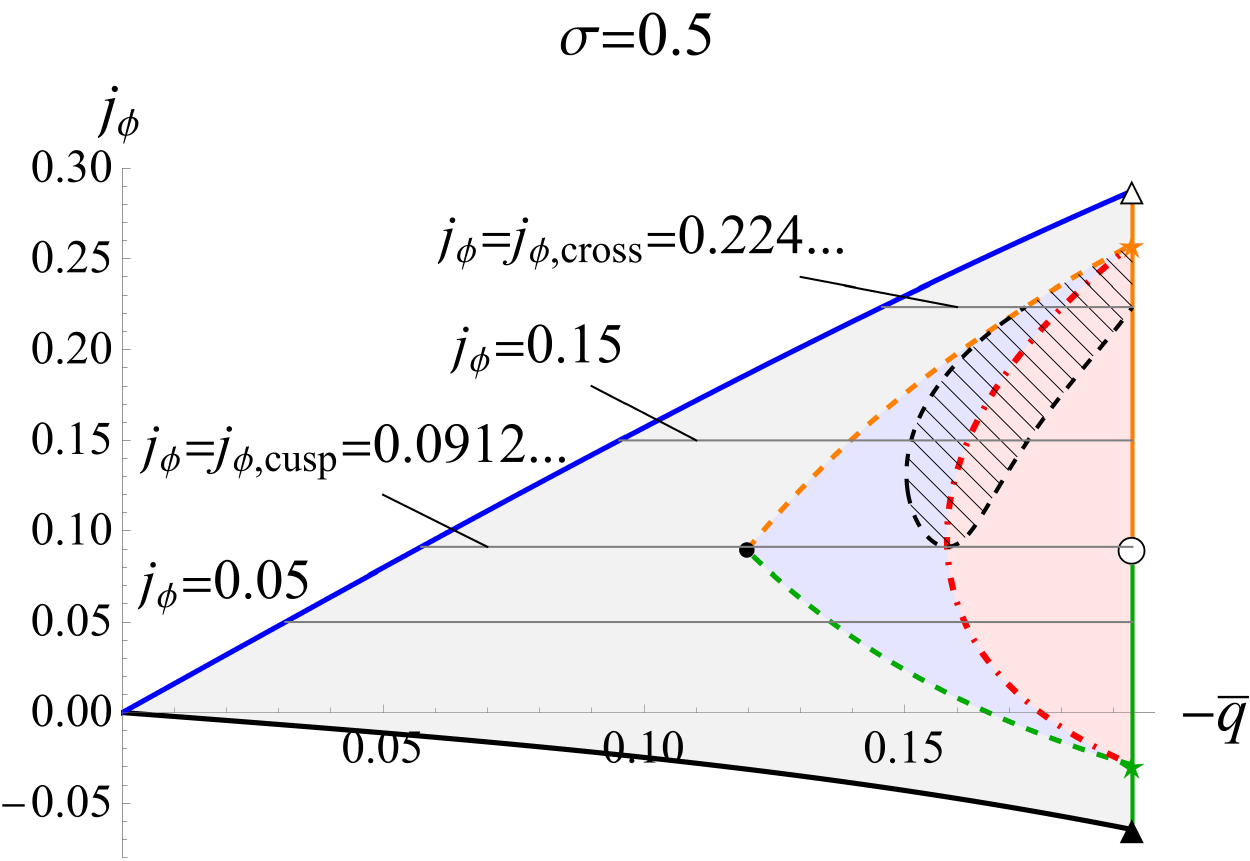}\\
\end{minipage}
\caption{
Entropy of the charged dipole black ring for $\sigma=0.5$ on some $j_\phi={\rm constant}$ surfaces (left and middle panels).
The black dots denote the boundaries of the phase diagram.
The fat and thin ring branches are illustrated by the red and blue curves, respectively, and the bifurcating points are represented by the black stars.
In the phase diagram (right panel), the fat ring branch (red colored region) has smaller entropy than the corresponding thin ring branch (blue colored region) for $j_\phi \leq  j_{\phi,{\rm cusp}}=0.0912\ldots$. However, for $j_\phi > j_{\phi,{\rm cusp}}$, a portion of the fat ring branch has larger entropy in the region near the critical curve (hatched region).
For $j_\phi \geq j_{\phi,{\rm cross}}=0.224\ldots$, the fat ring branch is thermodynamically favored compared to the corresponding thin ring branch. 
 \label{fig:jjplot-ex2}}
\end{center}
\end{figure}

The thermodynamic property of these two branches is not as simple as in the vacuum case where the thin branch has always larger entropy than the fat branch and hence thermodynamically more preferred~\cite{Elvang:2007hs}.
In the left and middle panels of Fig.~\ref{fig:jjplot-ex2}, we show the entropy of the $\sigma=0.5$ phase on several $j_\phi={\rm constant}$ lines, in which the thin ring and fat ring branches are represented by blue and red curves, respectively.
For sufficiently small $j_\phi$, the thin ring branch always has larger entropy than the fat ring branch as in the vacuum case (Fig.~\ref{fig:jjplot-ex2}: top left panel). The situation changes when $j_\phi$ becomes larger than a certain threshold $j_{\phi,{\rm cusp}}$, at which the $j_\phi={\rm constant}$ phase forms a cusp shape  (Fig.~\ref{fig:jjplot-ex2}: top middle panel).
For $j_\phi > j_{\phi,{\rm cusp}}$, a portion of the fat ring branch has larger entropy and becomes thermodynamically more preferred to the corresponding thin ring branch (Fig.~\ref{fig:jjplot-ex2}: bottom left panel).
When the value of $j_\phi$ reaches another threshold $j_{\phi,{\rm cross}}(>j_{\phi,{\rm cusp}})$,
the endpoint of the fat ring branch at $\lambda=1$ transitions on the thin ring branch (Fig.~\ref{fig:jjplot-ex2}, bottom middle panel).
For larger values of $j_\phi$, the fat ring branch always exhibits larger entropy than the thin ring branch with the same conserved charges.
It should be noted that for $\sigma=0.5$, these thresholds are specified as $j_{\phi,{\rm cusp}}=0.0912\ldots$ and $j_{\phi,{\rm cross}}=0.224\ldots$, and vary with the values of $\sigma$.
In the right panel of Fig.~\ref{fig:jjplot-ex2}, the hatched region represents the phases for the thermodynamically preferred fat ring branch and thin ring branch with the same conserved charges.

\medskip

\subsection{Neutral limit}
\begin{figure}
\includegraphics[width=7cm]{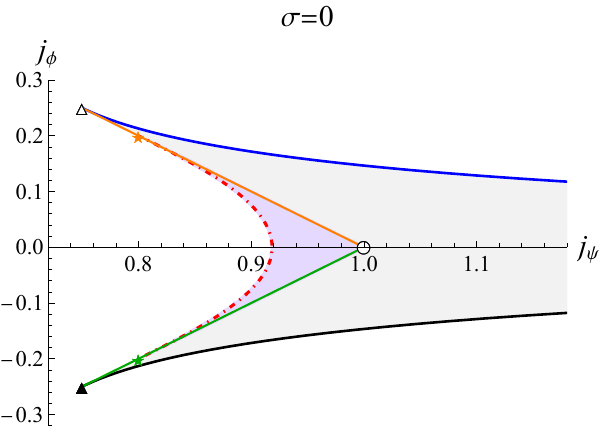}
\caption{
Phase diagram in the $(j_\psi,j_\phi)$ plane for $\sigma=0$.
The phase diagram exhibits symmetry about the $j_\psi$-axis, and 
each part of the $j_\phi \geq 0$ and $j_\phi \leq 0$ reproduces the phase diagram of the Pomeransky-Sen'kov solution presented in Ref.~\cite{Emparan:2008eg}, separately.  
The convention follows that in Fig.~\ref{fig:jjplot-ex1}.
 \label{fig:neutral-phase}}
\end{figure}

The solution includes the Pomeransky-Sen'kov solution~\cite{Pomeransky:2006bd} as the neutral limit $\sigma \to 0$. 
One can find that the phase diagram is identical to two copies of the Pomeransky-Sen'kov solution (the $j_\phi \geq 0$ region and  the $j_\phi\leq 0$ region in Fig.~\ref{fig:neutral-phase}, respectively).
Since this correspondence was shown  in our previous paper~\cite{Suzuki:2024eoz}, we do not mention the detail here.

\subsection{BPS limit}

The BPS limit can be obtained as $\sigma \to \pm 1$. 
In the $(j_\psi,j_\phi)$-plane,
the limit $\sigma \to \pm 1$ with any fixed $\nu$ and $\lambda$
results in the convergence $(j_\psi,j_\phi) \to (1/(2\sqrt{2}),\pm 1/(2\sqrt{2}))$, corresponding to the phase of the Breckenridge-Myers-Peet-Vafa (BMPV) black hole with equal angular momenta~\cite{Breckenridge:1996is}. 
This degeneracy can be resolved by taking the limit $\nu \to 0$ and $\lambda \to 0$  as $\sigma \to \pm 1$, or more precisely, by considering the following limit
\begin{align}\label{eq:bpslimit}
\sigma = \pm 1 \mp \fr{3} \epsilon,\quad \lambda = \chi_1 \epsilon,\quad \nu = \sqrt{\chi-\chi_1}\sqrt{\chi_1}  \epsilon,\quad \epsilon \to 0,
\end{align}
where $\chi,\chi_1$ are the parameters, which control the rate at which $\nu$ and $\lambda$ approach their limits. 
At this limit, regardless of the value of $\chi_1$, each dimensionless variable approaches specific values determined by $\chi$:
\begin{align}
j_\psi \to  \frac{1+\chi}{2\sqrt{2\chi(2+\chi)}},\quad j_\phi \to \pm \frac{3+\chi}{2} \sqrt{\frac{\chi}{2(2+\chi)^3}}, \quad a_H\to \sqrt{\frac{\chi}{(2+\chi)^3}},\quad 
\bar{q} \to -\fr{2}\sqrt{\frac{3\chi}{2(2+\chi)} }.
\end{align}
Hence,  at this limit, the phase space is reduced two parameter family, including the scale $\ell$.
The limit of each variable can be expressed in terms of $j_\psi$ as follows:
\begin{align}
j_\phi & = \pm \left(3 j_{\psi }-16 j_{\psi }^3+\frac{1}{\sqrt{2}}\left(8 j_{\psi }^2-1\right){}^{3/2}\right),\label{eq:jphi-jpsi-BPS}\\
a_H &= (16 j_\psi^2-1)\sqrt{8j_\psi^2-1}+4\sqrt{2} j_\psi(1-8j_\psi^2),\\
\bar{q}&= -\frac{\sqrt{3}}{8j_\psi+2\sqrt{2(8j_\psi^2-1)}},
\end{align}
where $ j_\psi > 1/(2\sqrt{2})$.

\medskip
Now, we compare the BPS limit $\sigma\to \pm1$ of our charged black ring with the well-known supersymmetric black ring solution found earlier by Elvang {\it et. al}~\cite{Elvang:2004rt}\footnote{We have multiplied $A$ in Ref.~\cite{Elvang:2004rt}  by the factor $-2$ so that it fits to our convention.}:
\begin{align}\label{eq:bps-sol-elvang}
\begin{split}
ds^2 &= - \hat{f}^2 (dt + \hat{\omega})^2 + \frac{\hat{f}^{-1} \sR^2}{(x-y)^2}\left[
\frac{dy^2}{y^2-1}+(y^2-1)d\psi^2+\frac{dx^2}{1-x^2}+(1-x^2)d\phi^2\right],\\
A &= -\sqrt{3}\left[ \hat{f} (dt+\hat{\omega})- \frac{\sq}{2} ((1+x)d\phi+(1+y)d\psi)\right],
\end{split}
\end{align}
where $\hat{\omega} = \hat{\omega}_\psi d\psi+ \hat{\omega}_\phi d\phi$ and
\begin{align}
\hat{f}{}^{-1} &= 1 + \frac{\sQ- \sq^2}{2\sR^2}(x-y) - \frac{\sq^2}{4\sR^2}(x^2-y^2),\\
\hat{\omega}_\phi & = - \frac{\sq}{2\sR^2}(1-x^2)[3 \sQ-\sq^2(3+x+y)],\\
\hat{\omega}_\psi &= \frac{3}{2}\sq(1+y) + \frac{\sq}{8\sR^2}(1-y^2)[3\sQ-\sq^2(3+x+y)],
\end{align}
and the solution is parametrized by three positive independent parameters $(\sR,\sQ,\sq)$ with the bound $\sR < (\sQ-\sq^2)/(2\sq)$.
The corresponding dimensionless variables are expressed as
\begin{align}
\hat{j}_\psi = \frac{\hat{\sq}(3(1+\hat{\sQ})-\hat{\sq}^2)}{4\sqrt{2} \hat{\sQ}^{3/2}},\quad
\hat{j}_\phi= \frac{\hat{\sq}(3 \hat{\sQ}-\hat{\sq}^2)}{4 \sqrt{2} \hat{\sQ}^{3/2}},\quad
\hat{a}_H = \frac{\sqrt{3} \hat{\sq} \sqrt{(\hat{\sq}^2-\hat{\sQ})^2-2 \hat{\sq}^2}}{2 \hat{\sQ}^{3/2}},\quad \hat{\bar{q}} = - \frac{\sqrt{3}\hat{\sq}}{2\sqrt{2\hat{\sQ}}},
\end{align}
where $\hat{\sq}:= \sq/(\sqrt{2}\sR)$ and $\hat{\sQ}:=\sQ/(2\sR^2)$.
\medskip

\begin{figure}
\includegraphics[width=7cm]{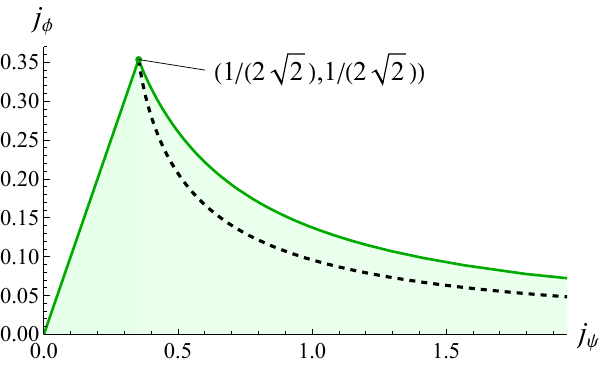}
\caption{
The BPS limit $\sigma \to 1$  of our solution (black dashed curve) compared with the BPS black ring solution in Ref.~\cite{Elvang:2004rt} (green shaded region bounded by the green curves).\label{fig:bps-comp}
}
\end{figure}

\medskip
First, we consider the limit $\sigma \to 1$ of our solution, which can be written as
\begin{align}\label{eq:bps-metric-1}
& ds^2 = -f^2 (dt + \bar{\omega}_\psi d\psi+\bar{\omega}_\phi d\phi)^2 + f^{-1} \frac{2\ell^2 }{(x-y)^2}\left[\frac{dx^2}{1-x^2}+\frac{dy^2}{y^2-1}+(1-x^2)d\phi^2+(y^2-1)d\psi^2\right],\\
 & A = - \sqrt{3} \left[ f (dt+ \bar{\omega}_\psi d\psi+\bar{\omega}_\phi d\phi)+\frac{\sqrt{3}}{2}\ell \chi ((1+x)d\phi+(1+y)d\psi)\right]+\sqrt{3} dt,
\end{align}
where 
\begin{align}
&f^{-1} = \lim_{\epsilon\to0} D = 1 + \frac{3\chi}{2}(x-y)-\frac{3\chi^2}{8}(x^2-y^2),\\
&\bar{\omega}_{\psi} = \lim_{\epsilon \to 0} \Omega'_\psi
= \frac{3\sqrt{3} \ell \chi}{16}(1+y)(8+6(1-y)\chi-(1-y)(x+y)\chi^2),\\
&\bar{\omega}_{\phi} = \lim_{\epsilon \to 0} \Omega'_\phi
= \frac{3\sqrt{3} \ell \chi^2}{16}(1-x^2)(-6+\chi(x+y)).
\end{align}
We can easily see that this coincides with the supersymmetric black ring solution~(\ref{eq:bps-sol-elvang}) by the parameter correspondence
\begin{align}
 \sQ = 3 \ell^2 \chi(2+\chi),\quad \sq = \sqrt{3} \ell \chi, \quad \sR = \sqrt{2} \ell. \label{eq:2pfamily-bpssol}
\end{align}
However, these three quantities are not independent but  are related via the relation $\sQ=\sq(\sqrt{6}\sR+\sq)$. Consequently, this limit only covers the two-dimensional subspace of the three-dimensional phase space of the supersymmetric black ring solution presented in Ref.~\cite{Elvang:2004rt}.
 Figure~\ref{fig:bps-comp} shows the allowed regions in the $(j_\psi,j_\phi)$-plane for the limit $\sigma \to 1$ of our solution and the supersymmetric black ring solution~\cite{Elvang:2004rt}.
This may be because our solution contains only four independent parameters, whereas the most general solution—should it exist—with five independent parameters, may exactly coincide with the supersymmetric black ring solution, which has three independent parameters, as described in Ref.~\cite{Elvang:2004rt}.
We also note that the three parameter charged black ring solution in our previous work~\cite{Suzuki:2024coe}, which is obtained from the current solution for $\beta=0$, already admits  the same two parameter subclass of the supersymmetric black ring in Eq.~(\ref{eq:2pfamily-bpssol})
by taking  the BPS limit in the same manner as in Eq.~(\ref{eq:bpslimit}).

\medskip

Another possibility for the BPS limit can be obtained by the limit $\sigma \to -1$,
yielding the metric in the form of Eq.~(\ref{eq:bps-metric-1}) with
\begin{align}
f^{-1}& = 1 - \frac{3\chi}{2}(x-y)+\frac{3\chi^2}{8}(x^2-y^2),\\
\bar{\omega}_\psi &= \frac{3\sqrt{3} \ell \chi}{16} (1+y)(8+ 6\chi (1-y)- (1-y)(x+y)\chi^2),\\
\bar{\omega}_\phi &= -\frac{3\sqrt{3} \ell \chi^2}{16}(1-x^2)(-6 + \chi(x+y)). 
\end{align}
However, one can easily see that this metric exhibits the negative mass, and hence cannot be regular from the positive mass theorem~\cite{SchenYau1981,Witten:1981mf}.

\subsection{Singly spinning phase ($\omega_\phi= 0$)}
\begin{figure}
\begin{center}
\begin{minipage}[b]{0.47\columnwidth}
\includegraphics[width=7cm]{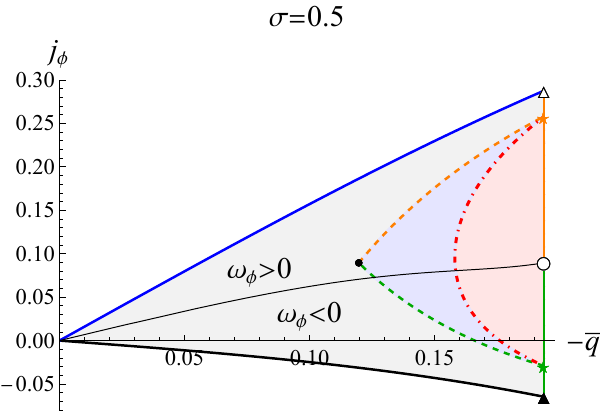}
\end{minipage}
\begin{minipage}[b]{0.47\columnwidth}
\includegraphics[width=7cm]{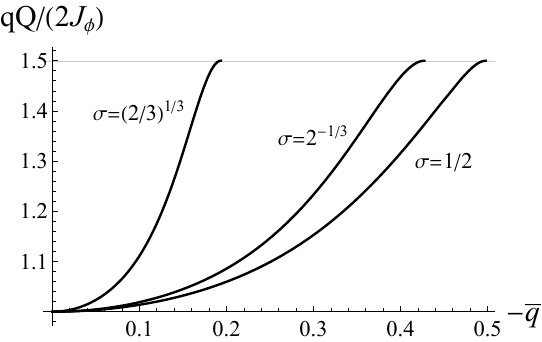}
\end{minipage}
\caption{ {\bf Left:} $\omega_\phi=0$ phase for $\sigma=0.5$ in the $(-\bar{q},j_\phi)$ plane.
For any $\sigma$, the $\omega_\phi=0$ curve (thin black curve) starts from the $(j_\phi,-\bar{q})=(0,0)$ and ends at $(j_{\phi,*},-\bar{q}_{\rm max})$ without touching $j_\phi=0$.
 {\bf Right:} $qQ/(2J_\phi)$ on the $\omega_\phi=0$ phase for different $\sigma$.  
 $qQ/(2J_\phi)$ monotonically increases and reaches the maximum of $3/2$ at $\bar{q}=-\bar{q}_{\rm max}$.
 \label{fig:singlespin}}\end{center}
\end{figure}

Another interesting case is the singly spinning phase with $\omega_\phi=0$.
In the charged black ring, the singly spinning case $\omega_\phi=0$ does not necessarily mean $j_\phi=0$ due to the influence of the rotating Maxwell field. 
In our solution, we find that the singly spinning phase always have $j_\phi \neq 0$ except in the large momentum limit $j_\psi\to\infty$, as shown in the left panel of Fig.~\ref{fig:singlespin}. 
This raises a question about whether the singly spinning phase in our solution overlaps with the known solution in Ref.~\cite{Elvang:2004xi}, which also features $\omega_{\phi}=0$. 
Verification can be easily achieved by checking the identity $(qQ)/(2J_\phi)=1$, as followed by the solution in Ref.~\cite{Elvang:2004xi}. 
It turns out that our solution with $\sigma\not =0$ always yields $(qQ)/(2J_\phi)>1$  for the $\omega_\phi=0$ phase, as depicted in the right panel of Fig.~\ref{fig:singlespin}. 
Therefore, our solution does not overlap with the black ring solution in Ref.~\cite{Elvang:2004xi} other than in the vacuum, singly spinning limit.

\subsection{$j_\phi=0$ phase}

Finally, we study the charged black ring solution characterized by a single angular momentum, $j_\psi\not =0$ and $j_\phi=0$. 
It is important to note,  as depicted in the left panel of Fig.~\ref{fig:singlespin}, that this does not mean the horizon angular velocity $\omega_\phi$ along the direction $\partial/\partial\phi$ vanishes, because the Maxwell field rotating outside the horizon contributes to $j_\phi$.  
As illustrated in the phase diagram in Figs.~\ref{fig:jjplot-ex1} and \ref{fig:n0-jjplots}, the occurrence of the $j_\phi=0$ phase varies depending the value of $\sigma$ as follows:

\begin{enumerate}
\item 
For $\sigma \geq \sigma_* = 0.579\ldots$ (top and middle panels in Fig.~\ref{fig:n0-jjplots}),  
 the $j_\phi=0$ phase does not exist, where $\sigma_*$ is a root of $1-5\sigma_*^3-\sigma_*^6+\sigma_*^9=0$,

\item 
For $\sigma_\infty = \sqrt[3]{\sqrt{10}-3} = 0.545\ldots < \sigma < \sigma_*$ (bottom panels in  Fig.~\ref{fig:n0-jjplots}),
the $j_\phi=0$ phase exists for
\begin{align}\label{eq:jphi0-jpsiminmax}
j_{\psi,{\rm min}} = \frac{1-\sigma^3}{(1+\sigma^2)^{3/2}}< j_\psi < j_{\psi,{\rm max}} = -\sqrt{\frac{\sigma ^3}{\sigma ^6+6 \sigma ^3-1}} \frac{\left(\sigma ^{12}+8 \sigma ^9-10 \sigma ^6-8 \sigma ^3+1\right)}{\left(\sigma ^8+\sigma ^6+\sigma
   ^2+1\right)^{3/2}},
\end{align}
where $j_{\psi,{\rm min}}$ is defined as the intersection of the $\lambda=0$ curve and the $j_\phi=0$ line.

\item 

For $0 < \sigma <\sigma_\infty$,
the $j_\phi=0$ phase exists for $j_\psi > j_{\psi,{\rm min}}$ and has the large momentum limit $j_\psi \to \infty$.
Here,  the lower bound $j_{\psi,{\rm min}}$ corresponds to the intersection point (given by Eq.~(\ref{eq:jphi0-jpsiminmax})) of the $\lambda=0$ curve and the $j_\phi=0$ line for $\sigma_*' (:=  (1/2) (7-\sqrt{33})^{1/3}=0.539\ldots)< \sigma < \sigma_{\infty}$, 
 and  the intersection point  of the critical curve and the $j_\phi=0$ line (e.g. the $j_\phi$ intercept of the red dot-dashed curve in Fig.~\ref{fig:jjplot-ex1}) for $0<\sigma<\sigma_*'$, respectively.

 \end{enumerate}

\subsection{Ergoregion}
\begin{figure}
\begin{center}
\includegraphics[width=6cm]{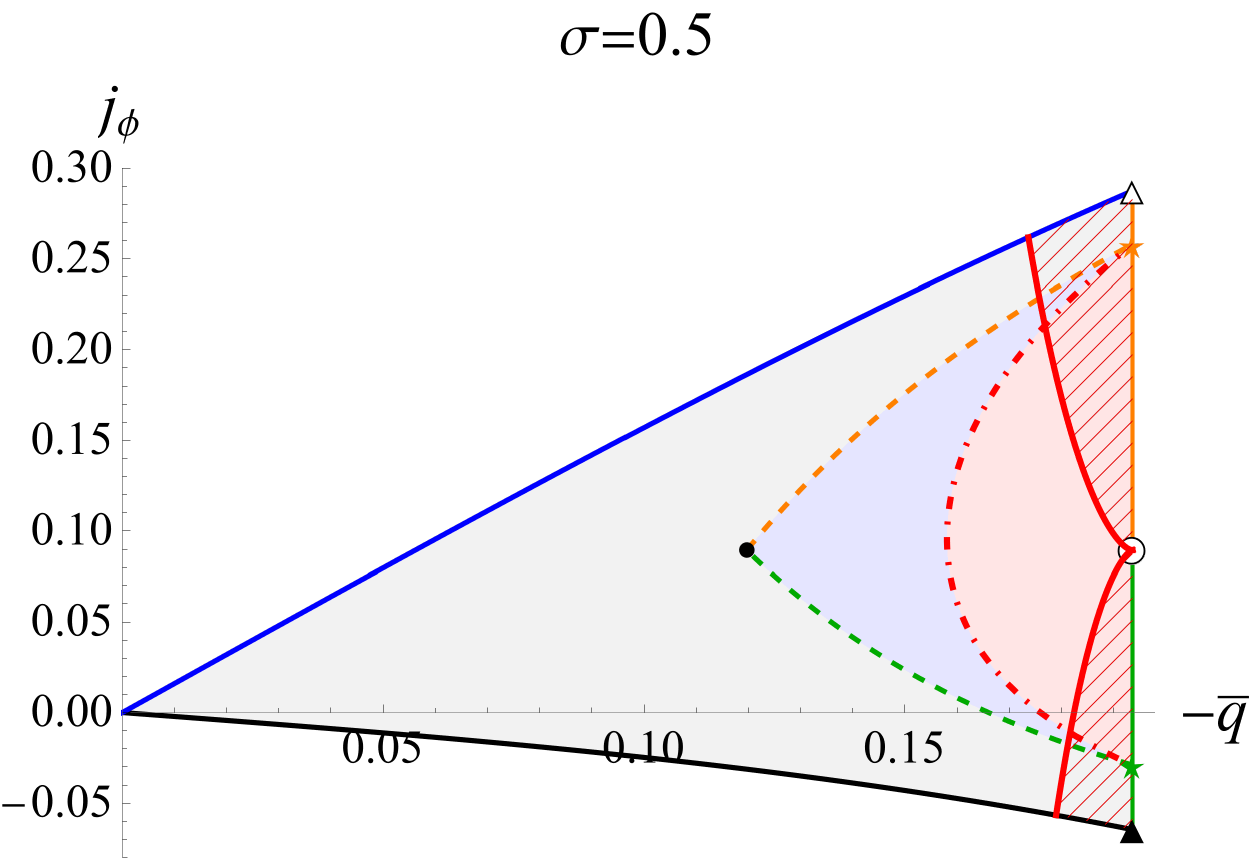}
\caption{The transition of the ergosurface in the $\sigma=0.5$ phase. The ergosurface consists of two $S^3$-surfaces in the red hatched region and a single $S^2 \times S^1$-surface in the other region.
 \label{fig:ergo}}\end{center}
\end{figure}

The ergoregion appears at the region where $H (y,x)<0$, which is bounded by the ergosurface defined by $H (y,x)=0$.
To study the topology of the ergosurface,
one can prove the following inequalities (see Appendix~\ref{sec:proof-ergo} for detailed calculations):
\begin{enumerate}
\item  $ H (y=-1,x=\pm1)>0,$
\item $ H (y=-1/\nu,x)<0\ {\rm for}\ x \in [-1,1],$
\item $\partial_x^2 H (y=-1,x)>0\ {\rm for}\ x \in[-1,1],$
\item  $\partial_y^2  H (y,x)<0 \ {\rm for} \ (x,y) \in [-1,1]\times [-1/\nu,-1],$
\item  $\partial_y  H (y=-1,x)+ H (y=-1,x)>0\ {\rm for} \ x \in [-1,1].$
\end{enumerate}
Following the discussion in Ref.~\cite{Suzuki:2024phv}, from these inequalities,
one can show that the ergosurface is classified into two types as in Ref.~\cite{Suzuki:2024coe}, where it is represented in the C-metric coordinates by
\begin{itemize}
\item[(i)] a single curve ending at $x=-1$ and $x=1$
\item[(ii)] two separate curves; one ends at $x=-1$ and $y=-1$, the other at $x=1$ and $y=-1$.
\end{itemize}
For the type (i), the spacetime has a single ergosurface of $S^2 \times S^1$-topology.
For the type (ii), the ergoregion exists between two ergosurfaces of $S^3$-topology, where the inner surface appears around the regular center at $(x,y)=(1,-1)$. 
These two types are partitioned by the marginal phase that satisfies the following condition
\begin{align}
 {}^\exists x \in [-1,1], \quad {\rm s.t.}\quad H (y=-1,x)=0\quad {\rm and} \quad \partial_x H (y=-1,x)=0.
\end{align}
From Fig.~\ref{fig:ergo}, one can see that the type (ii) only appears around $\lambda=0$ or $\lambda=1$ where the dipole charge becomes sufficiently large.
We note that non-BPS, extremal black ring phase also admits the ergoregion and exhibits the same transition between type (i) and type (ii) as shown in Fig.~\ref{fig:ergo}.

\section{Summary}\label{sec:sum}

In this paper, using the electric Harrison transformation~\cite{Bouchareb:2007ax}, we have presented a new exact solution of a non-BPS charged rotating black ring with all independent conserved charges in five-dimensional minimal supergravity. 
We have used the vacuum singular black ring solution with a Dirac-Misner string singularity inside the black ring that was constructed in our previous work~\cite{Suzuki:2024eoz}, as a seed for the Harrison transformation, deriving a charged solution that initially possesses the Dirac-Misner string singularity. 
After the Harrison transformation, we have imposed appropriate boundary conditions to remove the Dirac-Misner string singularity. 
The resulting solution has proven to be regular, without curvature singularities, conical singularities, Dirac-Misner string singularities, orbifold singularities on and outside the horizon, or CTCs.
The charged rotating black ring solutions previously found by Elvang, Emparan, and Figueras (EEF)~\cite{Elvang:2004xi}, and another black ring solution constructed in our previous work~\cite{Suzuki:2024coe}, have had the mass, two angular momenta, an electric charge, and a dipole charge, with only three of these quantities being independent. 
In contrast, our obtained black ring has possessed four independent conserved charges: its mass, two angular momenta, an electric charge, and a dipole charge, which is related to the conserved charges
\medskip

 This black ring solution, derived through the Ehlers-Harrison transformations, encompasses both well-known and unfamiliar physically significant solutions:
In the neutral limit, the solution coincides with the vacuum doubly spinning black ring solution, referred to as the Pomeransky-Sen'kov solution~\cite{Pomeransky:2006bd};
in the BPS limit, it yields the restrictive supersymmetric black ring~\cite{Elvang:2004rt}, in the sense that the electric charge (mass) and the dipole charge are not independent;
in an extremal limit, distinct from the BPS limit, it becomes the charged doubly spinning black ring with two nonzero angular velocities, including the extremal Pomeransky-Sen'kov black ring.
Furthermore, we wish to highlight the connections to two previously constructed black ring solutions: the EEF solution in Ref.~\cite{Elvang:2004xi} and the charged dipole black ring solution in Ref.~\cite{Suzuki:2024coe} in the five-dimensional minimal supergravity.
Both of these solutions carry the mass, two angular momenta, electric charge, and dipole charge, with only three of these physical quantities being independent, but they differ aside from a vacuum case.
At the limit $\beta\to 0$, our solution coincides with the latter, whereas at any limits, it does not coincide with the former as long as $\sigma\not= 0$.

\medskip

As mentioned previously in Ref.~\cite{Elvang:2004xi}, it is conceivable that a most general black ring solution exists, characterized by independent parameters: its mass $M$, two angular momenta $J_{\psi}, J_{\phi}$, electric charge $Q$, and dipole charge $q$. 
This solution appears to have been constructed in Ref.~\cite{Feldman:2014wxa}. 
However, given its considerable complexity and formal nature, it remains uncertain whether this solution adequately describes a regular black ring with all five independent parameters, including regularity, horizon topology, asymptotic flatness, and physical properties. 
Therefore, to further examine these aspects, we need a more streamlined form of the solution, such as the C-metric form.
We anticipate that applying our method within the context of five-dimensional minimal supergravity may enable the derivation of such an exact solution. 
In our forthcoming paper, we aim to delve into the construction of this black ring solution using advanced solution-generation techniques. 
Additionally, a promising area for future research involves extending our results to the domain of charged black holes with multiple horizons. 
This extension could involve constructing charged counterparts of exotic configurations like the black saturn~\cite{Elvang:2007rd} or black di-ring~\cite{Iguchi:2007is, Elvang:2007hs}.

\section*{Acknowledgement}
R.S. was supported by JSPS KAKENHI Grant Number JP24K07028. 
S.T. was supported by JSPS KAKENHI Grant Number 21K03560.

\appendix

\section{Coefficients $f_i$ and $g_i$}\label{sec:fi-gi}
The explicit forms of the coefficients $f_i$ and $g_i$ are shown below. 

\subsection{$f_1,\dots,f_{19}$}
$f_i$ are quadratic functions of $t_3$. $f_i$ for $16\leq i\leq 19$ does not appear in the metric functions but does only in the expressions of $g_i$.
\begin{subequations}
\begin{align}
\begin{split}
 f_1& = 1+2( a- b) \beta +\frac{\beta ^2}{\gamma^2 -\nu^2 } \biggr[(a-b)^2 \gamma ^2-(1-a)^2 \gamma  (1+2 b+(3+2 b) \nu )\\
&\quad\quad+\nu  \left((1-a)^2 (3+2 b)+\left(1-2 a+2 \left(1-a+a^2\right) b-b^2\right) \nu \right)\biggr],\\
 f_2 &= 1+2(1- a b) \beta +\frac{\beta ^2}{\gamma^2 -\nu^2 } \biggr[ \left((2-a) a-2 \left(1-a+a^2\right) b-(1- 2 a) b^2\right) \gamma ^2\\
&\quad \quad +(1-a)^2 \gamma  (1+2 b+(3+2 b) \nu   )-\nu  \left((1-a)^2 (3+2 b)+(1-b) (1+(1-2 a) b) \nu \right)\biggr],\\
 f_3 &= 1+2( a- b) \beta +\frac{\beta ^2}{\gamma^2 -\nu^2 } \biggr[ (a-b)^2 \gamma ^2-2 (1-a)^2 (1+a)+(1-a)^2 (1-2 a+2 b) \nu \\
&\quad\quad+ \left(1-2 a+2 \left(1-a+a^2\right)   b-b^2\right) \nu ^2+(1-a)^2 \gamma  (1+2 a-2 b-(1-2 a+2 b) \nu )\biggr],\\
 f_4 &= 1- \frac{2 \beta}{\gamma -\nu }  \left(a^2-1-a (a-b) \gamma +\nu(1 -a b) \right)+\frac{\beta ^2}{\gamma^2 -\nu^2 } \biggr[ 2 (1-a)^2 (1+a)-(1-2 a) (a-b)^2 \gamma   ^2\\
&\quad \quad -(1-a) (1-5 a+4 b) \nu -\left(1-2 a b-b^2+2 a b^2\right) \nu ^2-(1-a) \gamma  \left(1-a-4 a^2+4 a b-(1 -5 a  +4 b) \nu   \right)\biggr],\\
 f_5& = 1   - \frac{2 \beta }{c_2} \left(2 a   (-1+\gamma ) \nu +a b^2 (\gamma -\nu ) (1+\nu )+b \left(-1+a^2-a^2 \gamma -\left(2+a^2\right) (-1+\gamma ) \nu +\nu ^2\right)\right)\\
& \quad +\frac{\beta ^2}{(\gamma^2 -\nu^2 )  c_2} \biggr[ b \gamma  \left(2 (1-a)^2 (1+a)+(1-a) \left(-1+a+4 a^2-4 a b\right) \gamma +(-1+2 a) (-a+b)^2 \gamma ^2\right)\\
& \quad +\nu \biggr\{   \left(2 \left(-2+a+6 a^2-4 a^3\right)+\left(4-20 a+17 a^2-6   a^3\right) b+4 \left(2-3 a+2 a^2\right) b^2+(1-2 a) b^3\right) \gamma ^2\\
& \quad \quad 
+(-a+b)^2 (-4+2 a-b+2 a b) \gamma ^3-2   (1-a)^2 (1+a) (2+b)+2 (1-a)^2 \left(4+5 a+(-1+3 a) b-2 b^2\right) \gamma\biggr\}\\
& \quad    +  \nu ^2 \biggr\{  (1-a)   \left(2 \left(2-3 a+a^2\right)+\left(3-9 a+2 a^2\right) b +4 b^2\right)\\
& \quad \quad +\left(4 (a-2) (1-a)^2+\left(-10+26 a-21 a^2+4 a^3\right) b-2   \left(4-7 a+2 a^2\right) b^2+(1-2 a) b^3\right) \gamma\\
& \quad \quad +\left(2 (2-a) (1-a)^2+\left(5-14 a+10 a^2-2 a^3\right) b+\left(4-6 a+4 a^2\right)   b^2+(1-2 a) b^3\right) \gamma ^2\biggr\}\\
& \quad   - \left(2 a-\left(6-6 a+5 a^2\right) b+4 a b^2+(1-2 a) b^3\right) (1-\gamma ) \nu ^3+(1-b) b (1+(1-2 a) b) \nu ^4 \biggr],\\
 f_6&= 1-\frac{2 \beta}{(\gamma -\nu ) (1+\nu ) c_1}  \biggr[ a^2-1+\left(2-2 a+a^2+b-2 a b\right) \nu
 -(1-b) (1-2 a+b)   \nu ^2-(1-b) b \nu ^3\\
 &\quad\quad +(a-b)^2 \gamma ^2 (1+\nu )+\gamma  \left(1-2 a^2-b+2 a b-2   \left(1-a+a^2-2 a b+b^2\right) \nu +(1-b) (1-2 a+2 b) \nu   ^2\right)\biggr]\\
&\quad   +\frac{\beta ^2 }{(\gamma^2 -\nu^2 ) (1+\nu ) c_1}\biggr[
2 (1-a)^2   (1+a)-(1-a) \left(2-5 a-a^2+4 b\right) \nu \\
&\quad\quad+\left(-2-2 a+6 a^2-3 a^3+\left(3-4  a+3 a^2\right) b+(2-3 a) b^2\right) \nu ^2+\left(2-3 a+3 a^2 b-3 a b^2+b^3\right)
   \nu ^3+(1-b)^2 b \nu ^4\\
&\quad\quad +(a-b) \gamma ^2 \left(4 a^2-1-2a+(2-3 a) b+\left((2-a)^2-2 a b+b^2\right) \nu -\left(3 (1-a)^2+(2-a)b-b^2\right) \nu ^2\right)\\
&\quad\quad     -(a-b)^3 \gamma ^3 (1+\nu )+ (1-a) \gamma \left(a-2+5 a^2-4 a   b+\left(2-9 a-a^2+8 b\right) \nu\right)\\
&\quad\quad+\left(2+5 a-12 a^2+6   a^3-\left(8-12 a+7 a^2\right) b+3 a b^2-b^3\right) \gamma \nu ^2-\left(2-3 a+3 a^2 b-3 a   b^2+b^3\right) \gamma \nu ^3\biggr], 
\end{split}
\end{align}
\begin{align}
\begin{split}
 f_7&=  1-\frac{2 \beta }{b (\gamma -\nu ) (1+\nu )} \biggr[(b-1) b \gamma +\left(2-2 a+b-b^2\right) (1-\gamma ) \nu
   +(1-b) b \nu ^2 \biggr]\\
&   +\frac{\beta ^2}{b   (\gamma^2 -\nu^2 ) (1+\nu )}
 \biggr[b \gamma    \left((1-a)^2+\left((2-a) a-2 b+b^2\right) \gamma \right)-(1-b)^2 b \nu ^3\\
 &\quad
 +\left(-(1-a)^2   (4+3 b)+4 (1-a) (1-2 a+(2-a) b) \gamma  +\left(4 (1-a) a-\left(4-6   a+a^2\right) b-2 b^2+b^3\right) \gamma ^2\right) \nu \\
 &\quad
   +\left(4 (-1+a)+(2-3 a) a b+2   b^2-b^3+(1-a) (4-b-3 a b) \gamma \right) \nu ^2\biggr],\\
 f_8&= 1+\frac{2 \beta}{(a-b) (\gamma -\nu )}\biggr[1-a^2+(a-b)^2 \gamma -\left(1-2 a b+b^2\right) \nu\biggr]\\
&   -\frac{\beta ^2}{(a-b)   \left(\gamma ^2-\nu ^2\right)}\biggr[
2(-1+ a+ a^2- a^3)-a \gamma( 1  + 2 a - 3 a^2) -(a^3-b^3) \gamma ^2-b^2 \gamma    \left(2+2 a^2+a (-4+3 \gamma )\right)\\
&\quad
+b \left(-2 (1-a)^2   (1+a)+\left(1+4 a-7 a^2+2 a^3\right) \gamma +3 a^2 \gamma ^2\right)\\
&\quad
- \nu (1-a) (\gamma -1) \left((a-5) a+\left(5-3 a+2   a^2\right) b+2 (1-a) b^2\right) \\
&\quad
-\left(-2+a+\left(-1+6 a-2   a^2\right) b+\left(-2+a-2 a^2\right) b^2+b^3\right) \nu ^2\biggr],\\
 f_9&= 1+2 (a-b) \beta 
+\frac{\beta ^2}{\left(\gamma ^2-\nu ^2\right) c_1}\biggr[
-2 (1-a)^2 (1+a)-(a-b)^3 \gamma ^3+(1-a)^2  (4+a+2 b) \nu-(1-b) b (1-2 a+b) \nu ^3\\
& -\left(2-5 a+4 a^2+\left(1-2 a-a^2\right) b+a b^2\right) \nu   ^2
-(a-b) \gamma ^2 \left(1-a (2-b)-\left((1-a)^2-a   b+b^2\right) \nu \right)\\
&  -\gamma  \left((1-a)^2 (-2-3 a+2 b)+2 \left(2-3
   a+a^3\right) \nu +\left(-2+5 a-4 a^2+\left(-2+4 a+a^2\right) b-3 a b^2+b^3\right)
   \nu ^2\right)\biggr],\\
 f_{10}&=\frac{4 (1-a)  (1-\gamma ) d_1}{\gamma^2 -\nu^2 }\beta  (1+(a-b) \beta ),\\
 f_{11}&=1-\frac{2 \beta  (-1+a-a \gamma +b (-1+a+\gamma )+(-1+a b) \nu )}{\gamma +\nu   }\\
& +\frac{\beta ^2}{\gamma^2 -\nu^2 } \biggr[(a-b)^2 \gamma ^2
+ \nu  \left((1-a) \left(a-3+2  b^2\right)-(1-b) (1+b-2 a b) \nu \right)+(1-a) \gamma  \left(1+a-2  b^2+\left(3-a-2 b^2\right) \nu \right)\biggr],\\
f_{12}&=f_1+\frac{c_3 \left(f_1-f_3\right)}{(a+1) b (1-\gamma ) (\nu +1)},\\
f_{13}&=f_7+\frac{\left(b (\gamma-\nu^2) -(2+b) (1-\gamma ) \nu \right) f_{10}}{2 b (1+\nu ) d_1},
    \end{split}
\end{align}
\begin{align}
    \begin{split}
 f_{14}&=1-\frac{2 \beta }{d_2} \biggr[ b \left(-1+a^2+\gamma -2 a^2 \gamma -(1-2 a) b \gamma   +(a-b)^2 \gamma ^2\right)\\
 &\quad+\left(-2+\left(2+a^2\right) b+(1-2 a) b^2-2   \left(-1+b+a^2 b-2 a b^2+b^3\right) \gamma +b (-a+b)^2 \gamma ^2\right) \nu\\
&\quad   +\left(2-b-2 a b^2+b^3+\left(-2+b+(1+2 a) b^2-2 b^3\right) \gamma \right) \nu
   ^2+(-1+b) b^2 \nu ^3\biggr]\\
&   +\frac{\beta ^2}{\left(\gamma ^2-\nu ^2\right) d_2} \biggr[
b \gamma  \left(2 (1-a)^2   (1+a)+(1-a) \left(-2+a+5 a^2-4 a b\right) \gamma \right.\\
&\quad\quad\left. +\left(a \left(-1-2 a+4   a^2\right)+\left(1+4 a-7 a^2\right) b+(-2+3 a) b^2\right) \gamma ^2-(a-b)^3   \gamma ^3\right)\\
&\quad +\left(-2 (1-a)^2 (1+a) (2+b)+2 (1-a)^2 \left(5+6 a+4 a b-2   b^2\right) \gamma \right.\\
&\quad \quad\left. +\left(-6+4 a+16 a^2-12 a^3+\left(2-22 a+26 a^2-11 a^3\right)
   b+\left(9-16 a+11 a^2\right) b^2+(2-3 a) b^3\right) \gamma ^2\right.\\
&\quad\quad\left. -2 \left((3-2 a)
   a^2-3 a \left(2-2 a+a^2\right) b+\left(3-4 a+5 a^2\right) b^2-3 a b^3+b^4\right)
   \gamma ^3+b (-a+b)^3 \gamma ^4\right) \nu\\
&\quad  +\left((1-a) \left(2-6 a+4   a^2+\left(2-9 a+3 a^2\right) b+4 b^2\right)\right.\\
&\quad \quad
+\left(6 (1-a)^2 (-1+2   a)+\left(-12+37 a-34 a^2+8 a^3\right) b+\left(-13+20 a-5 a^2\right) b^2+(2-3 a)
   b^3\right) \gamma \\
&\quad \quad+\left(4-24 a+30 a^2-12 a^3+\left(18-39 a+32 a^2-7 a^3\right)
   b+2 \left(5-10 a+4 a^2\right) b^2\right) \gamma ^2\\
&\quad \quad\left.+\left(2 a \left(4-5 a+2   a^2\right)+\left(-8+13 a-10 a^2+2 a^3\right) b+\left(-3+4 a-3 a^2\right) b^2+(2+3
   a) b^3-2 b^4\right) \gamma ^3\right) \nu ^2\\
&\quad +\left(2 (a-2) a-\left(-10+14 a-10   a^2+a^3\right) b+\left(3-8 a+a^2\right) b^2+(-2+3 a) b^3\right.\\
&\quad \quad\left.
+2 \left(-1+4 a-2   a^2+\left(-8+14 a-10 a^2+a^3\right) b-\left(3-8 a+a^2\right) b^2-3 a
   b^3+b^4\right) \gamma \right.\\
&\quad \quad\left. +\left(2 (-1+a)^2-\left(-6+14 a-10 a^2+a^3\right)
   b+\left(5-8 a+a^2\right) b^2+(-2+3 a) b^3\right) \gamma ^2\right) \nu   ^3-(1-b)^2 b^2 \nu ^5\\
&\quad +\left(2-(2+a) b+\left(2-4 a+a^2\right) b^2+3 a b^3-b^4+\left(-2+(2+a)
   b-\left(1-4 a+a^2\right) b^2-(2+3 a) b^3+2 b^4\right) \gamma \right) \nu   ^4  \biggr],\\
f_{15}&=1-\frac{2 \beta }{(\gamma -\nu ) \left(c_3-(1-a b) (1-\gamma ) (1+\nu )\right)}\biggr[
1-a^2-\left(2+a^2+a \left(-2-2 b+b^2\right)\right) \nu \\
&+\left(1-2 (1-a) b-(1+a)   b^2+b^3\right) \nu ^2-(1-b) b^2 \nu ^3-(a-b) \gamma ^2 \left(a-b+b^2+\left(-2+a-b+b^2\right) \nu
   \right)\\
&+\gamma  \left(-1+2 a^2-2 a b+a b^2+2 \left((1-a)^2+b-2 a b+(1+a) b^2-b^3\right) \nu +\left(-1+2   (1-a) b+(2+a) b^2-2 b^3\right) \nu ^2\right)\biggr] \\
& +\frac{\beta ^2}{(\gamma^2 -\nu^2 ) \left(c_3-(1-a b) (1-\gamma )   (1+\nu )\right)}
 \biggr[ -2 (1-a)^2 (1+a) (1+b)\\
 &\quad-(1-a) \left(1+5 a-2 a^2+\left(-4+3 a-3 a^2\right) b-2 (1-a)   b^2\right) \nu \\
 &\quad +\left(4-6 a+3 a^2-\left(-2+8 a-4 a^2+a^3\right) b+\left(2-2 a+3 a^2\right) b^2-a b^3\right)
   \nu ^2 \\
 &\quad+\left(-1+(4-3 a) b+\left(-1+2 a+a^2\right) b^2-(2+a) b^3+b^4\right) \nu ^3+(-1+b)^2 b^2 \nu
   ^4 \\
 &\quad+(a-b)^2 \gamma ^3 \left(-1+2 a+(-2+a) b+b^2+\left(-3+2 a+(-2+a) b+b^2\right) \nu \right) \\
 &\quad+\gamma ^2
   \left(-1+2 a+4 a^2-6 a^3-\left(2+5 a-14 a^2+4 a^3\right) b+\left(4-10 a+3 a^2\right) b^2+a b^3\right.\\
   &\quad \quad \left. +\left(-2-4
   a+13 a^2-6 a^3+\left(8-22 a+18 a^2-5 a^3\right) b+\left(7-14 a+5 a^2\right) b^2+(2+a) b^3-b^4\right) \nu\right.\\
   &\quad \quad \left. 
   +\left(3 (-1+a)^2-(-2+a) (-1+a)^2 b+\left(1-4 a+2 a^2\right) b^2+2 b^3-b^4\right) \nu ^2\right) \\
 &\quad+\gamma 
   \left((-1+a) \left(-3+a+6 a^2+\left(-4-5 a+5 a^2\right) b-2 (-1+a) b^2\right)\right. \\
   &\quad \quad \left. +(-1+a) \left(-3-11 a+6
   a^2+\left(12-11 a+7 a^2\right) b-6 (-1+a) b^2\right) \nu\right.\\
   &\quad \quad \left.  -\left(7-12 a+6 a^2+\left(4-13 a+8 a^2-2 a^3\right)
   b+\left(5-6 a+5 a^2\right) b^2-(2+a) b^3+b^4\right) \nu ^2\right.\\
   &\quad \quad \left. +\left(1+(-4+3 a) b-\left(-1+2 a+a^2\right)
   b^2+(2+a) b^3-b^4\right) \nu ^3\right) \biggr],        
   \end{split}
   \end{align}
    \begin{align}
       \begin{split}
f_{16} &= 1-\frac{2 \beta  \left(-1+a^2+(a-b)^2 \gamma ^2+(1-2 a+b) \nu +(1-b) b \nu ^2-\gamma  \left(-1+2 a^2+b-2 a
   b+(1-2 a+b) \nu \right)\right)}{(\gamma +\nu ) c_1}\\
&   +\frac{\beta ^2}{\left(\gamma ^2-\nu ^2\right) c_1}\biggr[ 2 (1-a)^2 (1+a)-(a-b)^3 \gamma ^3-(1-a) \left(4+a^2-a (1+4 b)\right) \nu \\
&\quad+\left(2-3 a-\left(1-4 a+a^2\right) b+(-2+a) b^2\right) \nu   ^2+(1-b)^2 b \nu ^3\\
&\quad+(a-b) \gamma ^2 \left(-1-2 a+4 a^2+(2-3 a) b+\left((-1+a)^2+(2-3 a) b+b^2\right) \nu
   \right)\\
&\quad+\gamma  \left((1-a) \left(-2+a+5 a^2-4 a b\right)+2 (1-a) \left(2-a+a^2-4 a b+2 b^2\right) \nu\right)
\\
&\quad   -\left(2-3 a+(4-a) a b-(4-a) b^2+b^3\right)\gamma \nu ^2\biggr],\\    
 f_{17}& =1+\frac{2 \beta  (1-a+(a-b) \gamma +(1-b) \nu )}{\gamma +\nu }\\
&+\frac{\beta ^2 \left((a-b)^2 \gamma ^2
+ \nu    \left((1-a) (1-3 a+2 b)-(1-b)^2 \nu \right)+(1-a) \gamma  (1+a-2 b-\nu +3 a \nu -2 b \nu   )\right)}{\gamma^2 -\nu^2},       \\
f_{18}& =1+\frac{2 \beta  \left(1-a^2+a (a-b) \gamma +(2-a) (a-b) (1-\gamma ) \nu -(1-a b) \nu ^2\right)}{(1-\nu )
   (\gamma +\nu )}\\
&   + \frac{\beta ^2}{(\gamma^2 -\nu^2 ) (1-\nu )} \biggr[ 2 (1-a)^2 (1+a)-(1-a) \left(3+2 a^2-a (1+4 b)\right) \nu +\left((a-2)   a+2 (3-2 a) a b-(3-2 a) b^2\right) \nu ^2\\
&\quad   +(1-b) (1+(1-2 a) b) \nu ^3+(a-b)^2 \gamma ^2 (-1+2 a+(-3+2 a)   \nu )\\
&\quad+(1-a) \gamma  \left(-1+a+4 a^2-4 a b+\left(2-2 a+4 a^2-8 a b+4 b^2\right) \nu +\left(-1+a-4 a b+4   b^2\right) \nu ^2\right) \biggr],   \\
f_{19}&=1+\frac{2 \beta  (1-a+(a-b) \gamma -(1-b)\nu  )}{\gamma -\nu }+\frac{\beta ^2 \left(1-a^2-2 b (1-a (1-\gamma ))+a^2 \gamma +b^2 \gamma -(1-b)^2 \nu \right)}{\gamma -\nu   }.
   \end{split}
\end{align}
   \end{subequations}


\subsection{$g_1,\dots,g_{13}$ and $g_1'$}

 \begin{align}
 \begin{split}
g_1 & =f_{13} f_{14}-\frac{c_1 c_3 \left(f_{13} f_{14}-f_{16} f_{19}\right)}{2 (a-b) (1-\gamma )^2 \nu }+\frac{(1-\nu )
   c_3 f_3 f_{10}}{2 (a-b) (1-\gamma ) (1+\nu ) d_1},\\
g_1' &=  g_{1}+\frac{\nu c_3 f_3 f_{10} }{(1-\gamma ) d_1 (\nu +1) (a-b)} ,  \\
g_2&=\Delta+\frac{c_3 \left(\Delta-f_3 f_4\right)}{d_3}, \\
g_3 &=f_{14}f_{18}-\frac{d_1 \left(f_{14} f_{18}-f_2 f_{17}\right)}{(1-\gamma ) (1+\nu ) \left(a   c_1-1+\nu \right)}-\frac{b \nu  (\gamma -\nu ) f_{10} f_{12}}{(1-\gamma ) (1-\nu^2 ) \left(a   c_1-1+\nu \right)},\\
g_4 &=\Delta+\frac{1}{d_4}\biggr[-(1-\gamma )^2 (1-\nu )^2 \nu  d_1 \left(\Delta-f_1^2\right)
-\frac{8 \nu  d_1 \left(d_1-(1+a) (1-\gamma )   (1+\nu ) c_1\right) \left(\Delta-f_1 f_2\right)}{1+a}\\
&+2 (1-\gamma ) (1-\nu ) \nu  \left(-2-3 \nu +3 \nu   ^2\right) d_1 \left(\Delta-f_2^2\right)-\frac{1}{2} (1-\nu ) (3-\nu ) d_2^2
   \left(\Delta-f_{14}^2\right)\\
&-\frac{2 (\gamma -\nu ) (1-\nu )^3 c_3 \left(c_3+ b(1+a)  (1-\gamma ) (1+\nu   )\right) \left(\Delta-f_3 f_4\right)}{1+a}+2 (1-\gamma ) (1-\nu )^2 \nu  c_2^2   \left(\Delta-f_5^2\right)\\
&+\frac{1}{2} (\gamma -\nu ) (1+\nu ) \left(3-2 \nu +\nu ^2\right) c_1^2 c_3
   \left(\Delta-f_6^2\right)
   +(1-\nu ) \nu ^2 \left(c_3-2 (1-\gamma ) \nu \right) d_1 \left(\Delta-f_7^2\right)\\
   &+2
   (a-b)^2 (1-\gamma )^3 (1-\nu ) (\gamma -\nu ) \nu ^3 \left(\Delta-f_8^2\right)+\frac{(\gamma -\nu ) (-1+\nu )^2
   \nu ^2 (\gamma +\nu ) c_3 f_{10}^2}{(1+\nu ) d_1}\biggr], \label{eq:def-g4}\\
g_{5} & =\Delta-\frac{d_1 \left(\Delta-f_1 f_2\right)}{a d_1+\left(1-a^2\right) (1-\gamma ) (1+\nu ) c_1},\\   
g_6 &=f_8^2+\frac{2 (1-\gamma )^2 (1-\nu )^2 \nu  d_1 \left(f_1^2-f_8^2\right)}{(1+\nu ) d_5}+\frac{(1-\gamma )^2   (1-\nu )^4 c_3 \left(f_3^2-f_8^2\right)}{(1+\nu ) d_5},\\
g_7 &= g_8 +\frac{c_3 f_{10}^2 (\nu -1)^2 \nu  \left(\gamma ^2-\nu ^2\right)}{d_1 d_6 (\nu +1)},\\
g_8 &=f_{14}^2-\frac{2 (1-\gamma ) (1-\nu )^2 \nu  d_1 \left(f_{14}^2-f_2^2\right)}{d_6}+\frac{(\gamma -\nu )
   (1-\nu )^2 (1+\nu ) c_1^2 c_3 \left(f_{14}^2-f_6^2\right)}{2 d_6 \nu },\\
g_9&=f_8 f_{18} +\frac{(2+\nu ) f_1 f_{10}}{(a-b) (1-\gamma ) (1-\nu^2 )},\\
g_{10}&=f_{14} f_{18}-\frac{b f_{10} f_{12} (\nu +1) (\gamma -\nu )}{2 d_2 (\nu -1)},\\
g_{11}&=f_{16}  f_{15}-\frac{(1-\gamma ) \left(4+\nu -\nu ^2\right) f_1 f_{10}}{(1+\nu ) \left(2 (a-b) (1-\gamma )^2   (1-\nu ) \nu -b (1+\nu ) d_1\right)}-\frac{(a-b) (1-\gamma )^2 (1-\nu )^2 \left(f_{18} f_8-f_{16} f_{15}\right)}{2 (a-b) (1-\gamma )^2 (1-\nu ) \nu -b (1+\nu ) d_1},\\
g_{12} &= f_{5}f_{13}+\frac{c_3(f_4 f_{17}-f_{5}f_{13})}{2\nu (1-\gamma)(1-ab)},\\
   g_{13}&= f_8 f_{18}-\frac{f_1 f_{10}}{2 (1-\gamma ) (1-\nu ) (b-a)}.
\end{split}
\end{align}

\section{Redefined coefficients in the metric functions}\label{sec:prooffi}
Here we show the useful formula that is used to extend the parameter region beyond $\gamma=1$.
After the use of Eqs.~(\ref{eq:n0sol-b1}), (\ref{eq:n0sol-beta1}) and (\ref{eq:n0sol-a}), one can find that 
odd powers of $\sqrt{1-\gamma}$ do not appear in the following combinations of $f_i$
\begin{align}
   \begin{split}
\tilde{f}_1& := 
   \frac{\sqrt{1-\gamma } f_1}{\sqrt{1+\nu } f_2}=\frac{-2( \gamma - \nu) + \sqrt{2\nu(1-\nu^2 )  (\gamma -\nu )} (\gamma +\nu )}{2 \left(\nu
   +\nu ^3+\gamma  \left(-1+\nu ^2\right)\right)},\label{eq:f1-new}\\
   \tilde{f}_3& :=  \sqrt{-\frac{(1-\gamma ) (1-\nu ) (\gamma   -\nu ) \nu  \left(1-\sigma ^6\right) c_3}{(1+\nu ) (\gamma +\nu ) d_1}} \frac{f_3}{f_2}\\
   & = -\frac{2 \nu  (-\gamma +\nu ) \left(-1+\nu +\sigma ^3+\nu  \sigma ^3\right)+\sqrt{2\nu(1-\nu^2)(\gamma -\nu )}
 \left(-\nu  (\gamma +\nu ) \left(1-\sigma ^3\right)+(\gamma -\nu ) \left(1+\sigma
   ^3\right)\right)}{4 \left(\nu +\nu ^3+\gamma  \left(-1+\nu ^2\right)\right)},\\
   \tilde{f}_4& := 
   \sqrt{-\frac{(1-\nu ) (\gamma -\nu ) \left(1-\sigma ^6\right) c_3}{\nu  (\gamma +\nu ) d_1}}
   \frac{f_4}{f_2}\\
   &=-\frac{\left(\nu +\nu ^3+\gamma  \left(-1+\nu ^2\right)\right) \left(-1+\sigma ^3\right)-2 \nu ^2 \left(1+\sigma
   ^3\right)+\sqrt{2\nu  \left(1-\nu ^2\right)(\gamma -\nu ) } \left(1+\sigma ^3\right)}{2 \left(\nu +\nu
   ^3+\gamma  \left(-1+\nu ^2\right)\right)},\\
 \tilde{f}_5& :=  \sqrt{-\frac{1-\sigma ^6}{\nu  (\gamma +\nu ) d_1}} \frac{c_2 f_5}{f_2} =\frac{\left(\nu +\nu ^3-\gamma  \left(1-\nu ^2\right)\right) \left(-1+\sigma ^3\right)+2 \nu ^2 \left(1+\sigma
   ^3\right)- \sqrt{2\nu(1-\nu^2 )  (\gamma -\nu )} \left(1+\sigma ^3\right)}{2 \left(\nu +\nu
   ^3+\gamma  \left(-1+\nu ^2\right)\right)},\\
 \tilde{f}_6& := 
  \sqrt{-\frac{(\gamma -\nu ) (1+\nu ) \left(1-\sigma ^6\right) c_3}{(1-\gamma ) (1-\nu
   ) \nu  (\gamma +\nu ) d_1}} \frac{f_6 c_1}{f_2} = \tilde{f}_5,\\
 \tilde{f}_7& :=  \frac{b \sqrt{1+\nu } f_7}{\sqrt{1-\gamma } f_2}=\frac{4 \gamma  \nu -\sqrt{2} \nu  (\gamma +\nu ) \sqrt{2\nu  \left(1-\nu ^2\right)(\gamma -\nu ) }+\gamma ^2
   \left(-1+\nu ^2\right)-\nu ^2 \left(3+\nu ^2\right)}{2 (-\gamma +\nu ) \left(\nu +\nu ^3+\gamma  \left(-1+\nu
   ^2\right)\right)},\\
 \tilde{f}_8& := 
\frac{   (b-a) \sqrt{1-\sigma ^6} }{\sqrt{-\nu  (\gamma +\nu )d_1}} \frac{f_8}{f_2}\\
&=\frac{2 \nu  (-\gamma +\nu ) \left(1+\nu -\sigma ^3+\nu  \sigma ^3\right)+ \sqrt{2\nu (1-\nu^2) (\gamma -\nu )
  } \left((\gamma -\nu ) \left(1+\sigma ^3\right)+\nu  (\gamma +\nu ) \left(1-\sigma
   ^3\right)\right)}{4 (1-\gamma ) (\gamma -\nu ) \nu  \left(\nu +\nu ^3+\gamma  \left(-1+\nu ^2\right)\right)},\\
 \tilde{f}_9& := 
   \sqrt{-\frac{(\gamma -\nu ) \nu  \left(1-\sigma ^6\right) c_3}{(1+\nu ) (\gamma +\nu ) d_1}}
   \frac{ c_1   f_9}{f_2}=-(1-\gamma)(\gamma-\nu)\nu \tilf_8,\\
 \tilde{f}_{10}& :=  \sqrt{\frac{(1-\nu ) (\gamma -\nu ) c_3}{(1-\gamma ) (1+\nu )}}
   \frac{f_{10}}{d_1 f_2}=-\frac{\nu -2 \nu ^2+\nu ^3+ \sqrt{2(\gamma -\nu ) \nu  \left(1-\nu ^2\right)}+\gamma  \left(-1+\nu
   ^2\right)}{\nu +\nu ^3+\gamma  \left(-1+\nu ^2\right)},\\
 \tilde{f}_{11}& :=  \frac{f_{11}}{f_2}=\frac{- \sqrt{2\nu(1-\nu^2)(\gamma -\nu ) }+(1+\nu ) \left(-\gamma +\nu +\gamma  \nu +\nu
   ^2\right)}{2 \left(\nu +\nu ^3+\gamma  \left(-1+\nu ^2\right)\right)},\\
 \tilde{f}_{12}& :=  \frac{b f_{12}}{f_2}=\frac{-(\gamma -\nu ) \left(\gamma  \left(-1+\nu ^2\right)+\nu  \left(3+\nu ^2\right)\right)+ \nu  (\gamma   +\nu ) \sqrt{2\nu  \left(1-\nu ^2\right)(\gamma -\nu ) }}{2 (\gamma -\nu ) \left(\nu +\nu ^3+\gamma  \left(-1+\nu
   ^2\right)\right)},\\
 \tilde{f}_{13}& := 
   \frac{b \sqrt{1+\nu } f_{13}}{\sqrt{1-\gamma } f_2}=\frac{-2 (\gamma -\nu ) \nu  (1+\nu )
   +\sqrt{2\nu  \left(1-\nu ^2\right)(\gamma -\nu ) } ((-1+\nu ) \nu
   +\gamma  (1+\nu ))}{2 (\gamma -\nu ) \left(\nu +\nu ^3+\gamma  \left(-1+\nu ^2\right)\right)},\\
 \tilde{f}_{14}& :=  -\frac{\sqrt{1-\sigma ^6} d_2
   f_{14}}{\sqrt{-d_1(1-\gamma ) \nu  (1+\nu ) (\gamma +\nu )} f_2}=\tilf_4,\\
 \tilde{f}_{15}& :=  \frac{\sqrt{\nu(1-\sigma ^6)} \left(d_2-(1-\gamma ) (1-\nu ) (\gamma +\nu )\right) f_{15}}{\sqrt{-d_1(1-\gamma )   (1+\nu ) (\gamma +\nu )}  f_2}= - \tilf_3,\\
 \tilde{f}_{16}& := \sqrt{-\frac{(\gamma -\nu ) \nu  (1+\nu ) (\gamma +\nu
   ) c_3}{(1-\gamma ) (1-\nu ) d_1}} \frac{c_1 f_{16}}{f_2}=\frac{\nu  (\gamma +\nu ) \left((\gamma -\nu ) 
   \left(-1+\nu ^2\right)+ \nu  \sqrt{2\nu   \left(1-\nu ^2\right)  (\gamma -\nu )}\right) \sigma ^3}{(1-\nu ) \left(\nu +\nu ^3+\gamma  \left(-1+\nu ^2\right)\right)
   \sqrt{1-\sigma ^6}},\\
 \tilde{f}_{17}& :=  \sqrt{\frac{(1-\nu )
   (\gamma -\nu ) c_3}{(1-\gamma ) (1+\nu )}} \frac{f_{17}}{f_2}=\frac{2 (\gamma -\nu ) (-1+\nu ) \nu
    +\sqrt{2 \nu \left(1-\nu ^2\right)(\gamma -\nu ) }
   (\gamma  (-1+\nu )+\nu  (1+\nu ))}{2 \left(\nu +\nu ^3+\gamma  \left(-1+\nu ^2\right)\right)},\\
 \tilde{f}_{18}& :=  \sqrt{-\frac{(1-\nu )^3 (\gamma
   -\nu ) \left(1-\sigma ^6\right) c_3}{\nu  (\gamma +\nu ) d_1}}\frac{ f_{18}}{f_2}=\frac{(1-\nu)\sqrt{1-\sigma^6}}{\nu(\gamma+\nu)\sigma^3} \tilf_{16},\\
 \tilde{f}_{19}& := 
   \sqrt{\frac{(\gamma -\nu ) c_3}{(1-\gamma ) \left(1-\nu ^2\right)}} \frac{f_{19}}{f_2} = \frac{(\gamma +\nu ) \left(2 \nu ^2- \sqrt{2\nu   \left(1-\nu ^2\right)(\gamma -\nu )}\right)}{2 \left(\nu
   +\nu ^3+\gamma  \left(-1+\nu ^2\right)\right)}.
      \end{split}
\end{align}

\section{Some proofs for the ergoregion}\label{sec:proof-ergo}
\paragraph{Proof of $\partial_x^2  H (y=-1,x)>0$ and $ H (y=-1,x=\pm1)>0$.}
This follows from
\begin{align}
\begin{split}
\partial_x^2  H (y=-1,x) &= 4 (1-\nu)^2\nu^2(1+\nu) h_1 h_{-6}^2>0,\\
 H (y=-1,x=1)&=16 \nu^2 (1-\nu^2)^3(1-\sigma^6)(1-\lambda)^2\lambda^2>0,\\
 H (y=-1,x=-1)&=16 h_2   (1-\nu )^3 \nu ^2 (1+\nu ) \left(1-\sigma^6\right)(1-\lambda ) \lambda>0.
\end{split}
\end{align}

\paragraph{Proof of $ H (y=-1/\nu,x)<0$.}
This becomes evident by writing it in the form
\begin{align}
 H (y=-1/\nu,x)= -A_1 (1+x)^2-A_2 (1-x)^2 - A_3 (1-x^2),
\end{align}
where
\begin{align}
\begin{split}
&A_1 = \frac{1}{2} (1-\nu )^2 (1+\nu )^3 h_1 h_{+6}^2>0,\\
&A_2 = \frac{1}{2} (1-\nu )^2 (1+\nu ) h_1 h_{+5}^2>0,\\
& A_3=
   \frac{1}{2} (1-\nu )^2 (1+\nu ) \left(2 \left(1-\sigma^6\right) (1+\nu ) h_{-4}^2 h_{+4}^2+h_1 h_{+5}^2+(1+\nu )^2 h_1
   h_{+6}^2\right)>0.
   \end{split}
\end{align}

\paragraph{Proof of $\partial_y^2  H (y,x)<0$.}
This becomes evident by writing it in the form
\begin{align}
\partial_y^2  H (y,x) = - B_1 (1+x)^2- B_2 (1-x)^2 - B_3 (1-x^2),
\end{align}
where
\begin{align}
\begin{split}
&B_1 = (1-\nu ) \nu ^2 (1+\nu )^2 h_1 h_{-6}^2>0,\\
&B_2 = (1-\nu ) \nu ^2 h_1 h_{-5}^2>0,\\
&B_3 = \nu ^2 (1+\nu ) \left(2   \left(1-\sigma^6\right) (1-\nu ) h_{-4}^2 h_{+4}^2+h_1 h_{-5}^2+(1-\nu )^2 h_1 h_{-6}^2\right)>0.
\end{split}
\end{align}
\paragraph{Proof of $\partial_y  H (y=-1,x)+ H (y=-1,x)>0$.}
This becomes evident by writing it in the form
\begin{align}
\partial_y  H (y=-1,x)+ H (y=-1,x) = C_1 (1+x)^2+C_2 (1-x)^2 + C_3 (1-x^2),
\end{align}
where
\begin{align}
\begin{split}
&C_1 = (1-\nu ) \nu ^2 (1+\nu )^2 \left(2 (1-\lambda ) \lambda  (1+\nu ) \left(1-\sigma ^6\right) \left(2 (1-\lambda ) \lambda 
   (1-\nu )+h_1\right)+h_1 h_{-6}^2\right)>0,\\
&   C_2 = (1-\nu ) \nu ^2 \left(2 (1+\nu ) \left(1-\sigma ^6\right) \left(2 (1-\lambda ) \lambda
    (1-\nu )+h_1\right) h_2+h_1 h_{-5}^2\right)>0,\\
& C_3 = 4 \nu ^2 \left(1-\nu ^2\right) \left(1-\sigma ^6\right) \left(\lambda
   ^2+(1-\lambda ) \left((1-\lambda ) \lambda +\left(1+\lambda ^2\right) \nu ^2\right) h_2\right)>0.
\end{split}
\end{align}

\end{document}